\documentclass[prx,twocolumn,eqsecnum,notitlepage]{revtex4-1}
\usepackage{amsmath,amsthm,amsfonts,amssymb}
\usepackage{graphicx}
\usepackage{marginnote}
\usepackage{float}
\usepackage{array}
\usepackage{multirow}
\usepackage[dvipsnames]{xcolor}
\usepackage[normalem]{ulem}
\usepackage{upgreek}
\usepackage{wasysym}
\usepackage{sidecap}
\sidecaptionvpos{figure}{c}
\usepackage{url}
\usepackage{rotating}
\usepackage{gensymb}
\graphicspath{{./figures/}}
\usepackage{hyperref}

\renewcommand{\vec}[1]{{\boldsymbol{#1}}}
\newcommand{\mean}[1]{\left<{#1}\right>}

\renewcommand{\tensor}[1]{{\mathsf{#1}}}
\newcommand{\change}[1]{{\color{black}{#1}}}

\definecolor{dangreen}{RGB}{95, 183, 92}

\definecolor{mattcolour}{RGB}{255, 128, 0}

\renewcommand{\vec}[1]{{\boldsymbol{#1}}}
\renewcommand{\tensor}[1]{{\mathsf{#1}}}

\definecolor{oldgold}{rgb}{0.81, 0.71, 0.23}
\definecolor{ppurple}{rgb}{0.62, 0.0, 0.77}
\definecolor{ceruleanblue}{rgb}{0.16, 0.32, 0.75}
\definecolor{orangeplot}{RGB}{255, 102, 0}
\definecolor{greenplot}{RGB}{0, 100, 0}
\definecolor{greyplot}{RGB}{102, 102, 102}
\definecolor{lightblueplot}{RGB}{70,130,180}
\definecolor{purpleplot}{RGB}{75, 0, 130}
\definecolor{armygreen}{rgb}{0.29, 0.33, 0.13}
\definecolor{darkgreen}{rgb}{.29,.33,.13}

\begin{document}
\title{Cooperative intramolecular dynamics control the chain-length-dependent glass transition in polymers}
\author{Daniel L. Baker$^1$}
	\author{Matthew Reynolds$^1$}
	\author{Robin Masurel$^2$}
	\author{Peter D. Olmsted$^2$}
	\author{Johan Mattsson$^1$}
	\email{k.j.l.mattsson@leeds.ac.uk}
	\affiliation{$^1$School of Physics and Astronomy, University of Leeds, Leeds LS2\,9JT, United Kingdom}
	\affiliation{$^2$Department of Physics and Institute for Soft Matter Synthesis and Metrology, Georgetown University, Washington DC, 20057}
	
	\date{\today}

    \begin{abstract}
The glass transition is a long-standing unsolved problem in materials science. For polymers, our understanding of glass-formation is particularly poor due to the added complexity of chain connectivity and flexibility; structural relaxation of polymers thus involves a complex interplay between intra- and inter-molecular cooperativity. Here we study how the glass transition temperature $T_g$ varies with molecular weight $M$ for different polymer chemistries and chain flexibilities. We find that $T_g(M)$ is controlled by the average mass (or volume) per conformational degree of freedom, and that a `local' molecular relaxation (involving a few conformers) controls the larger-scale cooperative $\alpha$ relaxation responsible for $T_g$. We propose that dynamic facilitation where a `local' relaxation facilitates adjacent relaxations, leading to hierarchical dynamics, can explain our observations including logarithmic $T_g(M)$ dependences. Our study provides a new understanding of molecular relaxations and the glass transition in polymers, which paves the way for predictive design of polymers based on monomer-scale metrics.
    \end{abstract}
\maketitle

\section{Introduction}
As a polymer melt is cooled, the time-scale $\tau_{\alpha}(T)$ characterising its structural ($\alpha$) relaxation increases, eventually leading (in the absence of crystallization) to an arrested out-of-equilibrium amorphous solid called a \textit{glass} \cite{angell2000relaxation}; since dynamic arrest depends on cooling rate \cite{angell2000relaxation}, the glass transition temperature is conventionally defined by choosing $\tau_{\alpha}(T_g)\equiv100\,\textrm{s}$ \cite{dyre2006colloquium}. Variation in chain length and chain flexibility of polymers provide tremendous versatility, tuneability and processability. Thus, polymer glasses are ubiquitous and found in construction materials (aerospace, medical implants, additive manufacturing), in coatings, optical components, and in membranes for controlled transport of ions, gases or electrons. However, our fundamental lack of understanding of glass-formation in polymers often restricts our ability to design materials with optimised performance.

Glass formation is often attributed to the reduced available `free' volume for molecular motion \cite{doolittle1951studies,grest1981liquids,white2016polymer}, the increasing elastic energy required to create this volume \cite{dyre2006elastic}, or a decreasing configurational entropy \cite{adam1965temperature,dyre2006colloquium}. The earliest theory for polymer glasses, due to Fox and Flory \cite{fox1950second}, accounted for additional mobility in short-chain polymers due to the excess free volume around chain ends, leading to a smooth decrease in $T_g$ for decreasing polymer molecular weight; a similar argument was later propoposed in a lattice theory by Gibbs and DiMarzio \cite{Gibbs_1958} based on increasing configurational entropy near the chain-ends. The more recent so-called Generalized Entropy Theory is also a lattice model based on configurational entropy, originally designed for semi-flexible polymers, which includes main- and side-chain bending energies \cite{Dudowicz2005JPCB}; it also results in a Fox-Flory-type $T_g(M)$ behaviour. A similar $T_g(M)$ behaviour was also recently suggested based on the $M$-dependent non-affine contributions to displacements induced upon deformation of a glass \cite{zaccone2013PRL}. 

However, in 1975 Cowie \cite{Cowie1975some}, followed by others \cite{Hintermeyer2008molecular,claudy1983glass}, demonstrated that polymers show a more complex $T_g(M)$ behaviour dividing into three separate regimes, roughly corresponding to an oligomeric short-chain ($\lesssim$ 2 Kuhn steps), an intermediate ($\sim2\textrm{--}10$ Kuhn steps), and a long-chain $M$-independent regime; clearly, a different approach is required to understand these observations. 

A key difference betweeen polymeric and non-polymeric glass-formers is the presence of polymer-specific intramolecular dihedral barriers, which are absent from the theories discussed above. Indeed, computer simulations suggest that the dynamic arrest mechanism (and thus $T_g$) of polymers is significantly influenced by such barriers \cite{colmenero2015polymers,vogel2008macromol}. Moreover, in 1940, Kauzmann and Eyring \cite{kauzmann1940viscous} studied the viscosity of short-chain alkanes and inferred that viscous flow of polymers arise from a succession of elementary intramolecular movements within a `flow segment' with a typical characteristic size of $\sim 5-10$ bonds; this roughly corresponds to the size of the Kuhn `random walk' step size which controls the equilibrium chain statistics \cite{kauzmann1940viscous,siline2002length,jeong2015mass,ding2004does}. Studies later attempted to link the $\alpha$ relaxation, and thus $T_g$, to faster relaxations on the scale of the `flow segment' \cite{boyer1963relation,bershtein1985interrelationship,boyer1976mechanical,boyd1985relaxationMol,boyd1985relaxationExp,ngai1985relation,Roland2005RepProgPhys}, but there is still no consensus about this putative link. 

\change{The relative importance of intra- versus inter-molecular relaxation dynamics in polymers has been inferred from high pressure experiments, which can separate the effects of temperature and volume  \cite{Floudas2006JCP,Mpoukouvalas2009Macromol,Roland2005RepProgPhys,Casalini2007JCP}. The ratio $\mathcal{R}$ between the isochoric activation volume and isobaric activation enthalpy for the $\alpha$ relaxation in  polymers is typically $\mathcal{R}\sim 0.4-0.8$, and a correlation between $\mathcal{R}$ and the monomer volume \cite{Floudas2006JCP} has been identified. $\mathcal{R}=1$ implies that thermal energy, which regulates movements across intramolecular energy barriers and changes in cohesive energies, fully controls the dynamics, whereas $\mathcal{R}=0$ implies that the dynamics are controlled solely by volume changes. For example, $\mathcal{R}=0.73$ was found for poly(methyl methacrylate) (PMMA) \cite{Casalini2007JCP} and $\mathcal{R}=0.63$ for poly(styrene) (PS) \cite{Roland2005RepProgPhys}. Thus, both inter- and intramolecular motions play important roles in polymers, even though the balance between the two is system-dependent, and intramolecular degrees of freedom become increasingly important as the chain-length grows.}

Mode Coupling Theory (MCT) \cite{gotze1992relaxation} successfully captures some phenomenology of both non-polymeric \cite{gotze1992relaxation,Schweizer2007COIS} and polymeric \cite{Frey2015EPJE,Chong2002PRL} glass-formers for $T\gg T_g$, but fails near $T_g$ where thermal activation becomes important. Still, MCT-based analyses of experiments and simulations for $T\gg T_g$ suggest competing arrest mechanisms for polymers \cite{colmenero2015polymers,Frey2015EPJE}. Schweizer and co-workers \cite{saltzman2008large,mirigian2015dynamical} went beyond MCT to incorporate activation barriers for both segmental `cage escape' and elastic deformation of the segment-surrounding matrix \cite{dyre2006elastic}. They treated the polymer melt as a fluid of effective Kuhn sized hard spheres with multiple interaction sites, and based on this approach predicted a $T_g(M)$ smoothly growing with $M$ \cite{mirigian2015dynamical}. This model, however, lacks intramolecular barriers and the cooperativity necessitated by chain connectivity, and \change{relies on an unconventional assumption that properties of the Kuhn step depend on molecular weight.} 
Thus, there is presently neither a satisfactory phenomenological understanding of $T_g(M)$ and its related relaxation dynamics, nor any theory that incorporates intramolecular barriers, chain connectivity and the necessary $M$-dependent variation of inter- and intramolecular dynamics. 

Here, we present extensive experiments on the dependence of the glass transition and associated dynamics on polymer chain-length and chain flexibility, complemented by Rotational Isomeric State (RIS) simulations of chain dimensions. We propose a new framework for understanding the glass transition dynamics of polymers based on cooperative conformational rearrangements involving dihedral motion on a local (conformer) scale. For short chains, these rearrangements spread along the chain, resulting in a secondary ($\beta$) relaxation; for longer chains, chain-folding divides the chain into Rouse-like $\beta$-relaxation `beads'. The structural $\alpha$ relaxation (and thus $T_g$) results, in turn, from propagation of mobility through either inter- or intra-molecular dynamic facilitation \cite{keys2011excitations,ritort2003glassy,palmer1984models,zou2009packing} of the $\beta$ relaxations. The nature of this facilitated dynamic coupling varies with chain-length, separating $T_g(M)$ into three distinct dynamic regimes, as originally proposed by Cowie \cite{Cowie1975some}.

\begin{figure}
{\includegraphics[width=0.48\textwidth]{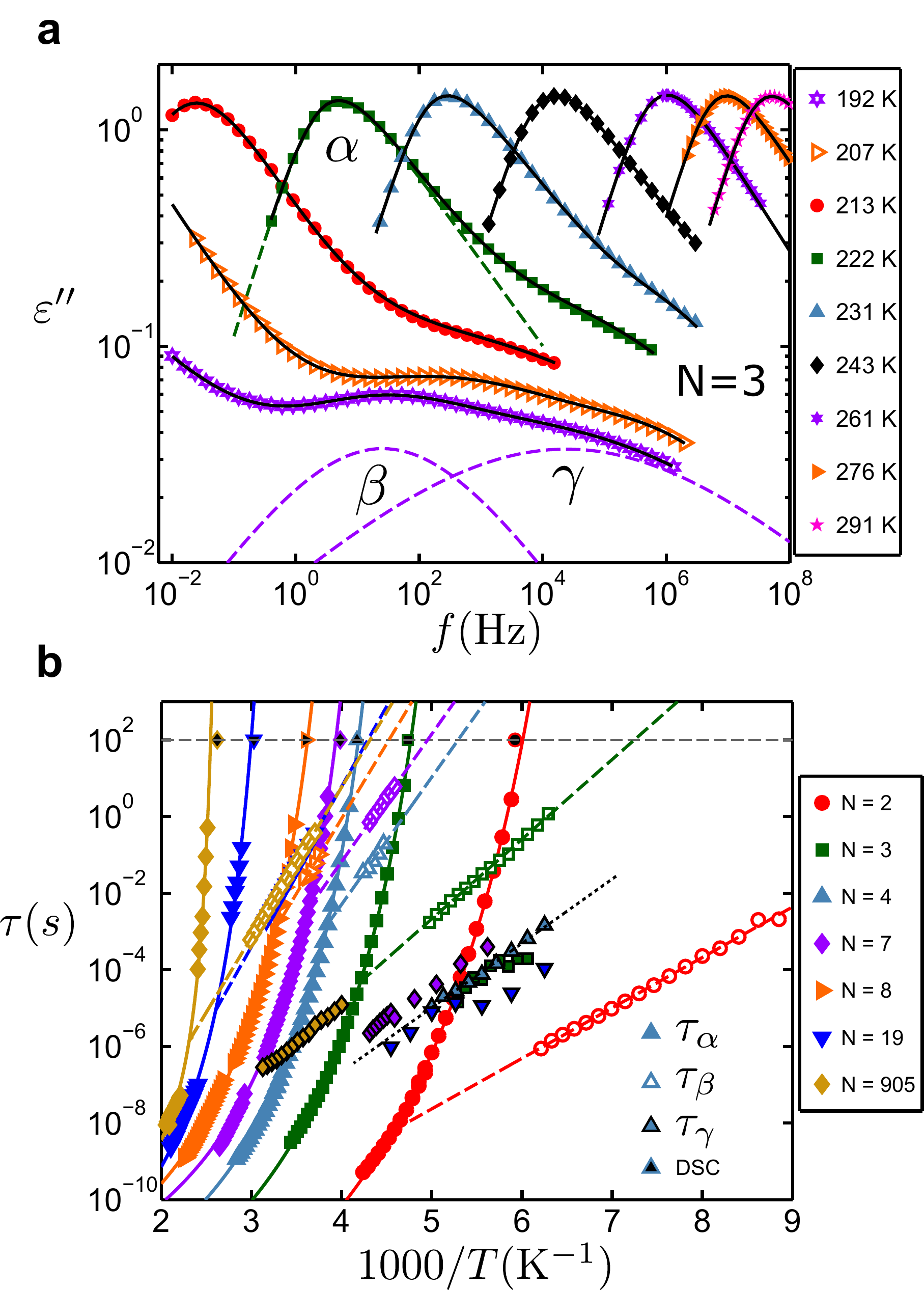}}
\caption{(a) The dielectric loss $\varepsilon''(f)$ for PMMA with $N = 3$ at different $T$. Dashed lines show $\alpha$, $\beta$ and $\gamma$ relaxations, while solid lines are fits to the spectra, as described in Appendix~\ref{app:Methods}. (b) Arrhenius plot showing the characteristic relaxation peak time-scales: $\tau_\alpha$ (filled symbols), $\tau_\beta$ (open symbols) and $\tau_\gamma$ (filled symbols, black outline) for PMMA with $N\in 2-905$. The solid lines are VFT fits to the $\tau_\alpha$ data, the dashed lines are Arrhenius fits to $\tau_\beta(T)$, and the dotted line is an Arrhenius fit to $\tau_\gamma(T)$ for $N=4$ (fits to other $N$ are omitted for clarity). All fitting parameters are tabulated in Table~\ref{tab:VFT_arr_params}. DSC data are shown in black-filled symbols at $\tau = 10^2$s (horizontal dashed line), which defines $T_g$.}
\label{fig:fig1}
\end{figure}

\section{Relaxation dynamics}
We determine the molecular weight ($M$) dependent relaxation dynamics and $T_g$ using Broadband Dielectric Spectroscopy (BDS) and Differential Scanning Calorimetry (DSC) (see Appendix~\ref{app:Methods}). The frequency-dependent dielectric loss $\varepsilon''(f)$ for oligomeric poly(methyl methacrylate) (PMMA) with a degree of polymerisation $N=3$ is shown in Fig.~\ref{fig:fig1}a. We observe three distinct relaxation processes (loss peaks): $\alpha$, $\beta$ and $\gamma$, where $\tau_{\alpha}> \tau_{\beta} > \tau_{\gamma}$. The $\alpha$ relaxation defines $T_g$, while the $\beta$ and $\gamma$ relaxations are typically assigned to molecular rearrangements that include both backbone and side-group rotations \cite{boyer1976mechanical,bershtein1985interrelationship,fried2007sub,kremer2012broadband,Smith2007JPolSci}. 

The peak relaxation times $\tau_p\equiv(2\pi f_p)^{-1}$ are plotted in Fig.~\ref{fig:fig1}b for PMMA with $N\in 2,\ldots,905$. The $\alpha$ relaxation time follows the empirical Vogel-Fulcher-Tammann (VFT) expression $\tau_{\alpha}=\tau_0\exp{DT_0/(T-T_0)}$ typically associated with glass-formation \cite{angell2000relaxation,dyre2006elastic}. Molecular relaxation times $\tau_{\beta},\tau_{\gamma}$ within the glassy non-equilibrium state, on the other hand (Fig.~\ref{fig:fig1}b), typically follow simple Arrhenius behaviour $\tau_i=\tau_{0i}\exp{(\Delta H_i/RT)}$, where $\Delta H$ is the activation enthalpy and $R$ is the gas constant. \change{The secondary relaxations are determined solely in the glassy state, where the analysis is straightforward and the behaviour is Arrhenius.} We determine $T_g(M)$ from the VFT fits by setting $\tau_{\alpha}(T_g)=100$ s, and from DSC by determining the onset of the heat capacity step for a heating rate of 10 K min$^{-1}$(Appendix~\ref{app:Methods}).

\begin{figure*}
\begin{center}{\includegraphics[width=1\textwidth]{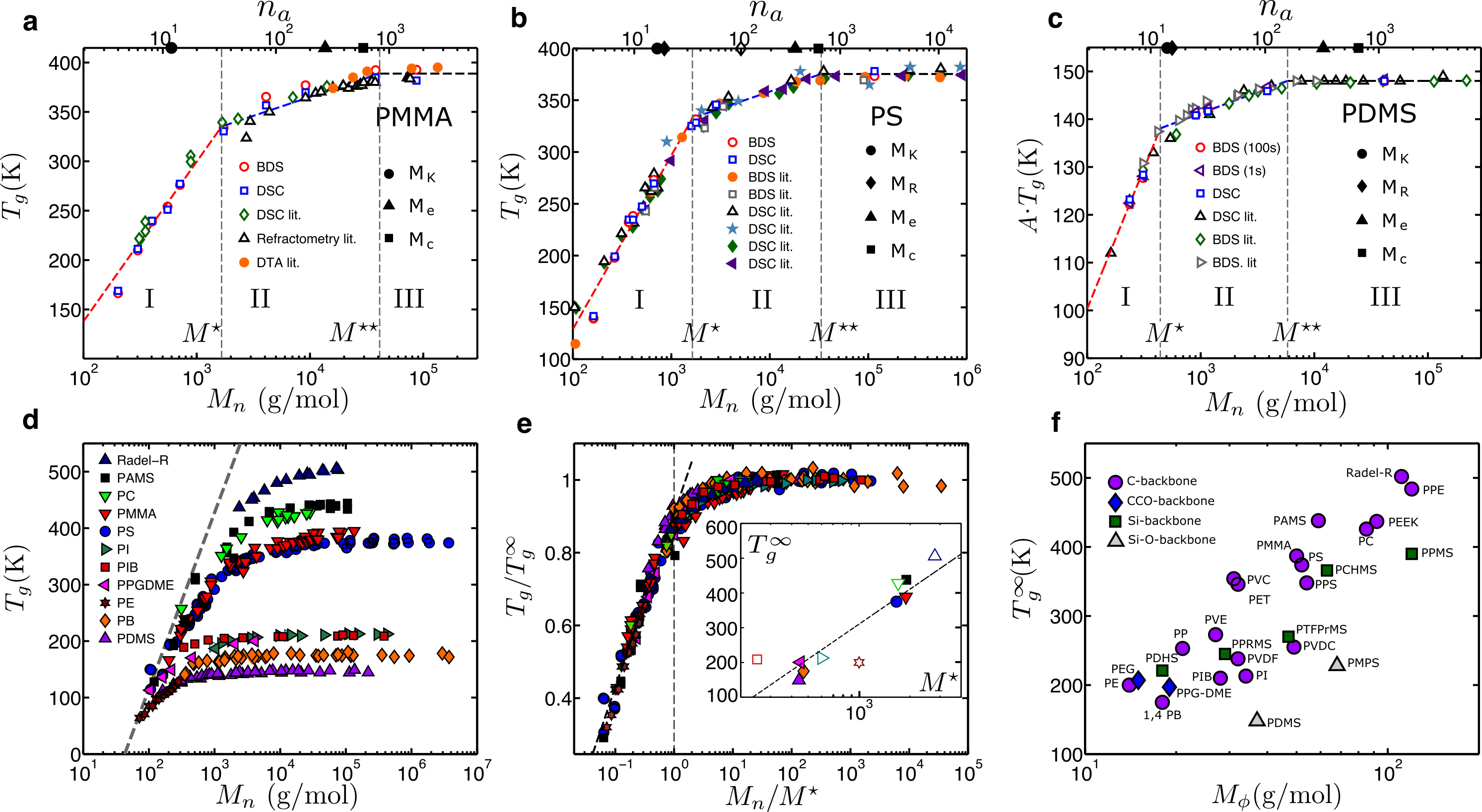}}\end{center}
\caption{\change{(a-c) $T_{g}$ as a function of the number average molecular weight $M_{n}$ and the number of backbone atoms $n_a$.} \change{We use $M_n$ throughout except for Radel-R and one literature data set for PS (Appendix~\ref{app:data}), for which only the weight-averaged molecular weight $M_w$ is known.} Data from BDS and DSC are combined with literature data for PMMA, PS, and PDMS (Appendix~\ref{app:data}). The experimental values of $T_g(M)$ vary slightly due to the nature of the experimental probe, $T_g$ definition, or polymer specification. For PDMS, these differences are more pronounced since $T_g$ is only weakly varying with $M$ ($\Delta T_g\sim 40$K); a scaling factor $A\sim1-1.03$ was thus introduced to collapse different PDMS data sets, where $A$=1 for the $T_g$ data from Ref. \cite{cowie1973molecular}. Since $T_g(M)$ for PDMS in Ref. \cite{ding2004does} was defined for $\tau_\alpha =1\,\textrm{s}$, we plot our $T_g$ data for PDMS both with the standard definition $\tau_\alpha =100\,\textrm{s}$ ($A= 1.024$) and $\tau_\alpha =1\,\textrm{s}$ ($A= 1.003$) to demonstrate that except for an absolute shift in $T_g$ this yields the same shape of $T_g(M)$. The symbols on the upper abscissa denote the Kuhn molecular weight ($\bullet$), the `dynamical' or Rouse molecular weight $M_{R}$ ($\blacklozenge$, or $\lozenge$ for an alternative $M_{R}$ definition \cite{ding2004does}), the  entanglement molecular weight $M_{e}$ ($\blacktriangle$), and the critical molecular weight $M_{c}$ ($\blacksquare$); all values are tabulated in Table~\ref{tab:Ngai_params}. The red, blue and black dashed lines are fits, respectively, to $T_{g}= A_{I,II} + B_{I,II}\log_{\scriptscriptstyle 10} M$ in regimes I and II, and $T_g=T_g^{\infty}$ in regime III. The vertical dashed lines at $M^{\star}$ and $M^{\star\star}$ denote the boundaries between the different regimes. (d) $T_g(M)$ for 11 different polymers (Appendix~\ref{app:data}) \cite{PENote}. The dashed line indicates $T_g(M)$ for `rigid' non-polymeric glass-formers, as discussed in the text. (e) $T_g/T_g^{\infty}$ vs $M/M^{\star}$, where $T_g^{\infty}=T_g(M\rightarrow \infty)$. The inset shows $T_g^{\infty}$ vs $\log_{\scriptscriptstyle 10} M^{\ast}$ 
(open symbols denote polymers with less certainty in $M^{\ast}$ due to data that do not  cover all three regimes). 
(f) $T_g^{\infty}$ vs the mass $M_{\phi}$ per conformational degree of freedom for polymers with different backbone chemistries (Appendix~\ref{app:data}), as signified by different colours and shown in the legend.}
\label{fig:Tg_values_PMMAPS}
\end{figure*}

\section{$T_g$ variation with polymer chain-length and chain flexibility} 

Traditionally, $T_g(M)$ for polymers is described using the Fox-Flory relation, $T_g^{\infty}-T_g\propto 1/M$, typically attributed to the dependence of `free volume' \cite{fox1950second} or configurational entropy \cite{Gibbs_1958} on the number of chain ends ($T_g^{\infty}$ is the long-chain limit of $T_g$). This relation often breaks down for oligomeric $M$ \cite{Gibbs_1958,beevers1960physical,Cowie1975some,bershtein1985interrelationship,novikov2013polymer,Boyer1974Macromol}. Cowie \textit{et al.} \cite{Cowie1975some} demonstrated that $T_g(M)$ can be divided into three regimes separated by molecular weights $M^{\star}$ and $M^{\star\star}$, where
\begin{equation}
T_g\simeq A_{I,II} + B_{I,II}\log_{\scriptscriptstyle 10} M
\end{equation}
in regimes $I$ and $II$, and  $T_g\simeq T_g^{\infty}$ in regime III. This behaviour is demonstrated for PMMA, poly(styrene) (PS), and poly(dimethyl siloxane) (PDMS) in Fig.~\ref{fig:Tg_values_PMMAPS}a-c. 

PMMA and PS are relatively rigid polymers with carbon-based backbones and bulky side-groups, whose $T_g$ values vary significantly with $M$ ($\Delta T_g>200\,\textrm{K}$ for $N\in (2\textrm{--}\infty)$). In contrast, the Si-O backbone of PDMS is much more flexible \cite{fetters2007chain} and has low rotational barriers \cite{agapov2010size}, leading to a much smaller variation in $T_g(M)$ ($\Delta T_g<40\,\textrm{K}$ for $N\in (2\textrm{--}\infty)$; Fig.\ref{fig:Tg_values_PMMAPS}c). Unlike PMMA and PS, PDMS can \textit{also} be described by the Fox-Flory relation (Fig.~\ref{fig:foxfloryfits}), suggesting a less pronounced regime behaviour for more flexible polymers. 
To demonstrate the generality of these observations, $T_g(M)$ data for 11 polymers (Table III; Appendix~\ref{app:data}) are shown in Fig.~\ref{fig:Tg_values_PMMAPS}e to collapse onto the scaling form $T_g/T_g^{\infty}=f(M/M^{\star})$, where $T_g^{\infty}$ and $M^{\star}$ (Appendix~\ref{app:data}) depend on chemistry \cite{Ding2004comment}. Due to the weaker variation in $T_g$ for flexible polymers, to differences in local segmental packing and interactions (e.g. due to side-groups), and to differences in tacticity and $M$-polydispersity, we do not expect this mastercurve to be perfect. We find evidence for a linear relationship (on average) between $T_g^{\infty}$ and $\log_{\scriptscriptstyle 10} M^{\star}$ (inset to Fig. 2e) across all chemistries. Such a relationship implies that \textit{a single chemistry-dependent molecular weight controls the full $T_g(M)$ behaviour} of each polymer.

\begin{table*}[t]
    \centering
    \begin{tabular}{cp{2.7truecm}p{13.0truecm}}
    \hline\hline \parbox{1.5truecm}{Molecular\\ weight $M$} & Type & Description\\
    \hline
        $M_o$ & monomer & Polymer repeat unit.\\[2truept]
        $M_{\phi}$ & conformer & $M$ per conformational degree of freedom.\\[2truept]
        $M_w$ & weight average & Weight-averaged molecular weight. \\[2truept]
        $M_n$ & number average & Number-averaged molecular weight. \\\hline
        $M_K$ & Kuhn & $M$ of a Kuhn step, defined by $L_c=N_K \ell_K, 6R_g^2=N_K \ell_K^2$, 
        and mass $M=N_K M_K$, where $N_K$ is the number of Kuhn steps of length $\ell_K$ and $M=M_K$ in a polymer 
        of contour length $L_c$ and total mass $M$. The experimental input is the polymer radius of gyration 
        $R_g^2$, see \cite{Ding2004comment}. \\[2truept]
        $M_R$ & Rouse (or dynamic bead size)& $M$ of the shortest time- and length-scale Rouse mode. Typically determined as a parameter in fits of chain relaxation spectra to the Rouse model (\textit{e.g.} from rheology, broadband dielectric spectroscopy, or quasi-elastic neutron scattering). \\[2truept]
        $M_e$ & entanglement & Mean $M$ between entanglements.\\[2truept]
        $M_c$ & critical & Minimum $M$ at which entangled dynamics are observed; ($M_c>M_e$).\\\hline
        $M_{\gamma}$ & $\gamma$ relaxation &  Characteristic $M$ of the $\gamma$ relaxation; the shortest 
        length (or mass) scale relaxation relevant to the glass-transition dynamics; involving a few cooperative conformational rearrangements.\\[2truept]
        $M^{\star}$ & $\beta$-relaxation; chain folding & The $M$ that separates regime I from regime II, as defined from the $T_g(M)$ behaviour; $M$ below which the $\alpha$ relaxation (glass transition) has mixed intra and intermolecular characteristic.\\[2truept]
        $M^{\star\star}$ & intra- to intermole\-cular $\alpha$-relaxation & $M$ above which $T_g$ is nearly constant; separates regime II from region III, as defined from the $T_g(M)$ behaviour. \\
        \hline\hline
    \end{tabular}
    \caption{Characteristic molecular weights.}
    \label{tab:MWs}
\end{table*}

Characteristic molecular weights for polymers (black symbols on the upper abscissa of Fig.~\ref{fig:Tg_values_PMMAPS}a-c) include the Kuhn molecular weight $M_K$ (which controls equilibrium flexibility) \cite{floryRIS}, the `dynamic' or Rouse bead molecular weight $M_R$ (which controls unentangled polymer dynamics \cite{ding2004does}), the entanglement molecular weight $M_e$, and the molecular weight $M_c$ at which entanglements become active \cite{fetters2007chain}; the characteristic molecular weights discussed in this work are summarised in Table~\ref{tab:MWs}. It is apparent that none of these molecular weights consistently match either $M^{\star}$ or $M^{\star\star}$. 

Earlier studies \cite{Weyland1970,matsuoka1997entropy,Schneider2005,schut2007glass} suggested a link between $T_g^{\infty}$ and a metric based on the polymer's conformational degrees of freedom (DOF). Accordingly, we determine the molecular weight $M_{\phi}$ per DOF (Appendix~\ref{app:data}) and plot the relation $T_g^{\infty}(M_{\phi})$ in Fig.~\ref{fig:Tg_values_PMMAPS}f for polymers with backbone chemistries based on C (purple), C-C-O (blue), Si (green), and Si-O (silver). These data suggest a linear relation between $T_g^{\infty}$ and $\log_{\scriptscriptstyle 10} M_{\phi}$ for C-based backbones. A simple interpretation is that $M_{\phi}$ parametrizes the displaced volume incurred in conformational motion, so that higher volume conformers correspond to higher $T_g$. Consistent with this, polymers with Si-based backbones have lower $T_g^{\infty}$ for the same $M_{\phi}$, which can be partially accounted for by the higher mass density of Si compared with C. For the Si-O-based polymers in Fig.~\ref{fig:Tg_values_PMMAPS}f (PDMS and PMPS), the larger mass density of Si and O compared with C cannot account for the entire discrepancy. The greater flexibility of the Si-O backbone, lower dihedral barriers, oxygen-specific interaction energies, or the fact that larger backbone angles (143$^{\circ}$ vs 110$^{\circ}$) incur larger volumes during dihedral rotation could all contribute. For the 11 polymer chemistries of Figs.~2d-e we find $M^*\approx 24 M_{\phi}$ (Fig.~\ref{fig:Tg_Vconf}), so that $T_g(M)$ for polymers roughly follows $T_g(M)\simeq T_g^{\infty}(M_{\phi})f(M/M_{\phi})$, where $f(x)$ is a chemistry-independent function.

\begin{figure*}
\begin{center}{\includegraphics[width=1\textwidth]{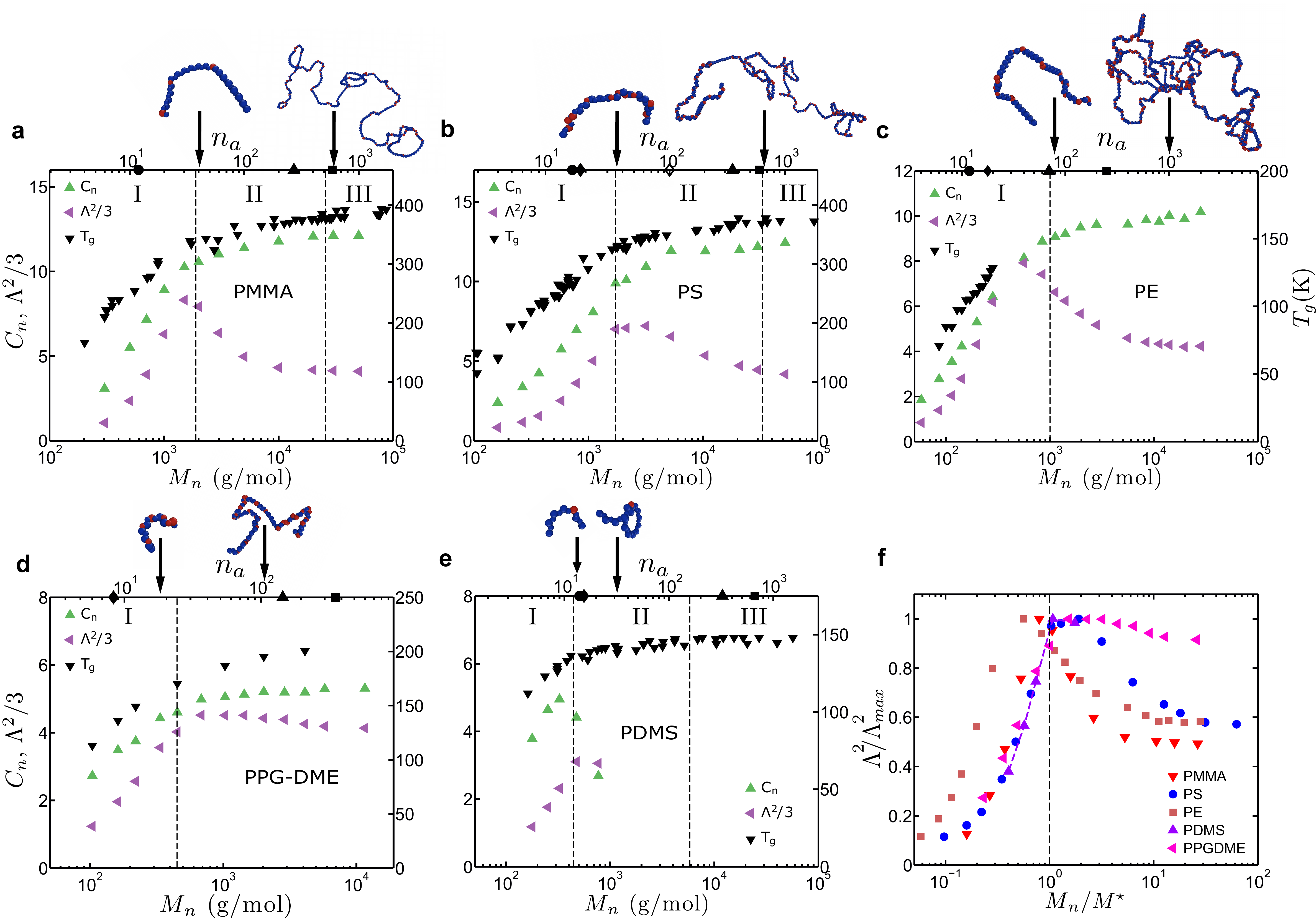}}\end{center}
\caption{(a-e) The glass transition temperature $T_g$ (inverted black triangles) from Fig.~\ref{fig:Tg_values_PMMAPS}; the Flory characteristic ratio $C_n$ (green triangles); the aspect ratio $\Lambda^2=\lambda_3^2/\lambda_1^2$, where $\lambda_3^2$ and $\lambda_1^2$ are respectively the largest and smallest eigenvalues of the average polymer conformational tensor (purple triangles) for: (a) PMMA (b) PS (c) PE, (d) PPG-DME, and (e) PDMS. $C_n$ and $\Lambda^2$ are determined from RIS simulations performed at $T_g(M)$ for PMMA, PS, PPG-DME, or at $T_g=200$ K for PE. The symbols on the top axes ($\bullet$, $\blacklozenge$, $\lozenge$, $\blacktriangle$, $\blacksquare$) identify characteristic molecular weights as in Fig.~2. For each polymer chemistry two typical molecular configurations are shown at the $M$ indicated with arrows; here, bonds in `excited' dihedral states are shown in red. The dashed lines identify the crossovers between regimes I-II (at $M^{\star}$) and II-III (at $M^{\ast\ast}$).(f) $\Lambda^2/\Lambda^2_{\textrm{max}}$ vs $M/M^*$, where
all maxima $\Lambda^2_{\textrm{max}}$  occur for  $M\simeq M^{\star}$.}
\label{fig:RIS_calcs}
\end{figure*}

\section{Chain structure and conformations}
The low-$T$ equilibrium conformation of a single polymer chain has a regular sequence of dihedral angles. For example, low-$T$ polyethylene (PE) is an all-\textit{trans} (rod-like) molecule, whereas low-$T$ isotactic PS is a rod-like helix with alternating \textit{trans} and \textit{gauche} conformations. At higher $T$ the activation of higher-energy dihedral sequences disorders the ground state so that longer polymers are more likely to be disordered and have the prolate ellipsoid shape of long flexible random coil polymers \cite{kreer2001monte}; hence, polymers show $M$-dependent variations in average chain configuration and thus chain shape \cite{NAKAMURA20125}.

To characterise the $M$-dependent variations in chain-structure at $T=T_g(M)$, we use the Rotational Isomeric State (RIS) method \cite{floryRIS} to calculate two metrics of chain structure \cite{jeong2015mass}: (i) the Flory characteristic ratio $C_n=R_{e}^2/n_bl_b^2$, where $R_{e}^2$ is the average squared chain end-to-end distance, $n_b$ is the number of backbone bonds, and $l_b$ is the average bond length; and (ii) the aspect ratio $\Lambda^2=\lambda_3^2/\lambda_1^2$\cite{jeong2015mass}, where $\lambda_3^2$ and $\lambda_1^2$ are respectively the largest and smallest eigenvalues of the average polymer conformational tensor (Appendix~\ref{app:Methods} and \ref{sec:RIS}). A small $C_n$ denotes a more flexible molecule, while $\Lambda^2$ parametrizes the chain shape. Both metrics are calculated at $T_g(M)$ for PMMA, PS, PDMS, and poly(propylene glycol)-dimethyl ether (PPG-DME), or at $T_g=200$\,K for poly(ethylene) (PE) \cite{PENote}.
(results for PE at different fixed $T$ are shown in Appendix~\ref{sec:RISstats}; Fig.~\ref{fig:deltaPE}).

The $M$-dependences of the two metrics and $T_g(M)$ are shown in Fig.~\ref{fig:RIS_calcs}; note that PPG-DME and PDMS are more flexible ($C_{\infty}=5.1; 6.3$) than PMMA, PS and PE ($C_{\infty}=8.2; 9.6; 8.3$). Also, the low energy state of PDMS comprises `loops' of $n_b\sim 24$ bonds \cite{birshtein1959,floryRIS}, which are prohibited for long chains due to steric repulsion; hence, we limit our RIS calculations for PDMS to $n_b\lesssim24$ \footnote{$T_g(M)$ for PDMS was previously compared to the jamming of granular bead chains, in which loops were proposed to control the packing \cite{zou2009packing}.}. We find that $C_n(M)$ for PMMA, PS, PE and PPG-DME behaves similar to $T_g(M)$ \cite{ding2004does,mirigian2015dynamical}, whilst $C_n(M)$ for PDMS has a maximum because of loop formation \footnote{For PDMS it has previously been suggested that $T_g(M)$ behaves similarly to $C_n(M)$ \cite{ding2004does,mirigian2015dynamical,chee1991japs}. In those studies $C_n(M,T)$ was determined for a constant $T>T_g^{\infty}$, where the many \textit{gauche} states inhibit loop formation, and thus the maximum of $C_n(M,T_g(M))$ for PDMS was not apparent.}.

All five polymers display a maximum (Fig.~\ref{fig:RIS_calcs}f) in $\Lambda^2$ near $M^{\star}$ (the maximum is less clear for the more flexible PPG-DME and PDMS; $M^{\star}$ for PE has a higher uncertainty, as discussed in Appendix~\ref{sec:RISstats}), which signifies a change in shape anisotropy, either due to the excited dihedral states leading to chain folding (PMMA, PS, PE, PPG-DME) or loops in the ground state (PDMS), as can be seen by the characteristic chain configurations shown in Fig.~\ref{fig:RIS_calcs}. Hence, the change in dynamical character of $T_g(M)$ at $M^{\star}$ is manifested in structural changes near $M^{\star}$ \cite{jeong2015mass}. Furthermore, the aspect ratio $\lambda^2$ approaches values characteristic of a Gaussian chain $\approx$11.9 \cite{kreer2001monte}, for $M\sim M^{\star\star}$, which also suggests a possible connection between $M^{\star\star}$ and equilibrium chain structure \cite{agapov2009does,Ding2004comment}.

\section{Comparison with $T_g(M)$ for non-polymeric glass-formers}
\begin{figure}[htb]
{\includegraphics[width=0.44\textwidth]{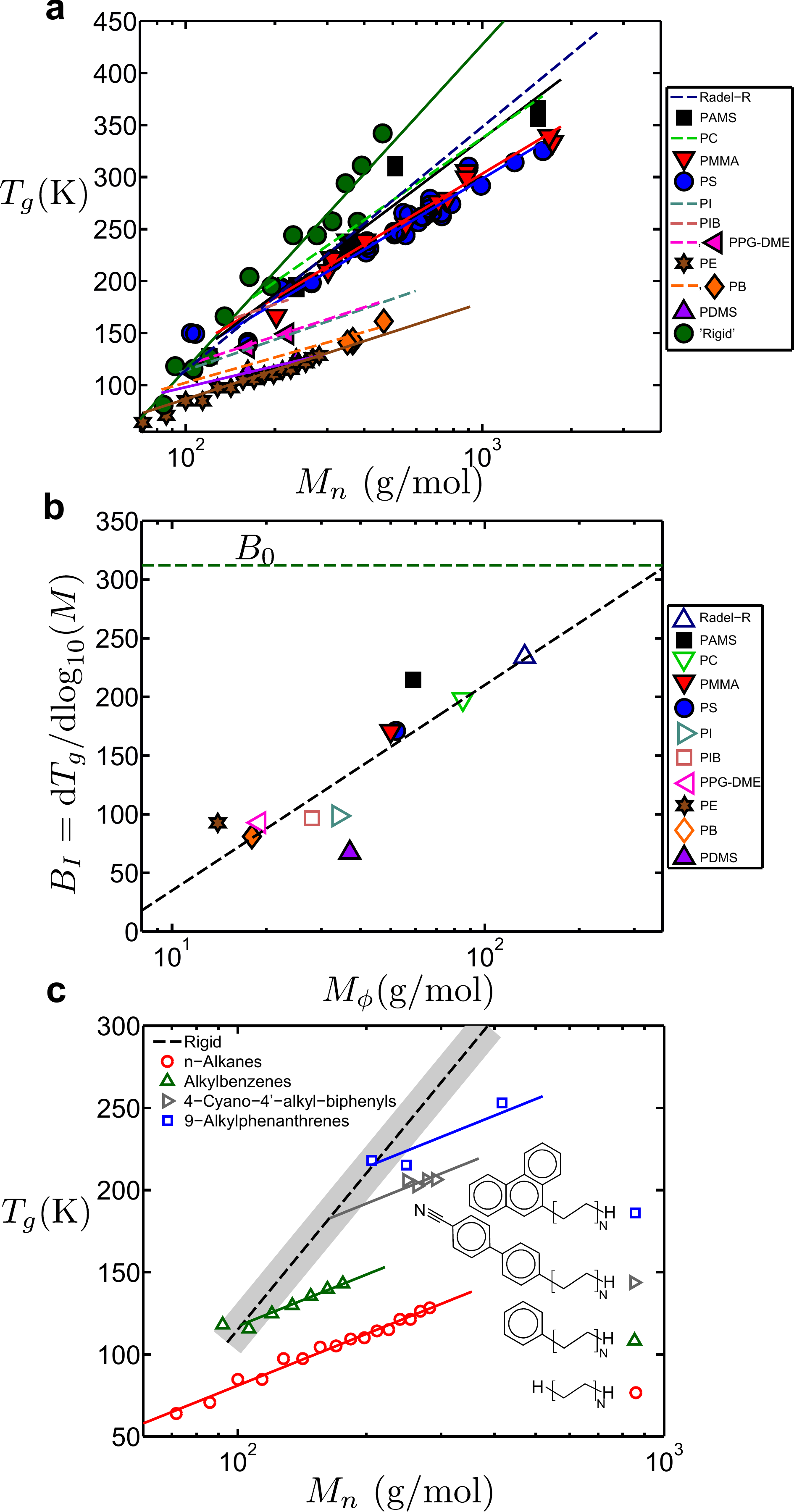}}
\caption{(a) Glass transition temperature $T_g$ vs molecular weight $M$ within regime I for polymers and `rigid' non-polymeric molecules. The solid lines are fits of the form $T_g=A_{I}+B_{I}\log_{\scriptscriptstyle 10} M$. For polymers with no (PC, PI, PIB, Radel-R), or sparse (PB, PPG-DME) data in regime I, the mastercurve in Fig.\ref{fig:Tg_values_PMMAPS}e is used to predict the regime I behaviour which is shown as dashed lines. (b) The chain mass sensitivity $B_{I}= dT_g/d\log_{\scriptscriptstyle 10} M$ vs $\log_{\scriptscriptstyle 10}(M_{\phi})$; $B_0=dT_g/d\log_{\scriptscriptstyle 10} M$ for the non-polymeric `rigid' molecules is shown as a horizontal dashed line. (c) $T_g$ for four different chain-series of n-alkanes (PE) with end-groups of different size. The dashed shaded line marks the $T_g(M)$ behaviour of the `rigid' non-polymeric molecules, as shown in green circles in (a).}
\label{fig:sim_params}
\end{figure}

For polymers, both chain-length and local bulkiness (i.e.~$M_{\phi}$) control $T_g(M)$ \cite{novikov2013polymer,Schmidtke2013JCP}. To separate these two effects, we compare the polymer data to $T_g(M)$ for non-polymeric, carbon-based, mainly aromatic, glass-formers with as few conformers as possible; we denote these as ``rigid'' (we mainly use data from Ref.~\cite{larsen_effect_2011}, see Table~\ref{tab:larsen}). As shown in Fig.~\ref{fig:sim_params}a (green circles), $T_g(M)$ for these non-polymeric liquids is well described by $T_g(M)\simeq A_0 + B_0\log_{\scriptscriptstyle 10} M$, similar to oligomeric glass-formers in regime I. However, the chain mass sensitivity $B_I\equiv dT_g/d\log_{\scriptscriptstyle 10} M$ for oligomers is smaller than $B_0$ for ``rigid'' molecules (green circles in Fig.~\ref{fig:sim_params}a). Moreover, $B_I$ is typically smaller for more flexible oligomers (Fig.~\ref{fig:sim_params}b), and increases with $M_{\phi}$. Thus, $M_{\phi}$ controls both $B_I$ and the absolute value of $T_g$, consistent with the scaling of Fig.~2e. Note that a semi-logarithmic $T_g(M)$ form does not necessarily apply for any system of `rigid' molecules; see Appendix~\ref{app:rigid} for a detailed discussion. In contrast, a change in mass of one of the polymer end-groups shifts the absolute value of $T_g$ \cite{zhang2016dramatic}. This is illustrated in Fig.~\ref{fig:sim_params}c, where n-alkanes (PE) are attached to end groups $i$ with four different masses. $T_g(M)$ of each series $i$ can be described as $T_{g,i}=A_i+B\log_{\scriptscriptstyle 10} M$, with $B$ determined by the conformational character of the alkane chains ($M_{\phi}$) and the intercept $A_i$ increasing with the anchor group mass. Thus, separate control of the absolute $T_g$ and the chain mass sensitivity $B_I$ can be achieved by varying the mass (or volume) of an anchor end-group. 

\section{$M$-dependent activation barriers} \label{sec:activation}
The $\alpha$-relaxation of non-polymeric glass-formers near $T_g$ involves correlated \textit{inter}molecular motion on length-scales $\sim 1\textrm{--}5\,\textrm{nm}$ \cite{hempel2000characteristic}. However, for polymers $T_g(M)$ is strongly linked to the properties of the conformer (Figs.~\ref{fig:Tg_values_PMMAPS}e-f), and the $\alpha$-relaxation has a significant \textit{intra}molecular contribution due to chain connectivity and dihedral motion within the polymer backbone \cite{bershtein1985interrelationship,boyer1963relation,colmenero2015polymers,vogel2008macromol,Paul2004RepProgPhys}. 
\change{The $\beta$ relaxation in polymers, in turn, has been interpreted as having a strongly intramolecular character, as demonstrated by its response to pressure \cite{Mpoukouvalas2009Macromol,Roland2005RepProgPhys}. Thus, we expect activation barriers for conformational relaxations to be of key importance for understanding glass-transition-related dynamics, and $T_g$.}

\begin{figure*}
\begin{center}
{\includegraphics[width=1\textwidth]{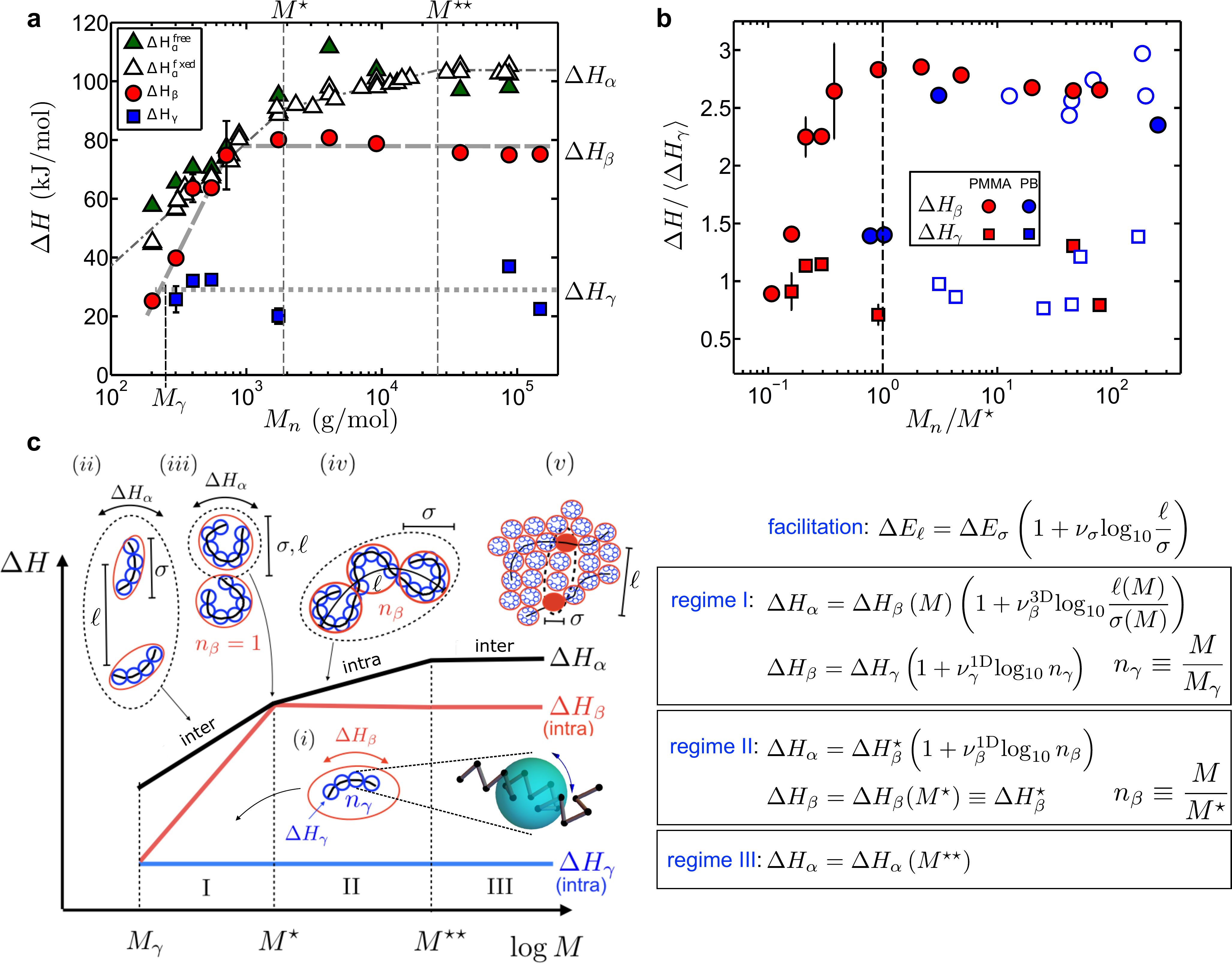}}
\end{center}
\caption{(a) Activation enthalpies $\Delta H_{\alpha,\beta,\gamma}(M)$ for PMMA. $\Delta H_{\alpha}(M)$ data were determined from $T_g(M)$ using a fixed $\tau_0=10^{-12}$s (open triangles), or from the VFT fits in Fig.~1b, as discussed in the text (filled triangles). $M_{\gamma}$, $M^{\star}$, and $M^{\star\star}$ (described in the text) are marked as vertical dashed lines. \change{The horizontal lines signify the average values of $\Delta H_{\gamma}$ (dotted, $M>M_{\gamma}$), $\Delta H_{\beta}$ (dashed, $M> M^{\ast}$), and $\Delta H_{\alpha}$ (dot-dashed, $M>M^{\ast\ast}$).}  
(b) $\Delta H_{\beta,\gamma}(M/M^{\star})$ for PMMA and PB, normalised by their respective averages $\left<\Delta H_{\gamma}\right>$; PB data from literature \cite{arbe1996study,hofmann1996secondary,deegan1995dielectric,richter1992decoupling,korber2017nature,lusceac2005secondary} are shown in open symbols. Error-bars in (a) and (b) represent the standard deviations obtained through least-mean-squares fitting of the secondary relaxation time data shown in Fig.~\ref{fig:fig1}b. (c) A sketch of the activation enthalpy behaviour for $\Delta H_{\alpha,\beta,\gamma}(M)$ for polymers. The equations describe the activation enthalpies due to facilitation in the different regimes, and the five cartoons (i-v) are described in detail in the text.}
\label{fig:fig5}
\end{figure*}

The activation enthalpies $\Delta H_{\beta,\gamma}$ for PMMA, determined from Arrhenius fits within the glassy state, are shown in Fig.~\ref{fig:fig5}a. We find that: 
\begin{enumerate}
\item[(i)] $\Delta H_{\gamma}$ is roughly $M$-independent;
\item[(ii)] $\Delta H_{\beta}\approx\Delta H_{\gamma}$ for $M\simeq M_{\gamma}$, suggesting that the $\beta$-relaxation originates from more `local' $\gamma$-relaxations acting on chain-sections of mass $M_{\gamma}\simeq200$ g/mol (4 backbone atoms, or $\sim4$ backbone conformers) \cite{boyer1976mechanical};
\item[(iii)] $\Delta H_{\beta}$ increases with $M$ for $M_{\gamma}<M<M^{\star}$, and is nearly $M$-independent for $M\geq M^{\star}$, suggesting that the $\beta$-relaxation in regimes II and III involves chain segments of size $\sim M^{\star}$. 
\end{enumerate}

For comparison, $\Delta H_{\beta,\gamma}$ for both PMMA and the more flexible polybutadiene (PB) are shown in Fig.~\ref{fig:fig5}b, normalised by the average $\left<\Delta H_{\gamma}\right>$ for each chemistry. For both polymers, the ratio $\Delta H_{\beta}/\left<\Delta H_{\gamma}\right>\sim 2\textrm{--}3$ in regimes II and III, suggesting a degree of generality. Furthermore, the absolute values of $\Delta H_{\beta,\gamma}$ are lower for the more flexible PB (Appendix~\ref{app:secondary}; Fig.~\ref{fig:fig5new}), consistent with the correlation between the conformational dihedral barrier height and $T_g$ observed in simulations \cite{colmenero2015polymers}. The general nature of the observed $\Delta H_{\beta}(M)$ behaviour is similar to $\Delta H_{\beta}(M)$ of PS, PAMS, PC and PDMS estimated from calorimetry experiments at varying heating rates following a temperature quench and subsequent glassy aging \cite{bershtein1994role,bershtein1985interrelationship}.  

To investigate how the $\alpha$ relaxation (and thus $T_g$) relates to the $\beta$ and $\gamma$ relaxations, we determine the activation enthalpy $\Delta H_{\alpha}(M)$ for the $\alpha$ relaxation at $T_g$ in two different ways, and obtain consistent results \footnote{In this paper we refer to the barrier that accompanies the $\alpha$ relaxation as an “enthalpic barrier”, determined from matching the VFT fit of $\alpha$ relaxation data to an Arrhenius form with a $T$-dependent enthalpy. Any entropic contributions to the barrier act as a pre-factor to the exponential.}. The $\Delta H_{\alpha}(M)$ data shown in open triangles in Fig.~5a were determined from $\tau_{\alpha}(T_g)=\tau_0 \exp{\left[\Delta H_{\alpha}(T_g)/{RT_g}\right]}$ by setting $\tau_0=\tau_0^{\textrm{micr}}=10^{-12}$s, where $\tau_0^{\textrm{micr}}$ is a microscopic time-scale. The data shown in filled triangles are instead determined by equating $\tau_{\alpha}(T_g)$ from the Arrhenius relation above to $\tau_{\alpha}(T_g)=\tau_0\exp{\left[(DT_0/(T_g-T_0)\right]}$ (where $\tau_0$, $D$, and $T_0$ are VFT fitting parameters, see Fig.~1b), which yields $\Delta H_{\alpha}\equiv DT_0RT_g/(T_g-T_0)$) \cite{casalini2011aging}. 

As shown in Fig. 5a, we find that $\Delta H_{\alpha}\approx \Delta H_{\beta}$ for $M\alt M^{\star}$, suggesting a similar nature of the two relaxations near $M^{\star}$. This result suggests that intramolecular rearrangements on the scale of $M^{\star}$ control the $\alpha$ relaxation for $M>M^{\star}$, where the chains are `folded', as shown in Fig.~\ref{fig:RIS_calcs} \cite{jeong2015mass}. Moreover, the complex $M$-dependent interrelationship between $\Delta H_{\alpha}(M)$ and $\Delta H_{\beta}(M)$ identified here for polymers (see Fig. 5a), strongly contrasts with the behaviour observed in non-polymeric glass-formers, where a fixed $M$-independent ratio of $\Delta H_{\alpha}$/$\Delta H_{\beta}$ is typically observed \cite{kudlik1998slow}. Importantly, the activation enthalpies $\Delta H_{\alpha}$ (regimes I and II) and $\Delta H_{\beta}$ (regimes I) appear to depend logarithmically on $M$ for oligomeric and intermediate $M$ chains, as shown in Fig. 5a. A similar logarithmic $M$-dependence has also been observed for the activation enthalpy of the high-$T$ viscosity $\Delta H_{\eta}(M)$ in both experiments and computer simulations \cite{jeong2015mass,siline2002length}. 

\section{Dynamic facilitation} 
 
Logarithmic activation barriers are a hallmark of hierarchical relaxations, and are observed in dynamic facilitation models, in which spatially asymmetric kinetic constraints control relaxation \cite{ritort2003glassy,keys2011excitations}. A simple example is the one-dimensional East model \cite{palmer1984models,ritort2003glassy,keys2013calorimetric}, which describes a chain of `spins' (or \textit{`relaxation beads'} in terminology appropriate for polymer relaxations) where each spin (or relaxation bead) can relax only when its neighbour on one side has relaxed. This simple asymmetric kinetic constraint gives rise to cooperative hierarchical dynamics and the main characteristics of glass-formation, including dynamic heterogeneities and a broad distribution of relaxation times \cite{keys2013calorimetric}. In this class of models, which have been successfully applied to intermolecular (3D) relaxation dynamics in non-polymeric glass-formers \cite{keys2013calorimetric}, relaxation on a length-scale $\ell(T)$ separating mobile spins (relaxation beads) of size $\sigma$ requires an activation barrier $\Delta E_{\ell}=\Delta E_{\sigma} \left[1+\nu \log_{\scriptscriptstyle 10} (\ell/\sigma)\right]$, where $\Delta E_{\sigma}$ is the barrier for a spin flip (bead relaxation). \change{The factor $\nu\sim\mathcal{O}(1)$ has been determined for several different models, and represents the nature and number of pathways available for facilitation \cite{Chleboun2013JStatMech,keys2011excitations,keys2013calorimetric}. For the East model of one-dimensional spin chains $0.72<\nu<1.4$ \cite{Chleboun2013JStatMech}; $\nu\simeq0.25$ was estimated for several small molecule glass formers based on modeling calorimetry data \cite{keys2013calorimetric}; and $\nu\simeq0.35-0.62$ was determined for simulated 2D and 3D glass-formers using several different interaction potentials \cite{keys2011excitations}.}

If we apply this picture to cooperative \textit{intra}molecular (1D) relaxation in polymers, then the activation barrier for relaxing a strand of  $n_{\textrm{bead}}$ beads is given by $\Delta E_{\textrm{strand}}=\Delta E_{\textrm{bead}} (1+\nu^{\textrm{1D}}_{\textrm{bead}}\,\log_{\scriptscriptstyle 10} n_{\textrm{bead}})$, where $\Delta E_{\textrm{bead}}$ is the barrier for relaxing a single bead \cite{keys2013calorimetric,ritort2003glassy}, Here, a ``bead" constitutes the part of the chain that undergoes cooperative rearrangements. The similarity (see Fig. 5a) between the $\beta$ and  $\gamma$ relaxation behaviour within regime I: $\Delta H_{\beta}=\Delta H_{\gamma} (1+\nu_{\gamma}^{\textrm{1D}}\,\log_{\scriptscriptstyle 10} n_{\gamma})$; and the $\alpha$ and $\beta$ relaxation behaviour within regime II: $\Delta H_{\alpha}=\Delta H_{\beta} (1+\nu_{\beta}^{\textrm{1D}}\,\log_{\scriptscriptstyle 10} n_{\beta})$ suggests that similar physical descriptions might be adopted in both cases. Hence, the nearly $M$-independent $\beta$ relaxation in regime II plays a role similar to that of the nearly $M$-independent $\gamma$ relaxation within regime I, where $n_\gamma\equiv M/M_{\gamma}$ and $n_\beta\equiv M/M_{\beta}$ are the numbers of relaxation beads per chain in either regime, and the parameters $\nu_{\gamma}^{\textrm{1D}}$ and $\nu_{\beta}^{\textrm{1D}}$ characterise the facilitation \change{kinetics, which depend on both the material and the facilitation mechanism (our data suggest $\nu_{\gamma}^{\textrm{1D}}\simeq2.0$ and $\nu_{\beta}^{\textrm{1D}}\simeq0.16$).}  

In regime I and III the $\alpha$ relaxation is controlled by intermolecular (3D) facilitation between $\beta$ relaxation beads, and $\Delta H_{\alpha}=\Delta H_{\beta} \left[1+\nu_{\beta}^{\textrm{3D}}\,\log_{\scriptscriptstyle 10} (\ell/\sigma)\right]$, where $\ell$ is the average distance between $\beta$ relaxation beads of size $\sigma$; in contrast, within regime II, the $\alpha$ relaxation is controlled by intramolecular (1D) facilitation between $\beta$ relaxation beads \footnote{Our proposed 3D-facilitation mechanism, where the $\alpha$ relaxation arises from facilitation of $\beta$ relaxations, should apply also to non-polymeric glass-formers. Recent computer simulation work used a swap Monte Carlo technique to reach temperatures closer to $T_g$ in equilibrium than previously achieved and observed a so-called excess wing on the high-frequency side of the $\alpha$ relaxation \cite{Guiselin2020arxiv}. This feature is often observed in molecular glass-formers and is typically interpreted either as a secondary $\beta$ relaxation in the vicinity of the $\alpha$ relaxation, or as a generic feature of the $\alpha$ relaxation itself \cite{Lunkenheimer2000ContPhys,mattsson2003chain}. Interestingly, this study found direct evidence that the relaxation contribution within the `excess wing' generates the $\alpha$ relaxation through dynamic facilitation. This observation thus appears to be consistent with the general nature of our proposed picture. We also note that dynamic phase transitions with links to dynamic facilitation have recently been suggested in computer simulation studies to explain interfacial effects in thin polymer films \cite{Ivancic2020PNAS}}.

We propose the following scenario for the observed hierarchy, as illustrated in Fig.~5c. In regime I, \textit{intra}molecular dynamic facilitation between $\gamma$ relaxation beads, governed by intramolecular barriers, induces the $\beta$ relaxation (sketch (i); Fig.~5c); while the $\alpha$ relaxation arises from \textit{inter}molecular facilitated dynamics, on a length-scale $\ell(T)$ set by the average distance between $\beta$ relaxations of size $\sigma$ that increase with $M$ (Fig.~5c(ii)). The semilogarithmic $M$ dependence of $T_g$ and thus $\Delta H_{\alpha}$ in regime I follows from the $\log_{\scriptscriptstyle 10}M$ dependence of $\Delta H_{\beta}$ modulated by the $M$ dependence $\ell(M)/\sigma(M)$. The ratio $\ell/\sigma$ decreases with $M$ and for $M\sim M^{\star}$, the data suggest $\ell\approx \sigma$, leading to effectively \textit{intra}molecular dynamics where $\Delta H_{\alpha}\approx \Delta H_{\beta}$ (Fig.~5c(iii)). Subsequently, within regime II the $\alpha$ relaxation arises from \textit{intra}molecular dynamic facilitation between $\beta$ beads (Fig.~5c(iv)), each with an essentially fixed size $\sim M^{\star}$ and activation barrier $\Delta H_{\beta}$\footnote{we note that the intramolecular facilitated relaxation mechanism suggested here could provide a much sought-for mechanism to explain dramatic effects on $T_g$ with film thickness (and $M$) in thin polymer films, as observed for both PS and PMMA \cite{forrest2001glass,napolitano2017}.}

For long enough chains, the \textit{intra}molecular $\alpha$ relaxation mechanism becomes kinetically unfavorable (at $M^{\star\star}$), so that within regime III the $\alpha$ relaxation occurs through effectively \textit{inter}molecular facilitation between the $\beta$ beads, akin to the $\alpha$ relaxation within regime I (Fig.~5c(v)). There is no general link between $M^{\star\star}$ and the onset of entanglements at $M_c$ for polymers \cite{agapov2009does}. However, for polymers with significant side-chains and thus large packing lengths \cite{fetters1999packing} such as PMMA and PS (Table.~\ref{tab:Ngai_params}, Fig.~\ref{fig:Tg_Vconf}), we speculate that the onset of entanglements is likely to hinder the \textit{intra}molecular $\alpha$ relaxation dynamics of regime II; such an effect would be consistent with the observations that $M^{\star\star}\approx M_c$ for PMMA and PS, as shown in Fig.~2a-b. We also note that $M^{\star\star}$ could be related to the onset of Gaussian chain statistics, as shown in Fig. 3.

\section{Conclusion and Outlook}

In this work, we have mapped out the relaxation dynamics and chain conformational structure, as a function of chain-length, for polymers characterised by different chain flexibilities and local packing properties. We show that the molecular weight ($M$) dependent glass-transition temperature $T_g(M)$ for polymers, can be collapsed onto a mastercurve using a \textit{single} chemistry-dependent parameter, which balances local conformational dynamics with packing. We find that the average molecular weight or volume per conformational degree of freedom, the \textit{conformer}, are relevant choices for this parameter. Moreover, a chain-length-dependent interplay between \textit{inter} and \textit{intra}molecular relaxation dynamics results in a delineation of $T_g(M)$ into three characteristic dynamic regimes, where two of these are well characterised by a logarithmic $M$-dependence. We find that the structural $\alpha$ relaxation, and thus $T_g$, is linked to more local secondary $\beta$ and $\gamma$ relaxations according to a hierarchical scheme, where the $\gamma$ relaxation, involving a few conformers, acts as a fundamental ``excitation". Finally, we demonstrate that Dynamic Facilitation can explain both the relation between the molecular $\alpha$, $\beta$, and $\gamma$ relaxations observed in polymers, and the observed logarithmic $T_g(M)$ behaviours, as a direct result of hierarchical relaxation dynamics arising naturally due to dynamic facilitation. 

The facilitation picture we propose suggests a new paradigm that couples local cooperative intramolecular motions on the scale of a few bonds to longer length-scale intramolecular and/or intermolecular motions, in turn resulting in structural relaxation. \change{We suggest that facilitation can occur \textit{along} the chain, as well as between chains, because of the cooperativity necessitated by substantial intramolecular barriers. Each form of facilitation can be expected to have different character and parameters (such as the facilitation exponent $\nu$), which depend on the details of the cooperative pathways available for the relaxation \cite{SollichEvansPRL1999}.  We note, however, that we do not expect prominent intramolecular facilitation in flexible polymers for which the activation barriers for main-chain bond reorientations are $\sim \mathcal{O}(k_B T)$, thus enabling smooth reorientation.} Moreover, our study has been limited to a relatively simple class of polymers; systematic variation of side-groups, inclusion of more complex backbones (such as conjugated polymers), or co-polymerization, form natural extensions to this work. Importantly, our results could pave the way for efficient predictive design of polymers based solely on the monomer structure that controls the dynamics on the conformer length-scale. 
\\
\\
Source data files are available at the University of Leeds Data Repository at \url{https://doi.org/10.5518/827}.

\acknowledgements
 We acknowledge the Engineering and Physical Sciences Research Council (EPSRC) for financial support (EP/M009521/1, EP/P505593/1, and EP/M506552/1). P.D.O. and R.M. thank Georgetown University and the Ives Foundation for support. We thank Johan Sj{\"o}str{\"o}m for discussions and for assistance with high-frequency BDS experiments. We also thank Ralph Colby, Juan P Garrahan and Peter Sollich for helpful discussions.

\appendix
\section{Methods}\label{app:Methods}
\subsection{Broadband Dielectric Spectroscopy}
Broadband Dielectric Spectroscopy (BDS) measurements were performed to determine the complex permittivity, $\varepsilon^*(f)=\varepsilon'(f)-i\varepsilon''(f)$ over a frequency range of $10^{-2}\leq f\leq10^{6}$ Hz using a Novocontrol Alpha-A dielectric analyser, and over a frequency range of $10^{6}\leq f\leq10^{9}$ Hz using an Agilent 4219B RF Impedance analyser. For the lower frequency range, the samples were measured between two circular electrodes (20 or 40 mm diameter) with a spacing of 100-200$\upmu$m, and for the higher frequency range between two circular electrodes (10 mm diameter) with a spacing of 100$\upmu$m. 
The temperature was controlled using a Novocontrol Quatro system with an accuracy of 0.1 K. The complex permittivity measured at a particular temperature was analysed using a sum of contributions from molecular relaxations as well as a contribution to the dielectric loss $\varepsilon''$ from ionic dc-conductivity ($\sigma_{\textrm{dc}}$) when observed within the experimental window, $\varepsilon^*= -i\sigma_{\textrm{dc}}/(2\pi f\varepsilon_0)$. Each relaxation contribution was described using the Havriliak-Negami (HN) expression \cite{kremer2012broadband}, 
\begin{equation}
\varepsilon^* = \varepsilon_{\infty}+\frac{\Delta\varepsilon}{(1+(i2\pi f\tau_{HN})^{m})^{n}},
\end{equation}
where $\Delta\varepsilon$ is the dielectric strength, $\varepsilon_{\infty}$ is the high-frequency permittivity, $\tau_{HN}$ is a characteristic relaxation time-scale. The parameters $m$ and $n$ describe the shape of the relaxation response; $m$ and $mn$ are the power law exponents the low- and high-frequency sides of the loss peak respectively. The $\beta$ and $\gamma$ relaxations were generally well-described using symmetrically stretched (Cole-Cole) loss peaks ($n=1$) for which the loss peak relaxation time is $\tau_p = \tau_{HN}$. The $\alpha$-relaxation loss peaks, as well as the $\beta$ relaxation for the highest $M$ PMMA, on the other hand, were asymmetrically stretched and $\tau_p$ was instead obtained from $\tau_{HN}$, $m$ and $n$ using a previously derived expression \cite{kremer2012broadband}. 

\subsection{Differential Scanning Calorimetry}
Differential Scanning Calorimetry (DSC) measurements were performed using a TA instruments Q2000 heat flux calorimeter, using a liquid nitrogen cooling system for the temperature control. The polymer samples (weight $\sim$ 10\,mg) were prepared in hermetically sealed aluminium pans, and measurements of the specific heat capacity as a function of temperature were performed for heating/cooling rates of 10 K/min. The glass transition is manifested as a step in the specific heat capacity, and the reported $T_{g}$ values were determined on heating from the onset temperatures corresponding to the steps.

\section{Rotational Isomeric State (RIS) formalism and calculations} \label{sec:RIS}
Flory's Rotational Isomeric State (RIS) theory \cite{floryRIS} is used to calculate conformational chain properties such as the average end-to-end distance ${R}_{e}$, the gyration radius $R_g$, and the gyration tensor $\boldsymbol{\tensor{Q}}$. Polymer chains comprise $n_a$ backbone atoms, $n_b=n_a-1$ backbone bonds, and $n_d=n_a-3$ dihedral angles. Polymer backbone bonds typically have one, two or three accessible dihedral angles $\phi_i$ per monomer; each $\phi_i$ is assumed to have discrete dihedral states. For example, in PE the nomenclature \textit{trans} commonly denotes a dihedral angle of $\phi=$180$^{\circ}$, leading to a planar zig-zag backbone for the ground state structure, while \textit{gauche} refers to $\phi=\pm60^{\circ}$, which leads to a non-planar backbone (a different convention for the dihedral angle is sometimes used, where \textit{trans} refers to $\phi=0^{\circ}$ and \textit{gauche} to $\phi=\pm120^{\circ}$). The conditional probability that a dihedral angle $\phi_i$ is followed by an angle $\phi_{i+1}$ is proportional to the matrix element $U_{\phi_i,\phi_{i+1}}$ of a so-called transfer matrix ${\boldsymbol{\tensor{U}}}$ \cite{floryRIS}, which is a square matrix with rank given by the number of dihedral angles for a given bond. By using this matrix, the probability ${\cal P}$ of finding an entire sequence of dihedral states $\Phi=\left\{\phi_1, \phi_2, \ldots,\phi_{N_d}\right\}$ can be calculated as the joint probability of finding $\phi_1$ next to $\phi_2$, $\phi_2$ next to $\phi_3$, and so on:
\begin{align}
{\cal P}_{\{\Phi\}}&=\frac{1}{\cal Z}U^{\ast}_{1,\phi_1} \prod_{i=1}^{n_d-1} U_{\phi_i,\phi_{i+1}}\label{eq:MCProb}\\
{\cal Z\/} &= \vec{q}_0\cdot\boldsymbol{\tensor{U}}^{\ast}\cdot\boldsymbol{\tensor{U}}^{n_d-1} \cdot\vec{q}_1\\
\vec{q}_0&=(1,0,\ldots),\qquad\vec{q_1}=(1,\ldots 1),
\end{align}
where ${\cal Z\/}$ is the partition function, $\boldsymbol{\tensor{U}^{\ast}}$ is the transfer matrix for the first dihedral angle, the vector $\vec{q}_0$ defines the plane of the initial two bonds to be the \textit{trans} plane, and the vector $\vec{q}_1$ ensures that all states are counted. 
The two first bonds of the chain define the initial plane from which subsequent bonds are accessed via the dihedral angles and bond angles for the specific polymer.

\begin{figure*}
\includegraphics[width=\textwidth]{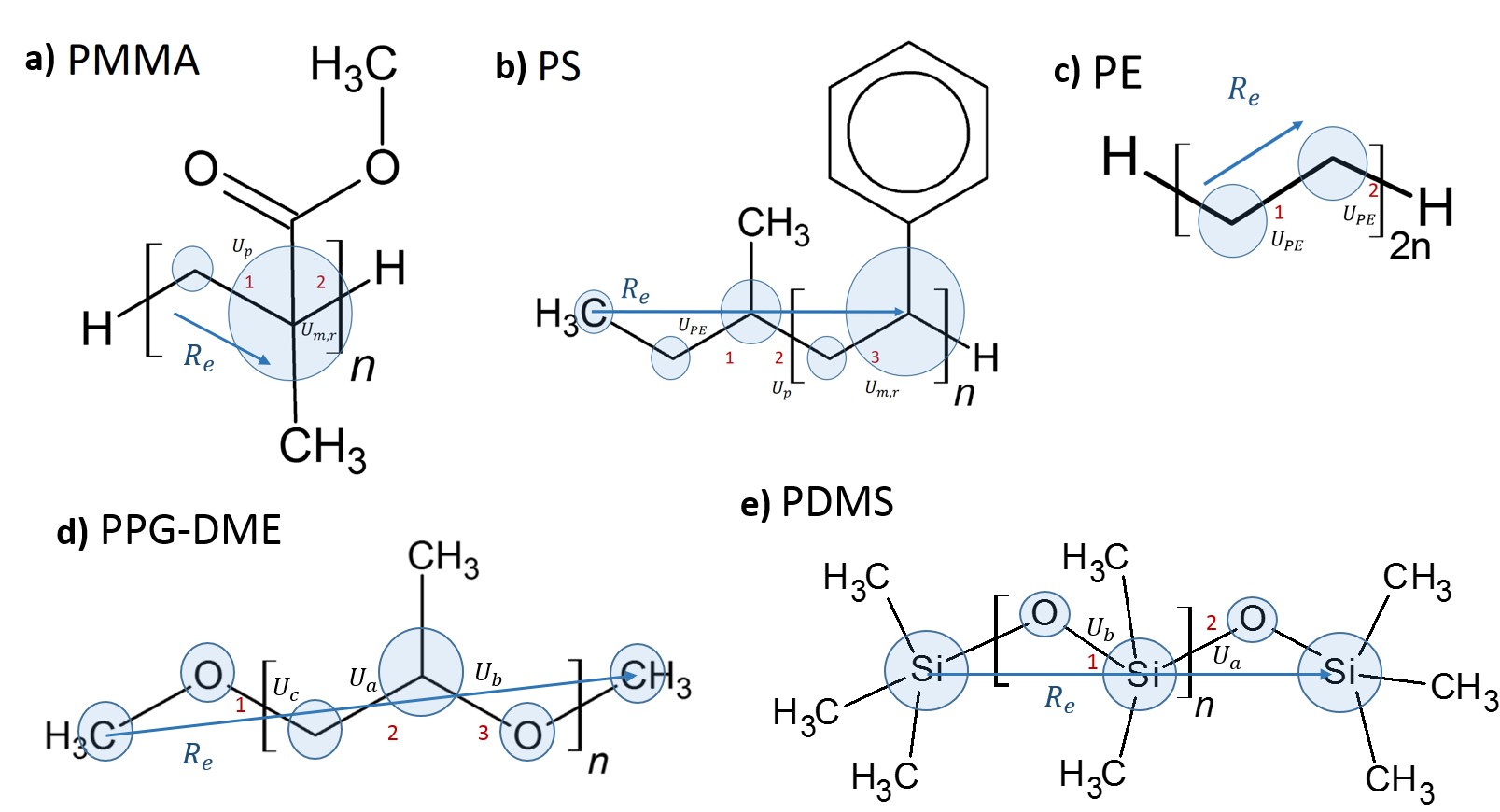}
\caption{The polymer structures used in the RIS calculations. The end-to-end vectors $\vec{R_e}$ are marked with blue arrows, red numbers label bonds about which dihedral angles rotate, and  blue circles show the atoms used to calculate the gyration tensor, with a spherical volume equal to the volume of all included atoms. The number of monomers is $n$. The volume of the backbone atom and its sidegroups is assumed to be localized on the backbone atom. For example, the blue circle centered on the Si atoms at the ends of PDMS represents the volume of Si and three CH$_3$ groups; while interior volumes comprise a Si and two CH$_3$ groups.}
\label{fig:Structures}
\end{figure*}

For polymers with multiple distinct dihedral angles per monomer this can be generalised using different transfer matrices $\boldsymbol{\tensor{U}}$ for each dihedral in the monomer; for a polymer such as PDMS one finds (see Fig.~\ref{fig:Structures})
\begin{align}
{\cal Z\/} &= \vec{q}_0\boldsymbol{\tensor{U}}^{\ast}\cdot\left(\boldsymbol{\tensor{U}}_b\boldsymbol{\tensor{U}}_a\right)^{N-1} \cdot\vec{q}_1,
\end{align}
where $N$ is the number of momomers in the chain (degree of polymerization).

The RIS method ignores interactions along the chain of longer-range than those between adjacent dihedral angles. Thus, very large side groups or charged polymers are poorly described, as well as excluded volume effects resulting from distant monomers, which is generally a good approximation due to the screening of excluded volume in melts \cite{rubinstein2003polymer}. A notable exception encountered here is the ground state configuration of PDMS, which involves a loop of about 24 bonds (12 monomers) in size, which will overlap in the ground state. As described in the main manuscript, at temperatures near $T_g$ for low $M$ these loops are not strongly disordered by  excited \textit{gauche} states \cite{floryRIS,birshtein1959}. 



Monte Carlo (MC) simulations are performed to sample the RIS distribution of dihedral angles along the chain. A MC step controls the transitions between rotational isomeric states $\Phi$. Each Monte Carlo step corresponds to flipping a randomly chosen dihedral angle to a different dihedral angle,  $\phi_i\rightarrow\phi'_i$ (e.g. from \textit{trans} to \textit{gauche}). The new dihedral conformation is then accepted or rejected using the Metropolis algorithm, with probabilities given by Eq.~\ref{eq:MCProb}. We typically perform 100,000 MC attempts in order to equilibrate a molecule at a given temperature, and conformational averages are  calculated using the next 200,000 steps. Since we use $n_d\simeq1-2000$ dihedral angles, the entire polymer is sampled from 100 to 200,000 times during the equilibration stage, depending on $n_d$ and thus the polymer length. 

The spatial position $\vec{r}_i$ of backbone atom $i$ is given by 
\begin{align}
\vec{r}_i&=\vec{r}_1+\sum_{j=1}^{i-1}\vec{b}_j,
\end{align}
where $\vec{r}_1$ is the first atom, and bonds are transformed along the chain by
\begin{equation}
\vec{b}_i=b_i\boldsymbol{\tensor{T}}_i\cdot\hat{\vec{b}}_{i-1},
\end{equation}
for $i>2$, where $b_i$ is the bond length, $\hat{b}$ is a unit bond vector, and the bond transformation matrix is given by
\begin{equation}
\boldsymbol{\tensor{T}}_i=\left( \begin{array}{ccc}
\cos\theta_i & \sin\theta_i & 0 \\
\sin\theta_i \cos\phi_i & -\cos\theta_i \cos\phi_i  & \sin\phi_i \\
\sin\theta_i \cos\phi_i  & -\cos\theta_i \sin\phi_i & -\cos\phi_i \end{array} \right),
\end{equation}
where $\theta_i$ is the bond angle, and $\phi_i$ is the dihedral angle. 

We study polymers whose repeat unit comprises two bonds (PS, PI, PDMS, \text{etc}), a single bond (PE), or three bonds (PPG-DME). Each distinct bond in a repeat unit has a bond length, a bond angle, and a set of dihedral angles.  The polymers PMMA, PS, and PPG-DME are stereoisomeric polymers and thus have \textit{tacticity}, \textit{i.e.} asymmetric sidegroups that lead to  local chiral symmetry (right or left handedness) depending on the sequence of sidegroups (Fig.~\ref{fig:Structures}). We study atactic polymers, which corresponds to a disordered mixture of chirality along the chain due to random right or left positions of the side groups. This tacticity can be quantified by either (i) the fraction of meso (two successive side groups in the same position) or racemic (two successive side groups in opposite positions) diads in polymers with a single atom between sidegroups (PMMA, PS); or (ii) the total proportion of right handed side groups in polymers separated by two atoms (PPG-DME). In the latter case a `meso' or `racemic' sequence does not change the RIS parameters because of the separation, but they do change the chain structure. We specify an average tacticity by $p_{\textrm{meso}}$ (the proportion of diads that are meso) or $p_{R}$ (the proportion of side groups that are right-handed). $n_{st}$ random stereochemical sequences are generated consistent with the average tacticity, with $n_{st}$ sufficiently large to lead to a statistically representative set of stereochemistries;  conformational averages are then performed for each stereospecific sequence. 

The polymer structures used in the RIS calculations are shown in Fig.~\ref{fig:Structures}; for each structure, the bonds which dihedral angles refer to are numbered in red and the corresponding transfer matrices $\boldsymbol{\tensor{U}}$ are tabulated in Table~\ref{tab:RISParameters}. We have used data from the literature, collected and referenced in  Table~\ref{tab:RISParameters}. The matrices $\boldsymbol{\tensor{U}}$ depend on parameters $(\eta,\sigma,\sigma', \omega, \dots)$, which are taken to have the Arrhenius form, e.g. $\eta=\Gamma_{\eta}e^{{-H_{\eta}}/{kT}}, \sigma=\Gamma_{\sigma}e^{{-H_{\sigma}}/{kT}}, \ldots$ at \textit{all} temperatures. The Arrhenius activation barriers e.g. $H_{\eta}$ were typically calibrated by optimization of $R_g^2$ and $\textrm{d}\ln R_g^2/\textrm{d}T$ for RIS-modeled chains to match the corresponding values from experimental data \cite{Rehahn1997} at a chosen calibration temperature $T_{\textrm{cal}}$ (Table \ref{tab:RISParameters}). In our MC simulations, we used RIS parameters calibrated using data on melts, for PS and PE, where such data are available; for PMMA, PDMS and PPG-DME, however, we instead use data calibrated on theta solutions. We expect the $M$-dependent trends to be the same for theta solutions and for melts, even though the overall chain dimensions vary slightly depending on the nature of the packing between the specific polymer and solvent(s) \cite{boothroyd1993temperature,krishnamoorti2002melt}. 
\begin{turnpage}
\begin{table*}
\begin{tabular}{p{0.6truein}p{2.7truein}p{1.0truein}p{1.1truein}p{2.25truein}cc}
\hline\hline
Polymer & \hfil Transfer Matrices\hfil & \hfil Enthalpies \hfil  & \hfil Bond lengths \hfil & \hfil Dihedral\hfil   & $T_{\textrm{cal}}$&Ref.\\
 & \hfil \hfil & \hfil {\small\phantom{x} kJ/mol} \,\, --  \hfil  & \hfil and angles \hfil & \hfil angles\hfil  & K&
 \\\hline\\[-20truept]
\raisebox{-0.3truein}{PE} & 
{\begin{align*}
\boldsymbol{\tensor{U}}&=\left( \begin{array}{ccc}
1 & \sigma & \sigma \\
1 & \sigma  & \sigma \omega \\
1 & \sigma \omega & \sigma \end{array} \right)
\end{align*}
}&
{\begin{align*}\begin{array}{ccc} 
\hline
& H & \Gamma\\\hline
\sigma & 2.1 & 1\\
\omega & 8.6 & 1 \\\hline
\end{array}
\end{align*}
}
&
{\begin{align*}
\textrm{C-C}&=153\,\textrm{pm}\\
\textrm{C-C-C}&=112^{\circ}
\end{align*}
}
&
{\begin{align*}
\begin{array}{ll}
t & =180^{\circ}\\
g^{+} &= +60^{\circ}\\
g^{-} &= -60^{\circ}\\
\end{array}
\end{align*}
}
&\raisebox{-1.0truecm}{433}&\raisebox{-1.0truecm}{\cite{hoeve1961unperturbed}}\\[-8truept]
\raisebox{-0.5truein}{PDMS} & 
{\begin{align*}
\boldsymbol{\tensor{U}}_a&=\left( \begin{array}{ccc}
1 & \sigma & \sigma \\
1 & \sigma'  & 0 \\
1 & 0 & \sigma' \end{array} \right)
&
\boldsymbol{\tensor{U}}_b&=\left( \begin{array}{ccc}
1 & \sigma & \sigma \\
1 & \sigma'  & \delta \\
1 & \delta & \sigma' \end{array} \right)
\end{align*}
}
&
{
\begin{align*}
\begin{array}{lccccccc}
\hline
 & H & \Gamma\\\hline
\sigma & 3.6 & 1\\
\sigma' & 3.6 & 1\\
\delta & 8.0 & 1\\ \hline
\end{array}
\end{align*}
}
&
{\begin{align*}
\textrm{Si-O}&=164\,\textrm{pm}\\
\textrm{O-Si-O}&=110^{\circ}\\
\textrm{Si-O-Si}&=143^{\circ}
\end{align*}
}
&
{\begin{align*}
\begin{array}{ll}
t & =180^{\circ}\\
g^{+} &= +60^{\circ}\\
g^{-} &= -60^{\circ}\\
\end{array}
\end{align*}
}
&\raisebox{-1.0truecm}{343}&\raisebox{-1.0truecm}{\cite{flory1964configuration}}\\[-12truept]
\raisebox{-0.7truein}{PS} & 
{
\begin{align*}
\boldsymbol{\tensor{U}}_p&=\left( \begin{array}{cc}
1 & 1 \\
1 & 0 \\
 \end{array} \right)
&\boldsymbol{\tensor{U}}_m&=\left( \begin{array}{cc}
\omega'' & 1/\eta  \\ 
1/\eta & \omega/\eta^2  \\
 \end{array} \right)\\[2truept]
\boldsymbol{\tensor{U}}_r&=\left( \begin{array}{cc}
1 & \omega'/\eta \\
\omega'/\eta & 1/\eta^2 \\
 \end{array} \right)&&
\end{align*}
}
&
{
\begin{align*}
\begin{array}{lccccccc}
\hline
 & H & \Gamma\\\hline
 \eta & -1.7 & 0.8\\
 \omega & 8.3 & 1.3\\
 \omega' & 8.3 & 1.3\\ 
 \omega'' & 9.2 & 1.8 \\
\hline
\end{array}
\end{align*}
}
&
\raisebox{-1.5truecm}{\parbox{1.0truein}{\begin{align*}
\textrm{C-C}&=153\,\textrm{pm}\\
\textrm{C-C$^{\alpha}$-C}&=112^{\circ}\\
\textrm{C$^{\alpha}$-C-C$^{\alpha}$}&=114^{\circ}\\
\end{align*}
}}
&
{\begin{align*}
\begin{array}{ll}
t & =175^{\circ}\\
g &= + 60^{\circ} (\textrm{meso})\\
g &= - 60^{\circ} (\textrm{racemic})
\end{array}
\end{align*}
}
&\raisebox{-1.0truecm}{300}&\raisebox{-1.0truecm}{\cite{yoon1975conformationalPS}}\\[-12truept]
\raisebox{-0.5truein}{PMMA} & 
{
\begin{align*}
\boldsymbol{\tensor{U}}_p&=\left( \begin{array}{cc}
1 & 1 \\
1 & 0 \\
 \end{array} \right)
&\boldsymbol{\tensor{U}}_m&=\left( \begin{array}{cc}
1 & \alpha  \\
\alpha & \alpha^2/\beta  \\
 \end{array} \right)\\[2truept]
 \boldsymbol{\tensor{U}}_r&=\left( \begin{array}{cc}
\beta & \alpha \\
\alpha & \alpha^2/\beta \\
 \end{array} \right)
&&
\end{align*}
}
&
{
\begin{align*}
\begin{array}{lccccccc}
\hline
 & H & \Gamma\\\hline
\alpha & 4.6 & 1.6\\
\beta & -2.5 & 1.4\\ \hline
\end{array}
\end{align*}
}
&
\raisebox{-1.2truecm}{\parbox{1.0truein}{\begin{align*}
\textrm{C-C}&=153\,\textrm{pm}\\
\textrm{C-C$^{\alpha}$-C}&=110^{\circ}\\
\textrm{C$^{\alpha}$-C-C$^{\alpha}$}&=122^{\circ}\\
\end{align*}
}}
&
{\begin{align*}
\begin{array}{ll}
t & =180^{\circ}\\
g &= + 60^{\circ} (\textrm{meso})\\
g &= - 60^{\circ} (\textrm{racemic})
\end{array}
\end{align*}
}
&\raisebox{-1.0truecm}{300}&\raisebox{-1.0truecm}{\cite{sundararajan1974configurational}}\\[-25truept]
\raisebox{-0.7truein}{PPG-DME} & 
{\begin{align*}
\boldsymbol{\tensor{U}}_a^R&=\left( \begin{array}{ccc}
1 & \alpha & \beta \\
1 & \alpha & \beta \omega \\
1 & \alpha \omega & 0
 \end{array} \right) &
\boldsymbol{\tensor{U}}_b^R&=\left( \begin{array}{ccc}
1 & 0 & 1  \\
1 & 0 & \omega  \\
1 & 0 & 1 \\
 \end{array} \right)\\
\boldsymbol{\tensor{U}}_c^R&=\left( \begin{array}{ccc}
1 & \sigma & 0 \\
1 & 0 & 0 \\
1 & 0 & \sigma \\
 \end{array} \right)&
\boldsymbol{\tensor{P}}&=\left( \begin{array}{ccc}
1 & 0 & 0 \\
0 & 0 & 1 \\
0 & 1 & 0 \\
 \end{array} \right)
\end{align*}}
&
{\begin{align*}
\begin{array}{lccccccc}
\hline
 & H & \Gamma\\\hline
\alpha & -1.26 & 1\\
\beta & 1.46 & 1\\
\sigma & 5.4 & 1\\
\omega & 1.7 & 1\\ \hline
\end{array}
\end{align*}}
&
\raisebox{-1.3truecm}{\parbox{1.0truein}{\begin{align*}
\textrm{C-C}&=153\,\textrm{pm}\\
\textrm{C-O}&=143\,\textrm{pm}\\
\textrm{C-O-C}&=111.5^{\circ}\\
\textrm{C-C-O}&=110^{\circ}
\end{align*}
}}
&
\raisebox{-1.0truecm}{\begin{tabular}{ccc}
\begin{tabular}{cc}
\multicolumn{2}{c}{\underbar{C-C}}\\
 t &$= 180^{\circ}$\\
 g &$= 60^{\circ}$
\end{tabular} 
& 
\begin{tabular}{cc}
\multicolumn{2}{c}{\underbar{C-O}}\\
 t &$= -160^{\circ}$\\
 g$^+$ &$= 60^{\circ}$\\
 g$^-$ &$= -80^{\circ}$\\
 \end{tabular}
&
\begin{tabular}{cc}
\multicolumn{2}{c}{\underbar{O-C}}\\
 t &$= 180^{\circ}$\\
 g$^+$ &$= 80^{\circ}$\\
 g$^-$ &$= -80^{\circ}$\\
\end{tabular}
\end{tabular}}
&\raisebox{-1.0truecm}{300}&\raisebox{-1.0truecm}{\cite{abe1979conformational}}\\[-12truept]\\
\hline
\end{tabular}

\caption{RIS data used for calculations. The transfer matrices $\boldsymbol{\tensor{U}}$ describe the relative weights of successive rotational isomeric states specified by the dihedral angles. The matrix indices label (\textit{t}, \textit{g}) or (\textit{t}, \textit{g$^+$}, \textit{g$^+$}) dihedral angles. Matrix elements have the form $\sigma=\Gamma_{\sigma}e^{-H_{\sigma}/k_BT},\omega=\Gamma_{\omega}e^{-H_{\omega}/k_BT},\ldots$. PE has a single unique bond, specified by $\boldsymbol{\tensor{U}}$, and is calculated at five different temperatures: 70, 110, 148, 200 and 237 K. All other polymers are calculated at $T_g(M)$ based on data as follows: PDMS, Ref.~\cite{Cowie1975some}; PMMA, experimental data from Fig.~2 of main manuscript; PPG-DME, interpolated from experimental data in Fig.~2; PS, experimental data from Fig.~2. PDMS has two distinct bonds (Fig.~\ref{fig:Structures}) specified by $\boldsymbol{\tensor{U}}_a$ and $\boldsymbol{\tensor{U}}_b$.
PS has two distinct bonds, the first specified by $\boldsymbol{\tensor{U}}_p$, and the second with a phenyl ring asymmetrically placed on either the right or left side relative to the chain direction. Similarly, PMMA has two distinct bonds and an asymmetric carbon. In each case $\boldsymbol{\tensor{U}}_m$ specifies the meso diad (two successive right or left asymmetric carbons) and $\boldsymbol{\tensor{U}}_r$ specifies the racemic diad (right/left or left/right sequence).
PPG-DME has three dihedral angles (\textit{a}, \textit{b}, \textit{c}) and one asymmetric carbon on the backbone chain per repeat unit. 
Matrices corresponding to a left (S) asymmetrical carbon are obtained from the right (R) matrices according to 
$\boldsymbol{\tensor{U}}_i^S=\boldsymbol{\tensor{P}}\boldsymbol{\tensor{U}}_i^R \boldsymbol{\tensor{P}}$.
}
\label{tab:RISParameters}
\end{table*}
\end{turnpage}

\subsection{Calculated quantities}
The gyration tensor $\boldsymbol{\tensor{Q}}_{\nu}$ for a given conformation $\nu$ is calculated using the position vectors $\vec{r}^i$  of the backbone atoms
\begin{align}
\tensor{Q}_{\alpha\beta,\nu}&=\frac{1}{n_a}\sum_{i=1}^{n_a}\left(r_{\alpha}^{i}-\bar{r}_{\alpha}\right)\left(r_{\beta}^i-\bar{r}_{\beta}\right),
\end{align}
where $\bar{\vec{r}}=\tfrac{1}{n_a}\sum_i\vec{r}^i$. Note that individual conformations $\nu$ rarely have spherical mass distributions $\boldsymbol{\tensor{Q}}_{\nu}$, but are usually anisotropic and have a biaxial shape similar  to a flattened prolate ellipsoid \cite{kreer2001monte}. This gyration tensor $\boldsymbol{\tensor{Q}}_{\nu}$ refers to point atoms. To calculate the physical gyration tensor we incorporate the finite volume of the backbone atoms and associated side groups. For simplicity, we center all side group volumes on the backbone atoms, and calculate the corresponding backbone atom volume $V_a$ as 
\begin{align}
V_a&=\frac43 \pi \sigma_{\textit{eff},a}^3 \\
 \sigma_{\textit{eff},a}^3&=\sum_{j=1}^{{m}_a} \sigma_{j,a}^3,
\end{align} 
where $\sigma_{j,a}$ is the van der Waals radius of the $j^{th}$ of ${m}_a$ non-hydrogen atoms in backbone atom group $a$ and its associated sidegroups. The volumes can be found in Ref.~\cite{batsanov2001van}. We ignore hydrogen atoms, which have small volumes and relatively small van der Waals energies. The position and respective size of the effective van der Waals volumes are shown as blue spheres in Fig.~\ref{fig:Structures}. The corrected gyration tensor $\boldsymbol{\tensor{Q}}^c$ is given by 

\begin{align}
\boldsymbol{\tensor{Q}}^c= \boldsymbol{\tensor{Q}} +  \boldsymbol{\tensor{I}}\,\tfrac{1}{3N}\,\sum_{a=1}^{N}\sigma_{\textit{eff},a}^{2},
\end{align} 
where $\boldsymbol{\tensor{I}}$ is the identity tensor. 

We quantify the shape of molecules by averages of the eigenvalues $\lambda^2_{i,\nu}$ of $\boldsymbol{\tensor{Q}}^c_{\nu}$ for given conformations $\nu$; 
\begin{align}
\boldsymbol{\tensor{Q}}^c_{\nu}&\equiv\left(\begin{matrix}
\lambda^2_{1,\nu}&0&0 \\
0 & \lambda^2_{2,\nu}& 0 \\
0& 0 & \lambda^2_{3,\nu}
\end{matrix}
\right),
\end{align}
and order the  eigenvalues of $\boldsymbol{\tensor{Q}}^c$  according to 
\begin{align}
\lambda^2_{1,\nu}<\lambda^2_{2,\nu}<\lambda^2_{3,\nu}.
\end{align}
We thus calculate the averages
\begin{align}
\lambda_i^2\equiv
\left<\lambda^2_{i,\nu}\right>=
\frac{1}{n_{\nu}}\sum_a^{n_{\nu}}\lambda^2_{i,\nu},
\end{align}
from many ($n_{\nu}\sim 10^5-10^6$), configurations obtained via MC calculations performed using the Metropolis algorithm to approximate a thermal average. For stereocomplex chemistries, we also average over many representative sequences $n_{st}$ to approximate a specified average tacticity. The radius of gyration is calculated as
\begin{align}
R_g^2&{=\mean{\textrm{Tr}\,\boldsymbol{\tensor{Q}}^c}}=\sum_{i=1}^3{\lambda^2_i}
\end{align}
\color{black}
and the average end-to-end distance is given by
\begin{align}
R_e^2&=\mean{\left|\vec{r}_N-\vec{r}_1\right|^2}.\label{eq:Ree}
\end{align}
The characteristic ratio $C_n$ of a chain with $n_b$ bonds is defined as 
\begin{align}
C_n=\frac{R_e^2}{n_b\,b_{\textit{eff}}^2} \label{eq:Cn}
\end{align}
where the effective bond size 
$b_{\textit{eff}}=\sqrt{\sum_{j=1}^Jb_j^2}$ is the harmonic mean over $b_j$ for each bond $j$ in the repeat unit (\textit{i.e.} the monomer). There are typically $J=1, 2$, or $3$ bonds per monomer. There are several conventions for defining $b_{\textit{eff}}$ for polymers with multiple bonds per monomer; an alternative choice \cite{fetters2007chain} is $b_{\textit{eff}}=\sqrt{\sum_jb_j^2\cos^2\theta_j}$ (the harmonic mean over $b_j\cos\theta_j$ ).

We quantify the shape anisotropy using the aspect ratio $\Lambda^2$, and here we also study a second anisotropy measure $\delta$ which captures the non-Gaussianity of finite length polymers:
\begin{subequations}
\begin{align}
\Lambda^2&=\frac{\lambda_3^2}{\lambda_1^2}&&\textrm{aspect ratio}\\
\delta&=\frac{R_e^2 }{6 R_g^2}&& \textrm{non-Gaussianity}\label{eq:delta}
\end{align}
\end{subequations}
For a Gaussian chain, $\Lambda^2=11.87$ \cite{kreer2001monte, janszen1996bimodality}, while for a thin rod $\Lambda^2=\frac{3L^2}{4D^2}$ where $D$ is the diameter of the rod and $L$ is its length. The non-Gaussianity $\delta=1$ for a Gaussian chain where $R_e^2\sim R_g^2 \sim n_b$. For smaller chains for which the persistence length is not vanishingly small compared to the contour length the end to end length typically scales as $n_b$ rather than $n_b^{1/2}$ and $\delta=1$ breaks down. For example, and $\delta=2/(1 + 1.5/n_a+ 1.7/n_a^2)$ for a string of $n_a$ (even) close-packed spheres in a linear array ($\delta=0.476$ for 2 spheres and $\delta=2$ in the infinitely rigid polymer limit). For a wormlike chain (WLC) with persistence length $\ell_p$ one finds \cite{yamakawa1971modern}
\begin{align}
    \left<R_e^2\right> &= 2\ell_p L - 2\ell_p^2 \left(1-e^{-L/\ell_p}\right)\label{eq:wlc_Re}\\
    \left<R_g^2\right> &= \tfrac13 \ell_p L - 2\ell_p^2\left[1 - \frac{\ell_p}{L} + \left(\frac{\ell_p}{L}\right)^2 \left(1-e^{-\ell_p/L}\right)\right], \label{eq:wlc_Rg}\\
    \noalign{\noindent so that}\nonumber\\
    \delta\left(\frac{L}{\ell_p}\right) &= \frac{1 - \frac{\ell_p}{L}\left(1-e^{-L/\ell_p}\right)}{1 - 6\frac{\ell_p}{L}\left[1 - \frac{\ell_p}{L} + \left(\frac{\ell_p}{L}\right)^2 \left(1-e^{-\ell_p/L}\right)\right]}.\label{eq:wlc_delta}
\end{align}
In the flexible limit the WLC corresponds to a Gaussian chain with Kuhn step $\ell_K=2\ell_p$.
 
\subsection{Details for specific polymers}

The RIS calculations were performed and analyzed in \textsc{Matlab}, implementing the standard procedure described in Flory’s seminal papers \cite{floryRIS}.  The RIS simulations for PMMA, PS, PPG-DME and PDMS were performed at temperatures $T_g(M)$, corresponding to the specific molecular weights. Since the full $T_g(M)$ for PE is not known, we perform MC-simulations at five different fixed temperatures to investigate the effect of temperature on the structural metrics; we note that the value of $C_\infty=9.0$ calculated from our simulations for PE at $T=237\,\textrm{K}$ is only slightly larger than $C_{\infty}=8.3$ determined from experiments at $T=298\,\textrm{K}$ \cite{fetters2007chain}. 

For stereospecific polymers we use $n_{st}=10$ for PMMA, $n_{st}=20$ for PPG-DME, and $n_{st}=30$ for PS; $p_{\textrm{meso}}=0.5$ for PS and PMMA; and  $p_R=0.5$ for PPG-DME.

\begin{figure}
     \includegraphics[width=0.5\textwidth]{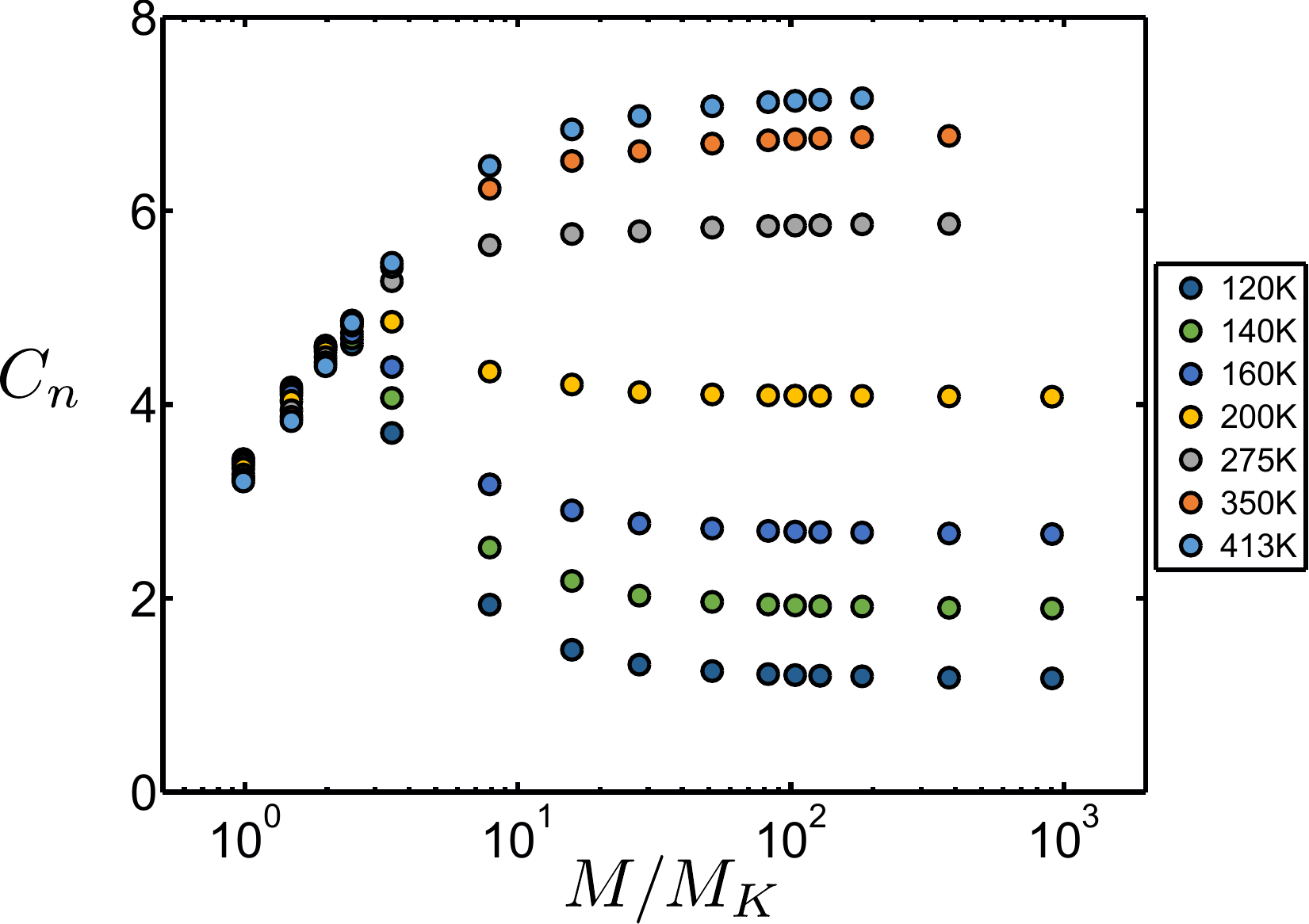}
     \caption{$C_n(M,T)$ of PDMS calculated using the RIS model for different $T$, as a function of molecular weight. The maximum at lower $T$ is because the ground state of PDMS in the RIS model is a circle (degenerate helix) \cite{birshtein1959,floryRIS}. Note that the experimental $T_g^{\infty}=148\,\textrm{K}$, at which point the RIS calculations predict a high prevalancy for loops. Experimental chain dimensions have been measured at  $T=298\,\textrm{K}\,\, (C_{\infty}=5.8)$ and $T= 413\,\textrm{K}\,\, (C_{\infty}=6.3)$ \cite{fetters1994connection}, and the RIS parameters were calibrated at $T=343\,\textrm{K}$ \cite{Rehahn1997}.}
     \label{fig:CnT_PDMS}
\end{figure}

PDMS is different from the other polymers because the Si-O-Si and the O-Si-O angles (Table~\ref{tab:RISParameters}) lead to a ground state conformation of a planar loop with circumference of approximately $n_b=24$ bonds \cite{birshtein1959,floryRIS}. At the low temperatures near $T_g$ 
only few, if any, \textit{gauche} states are excited which means that RIS calculations are only reliable for high temperature, or $n_b<24$ corresponding to $M\lesssim88\,\textrm{g/mol}$. The calculation of $C_n(M,T)$ at low $T$ thus has a predicted maximum corresponding to the molecular weight at which the ring starts to bend back on itself, as shown in Fig.~\ref{fig:CnT_PDMS}. Recall that excluded volume beyond the closest 4 monomers (two on each side) is not accounted for in RIS models \cite{floryRIS}, so a fully degenerate all-\textit{trans} state is an allowable configuration for RIS calculations, which is unphysical. Since  \textit{gauche} states take the conformation out of the plane, only temperatures low enough to allow very few \textit{gauche} states lead to unphysical configurations that overlap.
\section{Conformational statistics from RIS calculations}\label{sec:RISstats}
\begin{figure*}[htbp]
\begin{center}
{\includegraphics[width=0.9\textwidth]{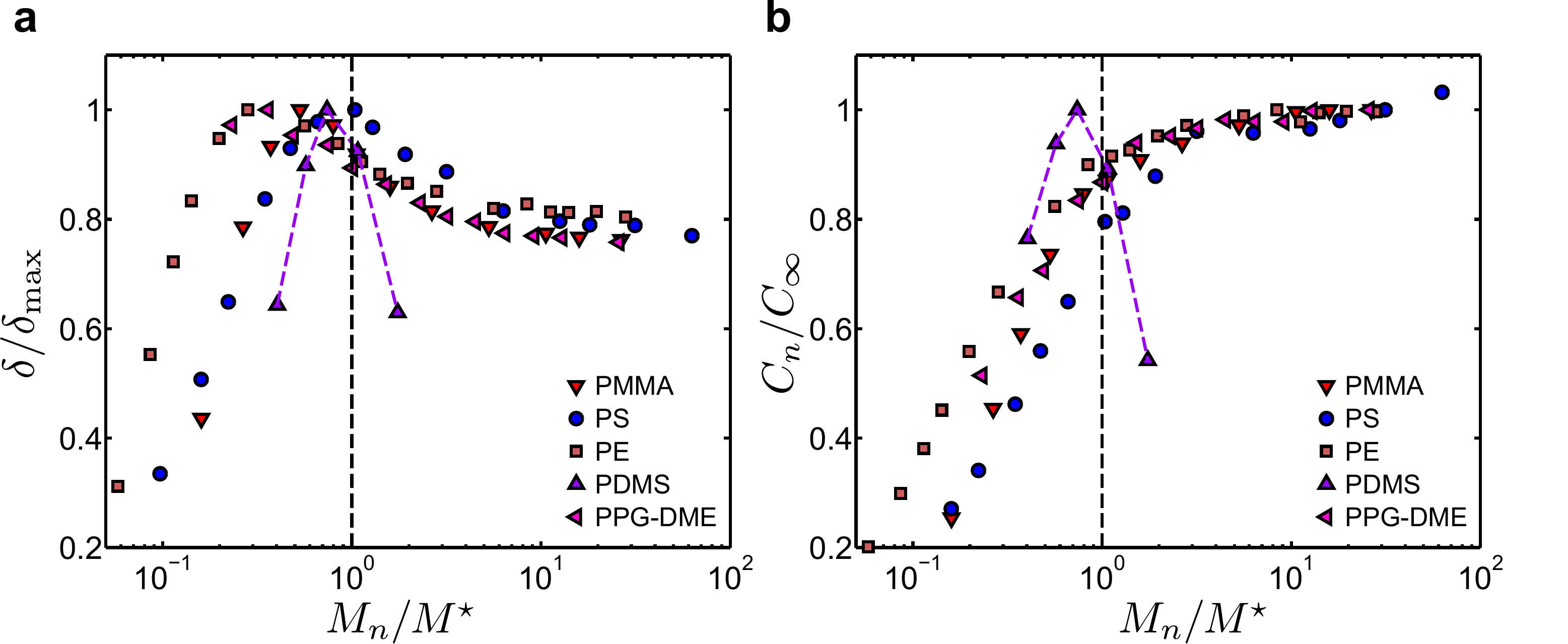}}
\end{center}
\caption{(a) Calculated non-Gaussianity $\delta$ as a function of normalised molecular weight $\delta(M/M^{\star})$, normalised by the maximum value $\delta_{\textrm{max}}$ from RIS simulations for PMMA, PS, PE, PDMS, and PPG-DME, where $M^{\star}$ is determined by the scaling in Figure 2. (b) Calculated characteristic ratio $C_n$ as a function of normalised molecular weight, normalised by $C_{\infty}$, which we take as the high-$M$ maximum value $C_n^{\textrm{max}}\simeq C_{\infty}$ (for PS, the maximum value is taken as the average of the three highest M data points). The data for PDMS are qualitatively different from those of the other polymers due of its unusual ground state energy loop structure; hence, a dashed line has been added to the PDMS data as a guide to the eye.}
\label{fig:delta_CN_SI}
\end{figure*}
As shown in Fig.~3f, the aspect ratio $\Lambda^2$ shows a maximum as a function of $M$ which is well-correlated with the molecular weight $M^\star$ that marks the change in $M$-dependence of $T_g(M)$. Similarly, the non-Gaussianity $\delta$ also shows a maximum as a function of $M$, observed at $M\approx M^*$, as shown in Fig.~\ref{fig:delta_CN_SI}.
Fig.~\ref{fig:delta_CN_SI}b shows the normalised values of the Flory characteristic ratio $C_n(M/M^{\star})/C_{\infty}$, where $C_{\infty}$ represents the high-$M$ value. As shown in Fig.~\ref{fig:RIS_calcs}, we find that $C_n(M)$ and $T_g(M)$ demonstrate similar behaviour for PMMA, PS, PE, and PPG-DME. Mirigian and Schweizer \cite{mirigian2015dynamical} reasoned that a polymer glass can be effectively treated as a hard sphere glass with a number of interaction sites that depend on the the conformation and more specifically $C_n(M)$. This  led them to conclude that $C_n(M)$ and $T_g(M)$ could show similar behaviour. We note, however, that similarity is not observed for PDMS due to the loop formation at low temperatures near $T_g(M)$ \cite{birshtein1959}. 
\begin{table*}[htbp]
\begin{tabular}{lccrcrcrrrcp{2.5truecm}}
\hline\hline
& $T_g^{\infty}$ & $T_{\textrm{cal}}$ & $M(\delta^{\textrm{max}})$ & $n_d({\delta^{\textrm{max}}})$ & $M({\Lambda^2_{\textrm{max}}})$ &
$n_d({\Lambda^2_{\textrm{max}}})$ &$M^{\star}$ & $n_d^{\star}$ & $C_\infty^{\textrm{RIS}}$ & $\ell_K$&\hfil$\ell^{\textrm{exp}}_K(T)$\hfil\\
& K & K&  g/mol&&g/mol&&g/mol&&&nm&\hfil nm\hfil\\\hline
PDMS& 148 & 343 & 326& 6 &	474& 10 &	441& 11 &  & -- & \hfil 1.14 (298 K) \hfil\\
PPG-DME & 197 & 300 & 162& 6 &	684& 33 &	450& 14 & 5.3 & 0.8& \hfil --\hfil\\
PMMA & 	387 & 300 & 1002& 7 &	1502& 27 &	1889& 38 & 12.1 & 1.9 &\hfil1.53 (490 K)\hfil\\
PE & 200 & 433 & 282 & 17  &	562 & 37 &1000 &  107 & 9.0 & 1.4&\hfil1.54 (298 K)\hfil\\ 
PS & 374 & 300 & 1721& 32 &	3177 & 60 & 1661& 31 & 12.8 & 2.0&\hfil 1.78 (413 K)\hfil\\
\hline\hline
\end{tabular}
\caption{$T_g^\infty$ and calibration temperature $T_{\textrm{cal}}$ for the RIS parameters. Molecular weights and number of dihedrals ($n_d$) corresponding to the maxima of $\delta$ and $\Lambda^2$ at $M^{\star}$, from RIS calculations. We use $M^{\star}$ as determined in Fig.~2e. Also shown are $C_\infty^{\textrm{RIS}}$,  the Kuhn length $\ell_K\equiv C_{\infty}^{\textrm{RIS}}b_{\textrm{eff}}$ calculated from the simulations at $T_g^{\infty}$, and the experimental Kuhn length $\ell_K^{\textrm{exp}}$ reported in the literature \cite{fetters2007chain} (typically for $T\gg T_g$). We have not found a reliable estimate for $\ell_K$ of PPG-DME. Excluded volume prohibits a reliable calculation of $C_{\infty}^{\textrm{RIS}}$ for PDMS at $T_g$, as discussed in the text.}
\label{tab:max_delta}
\end{table*}

We have shown (Fig.~\ref{fig:RIS_calcs}f) how the maximum in $\Lambda^2$, and the corresponding change in $C_n(M)$ for PMMA, PS, PPG-DME and PE, are due to chain folding when the molecular weight exceeds $M^*$. The chain folding is also reflected in the maximum in $\delta(M)$, which occurs near but somewhat below $M^*$, as shown in Fig.~\ref{fig:delta_CN_SI}a. However, the data for $\Lambda^2$, $\delta(M)$ and $C_n(M)$ for PDMS do not follow the same trends, and the observed maxima are instead due to the formation of loops; these maxima are located close to $M^*$ for all three metrics. Table \ref{tab:max_delta} provides a summary of our RIS simulation data together with a comparison with the corresponding experimental results. The table lists the molecular weights and the number of dihedrals $n_d$ corresponding to $\delta_{\textrm{max}}$ and $\Lambda^2_{\textrm{max}}$, together with the respective values at $M^{\star}$; our calculated values of the Flory characteristic ratio $C_{\infty}^{\textrm{RIS}}$ and the Kuhn length $l_K$ together with the experimentally determined $l_K$; and the experimental calibration temperature. 

\begin{figure*}[htbp]
\begin{center}
{\includegraphics[width=1\textwidth]{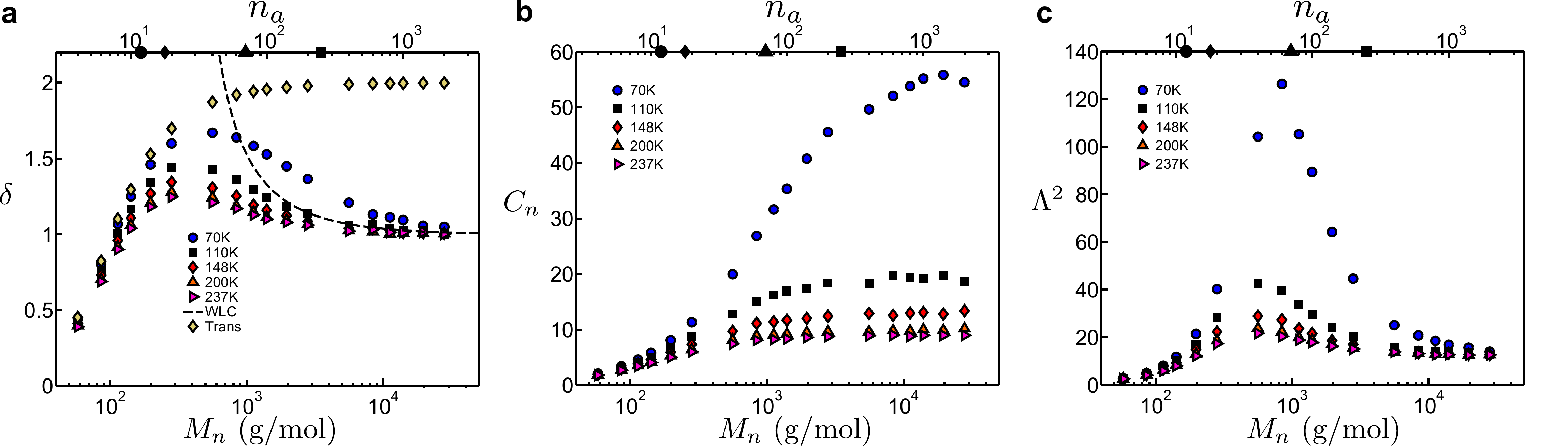}}
\end{center}
\caption{Chain conformation metrics $\delta, C_n, \Lambda^2$ for PE at different temperatures. (a) also includes calculations for the all-trans state of PE and a wormlike chain model with persistence length $\ell_p=\ell_K/2$, with Kuhn step $\ell_K=1.54\,\textrm{nm}$, corresponding to $T=298$\,K. The experimentally-measured values are $C_{\infty}(298\,\textrm{K})=8.26, C_{\infty}(413\,\textrm{K})=7.38.$ \cite{fetters2007chain}. }
\label{fig:deltaPE}
\end{figure*}

To illustrate the effects of $M$ and temperature on chain conformations we study PE by calculating $\Lambda^2(M)$, $C_n(M)$, and $\delta(M)$ for a range of temperatures $T$. Fig.~\ref{fig:deltaPE}a shows that $\delta(M)$ at low $T$ follow the results for an all-trans chain configuration with few or no excited \textit{gauche} states. The semiflexible worm-like chain model for $T=298\,\textrm{K}$ shows that the high-$M$ behaviour approaches the flexible chain limit $\delta=1$ for a Gaussian chain. In between the limiting rod-like and flexible regimes $\delta$ has a maximum, indicating how chains fold due to excited (\textit{gauche}) dihedral states. In the high-$M$ limit, $\delta\rightarrow1$ and $\Lambda^2 \rightarrow 11.9$, characteristic of a Gaussian chain \cite{kreer2001monte}. The maxima in $\delta(M)$  shift slightly to larger $M$ for lower temperatures, as expected.  Fig.~\ref{fig:deltaPE}b shows the behaviour of $C_n(M)$; the increasing fraction of trans states at the lowest temperatures leads to a significant increase in $C_n$ within regimes II and III. Fig.~\ref{fig:deltaPE}c shows that the aspect ratio $\Lambda^2$ behave similarly for the three highest $T$, including an increase at low $M$ within regime I; a maximum near the regime I-II crossover where the chains starts to fold; and a decrease towards the Gaussian limit where $\Lambda^2=11.87$ \cite{baschnagel1992monte} for high $M$. The maximum is more prominent for the lowest $T$, and occurs at larger values of $M$ due to the corresponding higher chain stiffness. 

\section{Fox-Flory description of $T_g(M)$ for polymeric glass-formers}\label{app:FF}
\begin{figure*}[htbp]
{\includegraphics[width=1\textwidth]{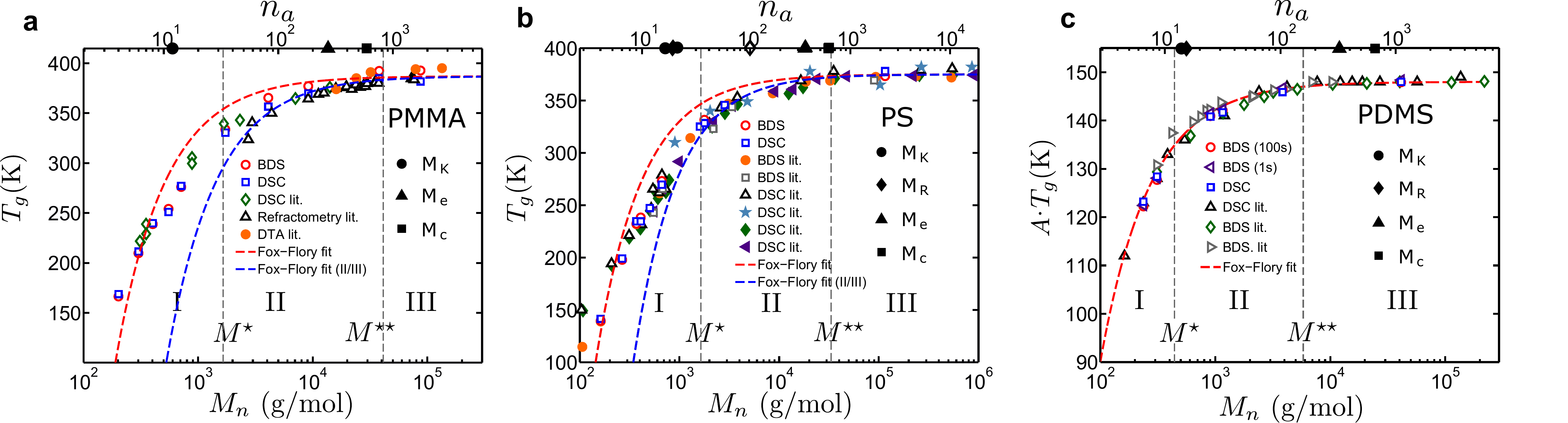}}
\caption{$T_{g}$ as a function of number average  molecular weight $M_n$ and the number of backbone atoms $n_a$. Data from broadband dielectric spectroscopy (BDS), differential scanning calorimetry (DSC) and rheology are combined with literature data for PMMA \cite{Thompson1966dependence,o1991chain,beevers1960physical}, PS \cite{Hintermeyer2008molecular,claudy1983glass,Cowie1975some,bartenev1990polymer,santangelo1998molecular} and PDMS \cite{Hintermeyer2008molecular,cowie1973molecular,kirst1994molecular}. The absolute value of $T_g(M)$ can vary slightly between different studies due to the variation in experimental techniques, $T_g$ definition, or polymer specification. For PDMS these differences are more pronounced since
$T_g$ is a weaker function of $M_w$ ($\Delta T_g\simeq 40$K) than for PMMA or PS ($\Delta T_g\simeq 200$K). Thus a scaling factor $A\sim1-1.03$ was used to collapse different data sets onto $A T_g(M)$. The PDMS data incorporate $T_g$ determined from two different definitions, $\tau_\alpha = $100s and 1s, which slightly change  $T_g$ without significantly changing $T_g(M)$. The symbols on the upper abscissa denote the Kuhn molecular weight ($\bullet$; PMMA: \cite{sacristan2008role}, PS: \cite{ding2004does}), the `dynamical' or Rouse molecular weight $M_{R}$ ($\blacklozenge$; PS and PDMS \cite{ding2004does}, $\lozenge$ PS; using an alternative $M_R$ definition \cite{Ding2004comment}), the  entanglement molecular weight $M_{e}$ ($\blacktriangle$; PMMA \cite{fetters2007chain}, PS and PDMS \cite{fetters1999packing}), and the critical molecular weight $M_{c}$ ($\blacksquare$; PMMA \cite{tadano2014new}, PS and PDMS \cite{fetters1999packing}. Dashed lines are fits to the Fox-Flory expression $T_g=T_g^{\infty}-a/M$, using different ranges for the fits. $n_a$ denotes the number of backbone atoms and $M^{\star}$ and $M^{\ast\ast}$ separate regimes I, II, and III.}
\label{fig:foxfloryfits}
\end{figure*}
\begin{figure*}[htbp]
\begin{center}
{\includegraphics[width=0.8\textwidth]{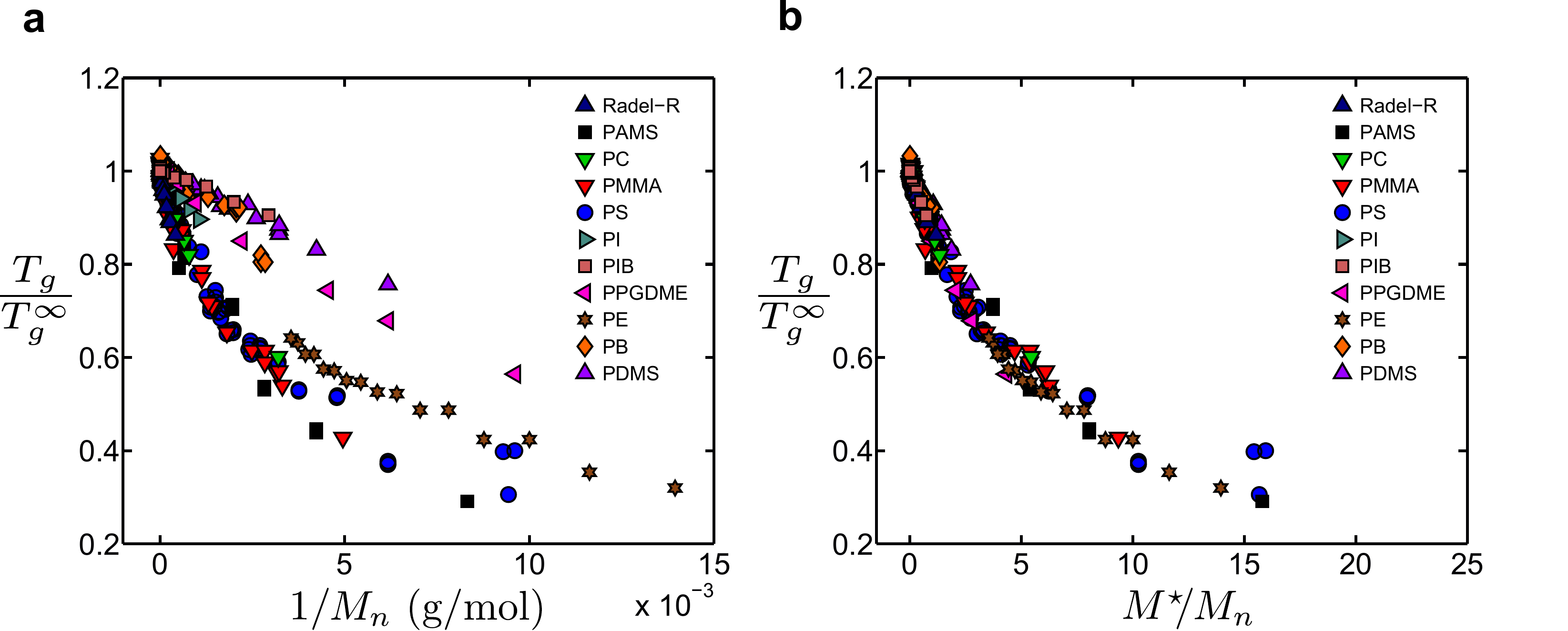}}
\end{center}
\caption{Fox-Flory (FF) plot for 11 polymers showing (a) $T_g/T_g^{\infty}$ vs $1/M$, and (b) 
$T_g/T_g^{\infty}$ vs $M^{\star}/M$.  FF behaviour would imply a straight line $T_g-T_g^{\infty}\sim 1/M$.}
\label{fig:Tg_vs_1overM}
\end{figure*}

Fig.~\ref{fig:foxfloryfits} shows $T_g(M)$ data for PMMA, PS and PDMS (also shown in Fig.~1a-c) together with fits to the standard Fox-Flory expression, $T_g=T_g^{\infty}-a/M$. For each polymer, the data were either fit over the full data range (regimes I-III; dashed red line), or over a limited data range (regimes II-III; dashed blue line). We find that a Fox-Flory expression cannot describe either PMMA or PS across all three regimes; a Fox-Flory expression can reasonably approximate regimes II and III, even though a semi-logarithmic fit ($T_g=A_{II}+b_{II} \log_{10}(M)$) provides a better fit within regime II. The more flexible PDMS can also be described  within regimes I and II by semi-logarithmic fits as shown in Fig.~1c, but contrary to the behaviour of the less flexible polymers PS and PMMA, a Fox-Flory expression can alternatively describe PDMS adequately across all three regimes I-III, as shown in panel~c. Since data in the literature are often plotted as $T_g$ vs $1/M$, we also illustrate the behaviour for the 11 polymers in this representation in Fig.~\ref{fig:Tg_vs_1overM}a, where $T_g$ has been normalized by its long chain-length value $T_g^{\infty}$ to facilitate comparisons. The same data are also showed in Fig.~\ref{fig:Tg_vs_1overM}b with the abscissa re-scaled by $M^{\star}$ to aid the comparison to Fig. 2e. 

\section{Polymer data and literature references}\label{app:data}


\begin{table*}[htbp]

\begin{center}

\begin{tabular}{llccccccccccccccp{1.5truecm}ccc}
\hline\hline\\[-6truept]
\#$^{a}$ & Polymer &  & $M_o$ & $M_{\phi}$ & $\displaystyle\frac{V_{\phi}}{\textrm{\AA}^3}$ & $M_{\textrm{end}}$ & $M_K$ &  & $M_R$ &  & $\displaystyle\frac{M_e}{10^3}$ &  & $\displaystyle\frac{M_c}{10^3}$  &  & 
$\displaystyle\frac{T_g^{\infty, b}}{\textrm{K}}$ & 
$\displaystyle C_{\infty}^c \left(\frac{T}{\textrm{K}}\right)$ &  &
$\displaystyle\frac{M^{\star}}{10^3}$ & $\displaystyle\frac{M^{\star\star}}{10^3}$ \\[10.0truept]\hline
1 & {PE} & \includegraphics[width=1.2truecm]{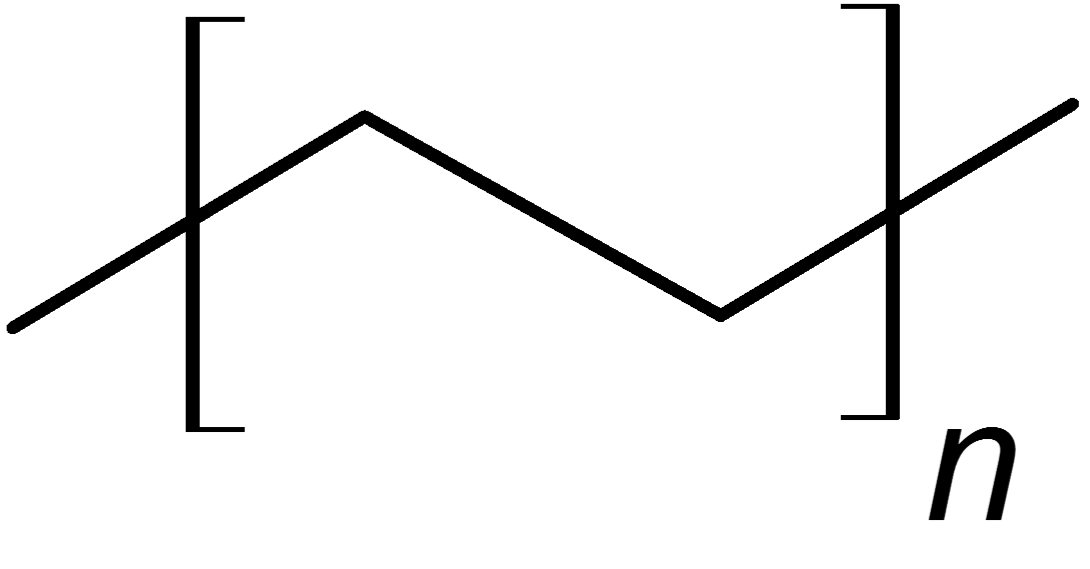} & 28 & 14 & 17 & 2 & 168 & \cite{fetters2007chain} & 252 & \cite{brereton1991nature} & 0.98 & \cite{fetters2007chain} & 3.5 & \cite{fetters2007chain} & 200 & \parbox{1.4truecm}{8.3~(298) \\ 7.4~(413)} & \cite{fetters2007chain} & 1.0 & -- \\
2 & 1,4-PB & \includegraphics[width=1.6truecm]{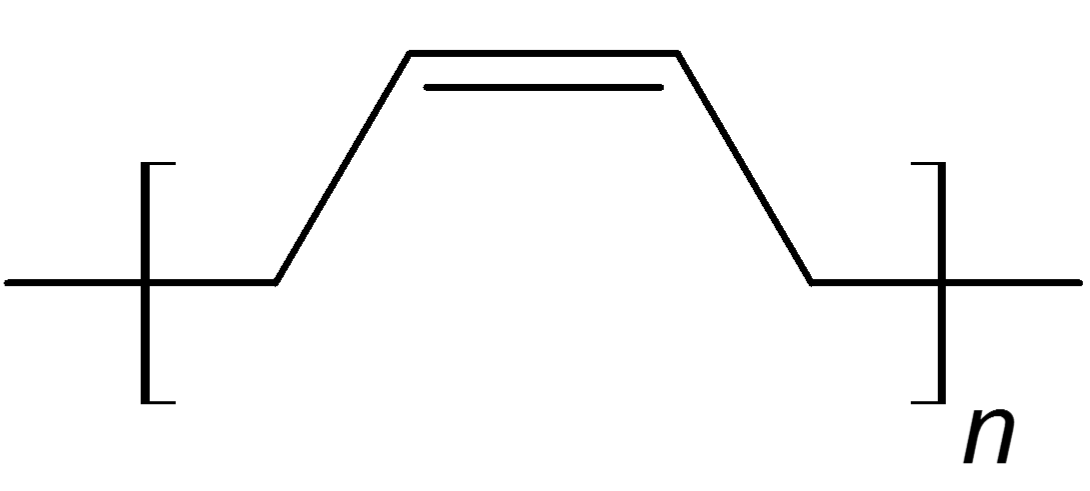} & 54 & 18 & 21 & 30 & 113 & \cite{rubinstein2003polymer} & 500 & \cite{hofmann2012glassy} & 2.9 & \cite{fetters2007chain} & 4.5 & \cite{graessley1981entanglement} & 175 & 4.6 (298) & \cite{fetters2007chain} & 0.47 & 2.3 \\
20 & {PPG-DME} & \includegraphics[width=1.6truecm]{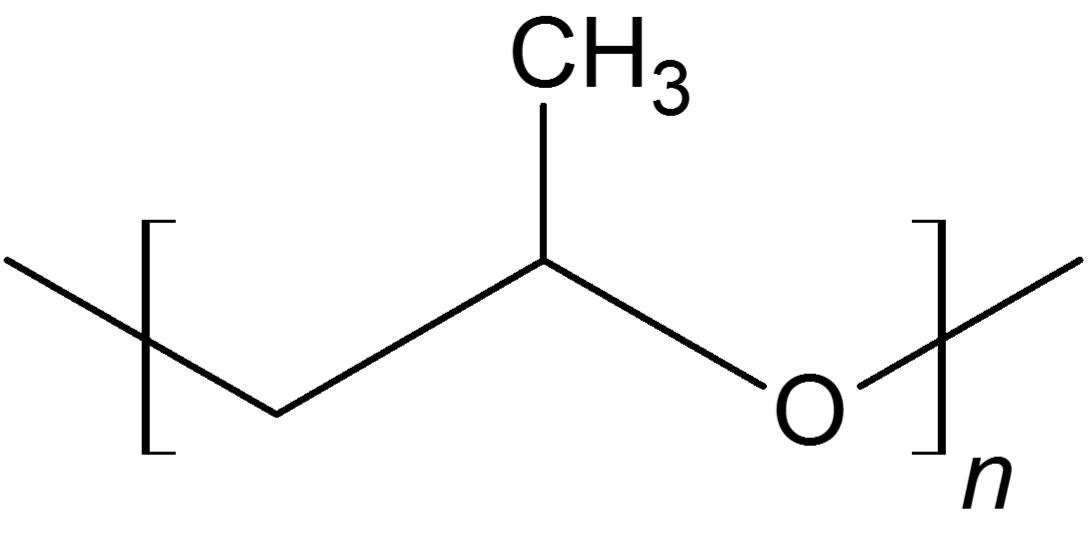} & 58 & 19 & 20 & 46 & -- &  & 150 & \cite{hofmann2012glassy} & 2.8 & \cite{fetters1994connection} & 7 & \cite{smith1985polymer} & 197 & 5.1 (298) & \cite{gainaru2010dielectric} & 0.45 & 6.0 \\
5 & {PIB} & \parbox[c]{1.2truecm}{\includegraphics[angle=0,width=1.2truecm]{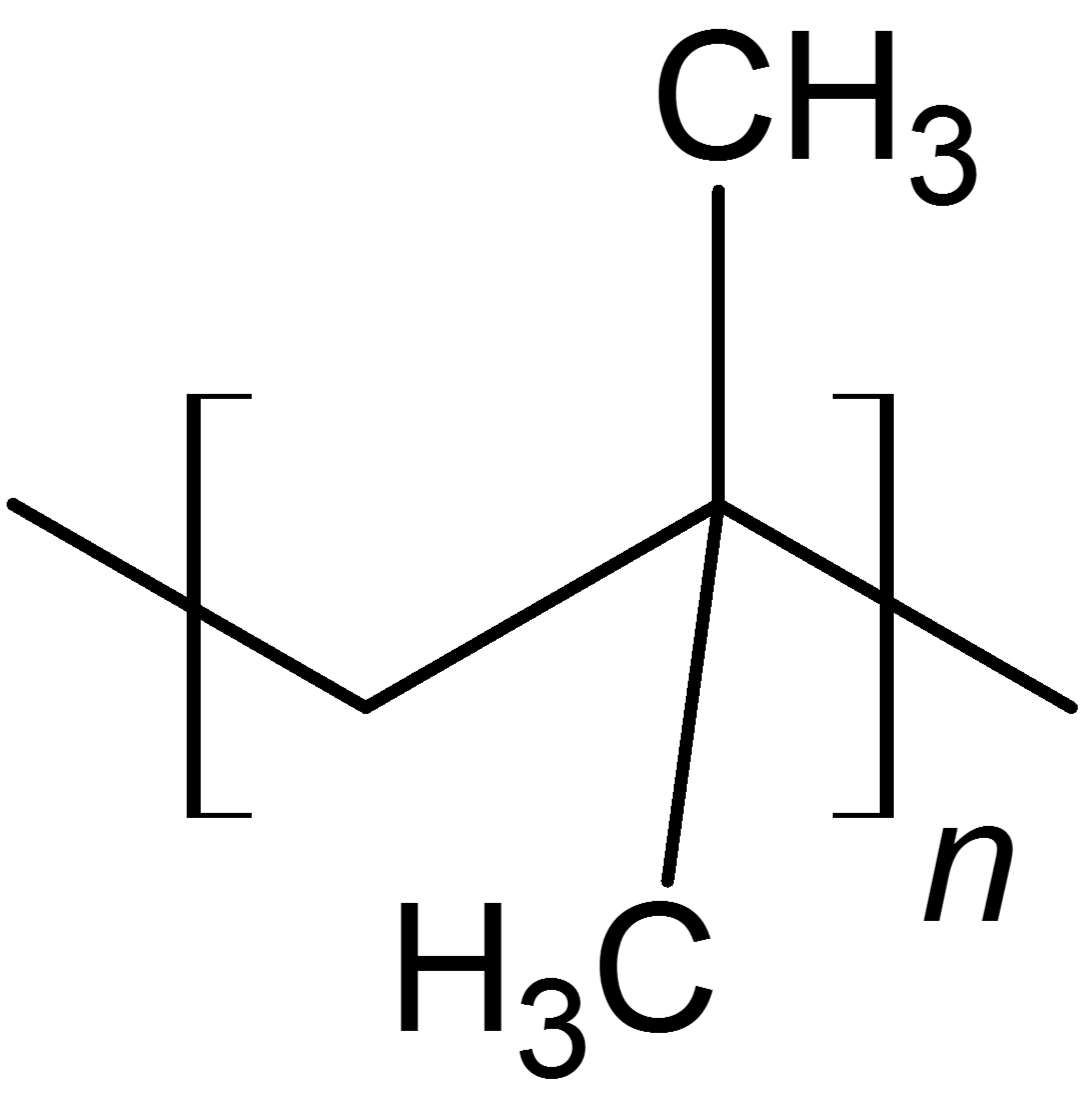}} & 56 & 28 & 34 & -- & 274 & \cite{fetters2007chain} & 200 & \cite{inoue1996role} & 6.9 & \cite{fetters2007chain} & 13 & \cite{fetters2007chain} & 210 & \parbox{1.4truecm}{6.7~(298)  6.6~(413)} & \cite{fetters2007chain} & 0.25 & 30.9 \\
9 & {1,4-PI} & \includegraphics[width=1.6truecm]{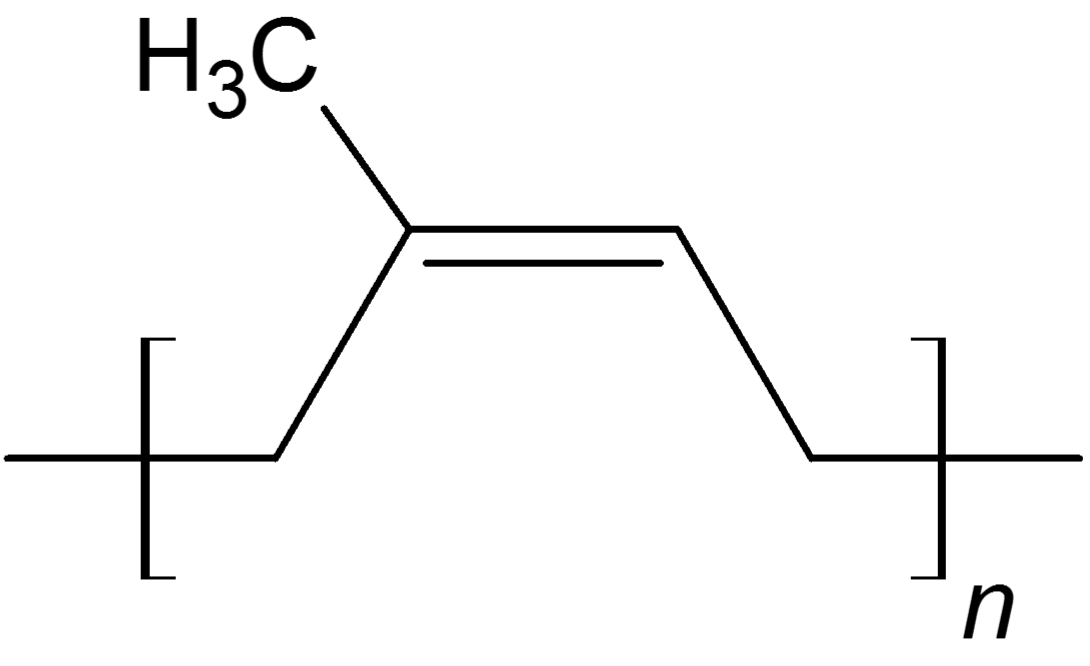} & 68 & 34 & 26 & -- & 129 & \cite{fetters2007chain} & 1000 & \cite{hofmann2012glassy} & 3.9 & \cite{fetters2007chain} & 10 & \cite{graessley1981entanglement} & 213 & 5.2 (298) & \cite{fetters2007chain} & 0.6 & 100.0 \\
11 & PMMA & \parbox[c]{1.0truecm}{\includegraphics[angle=0,width=1.0truecm]{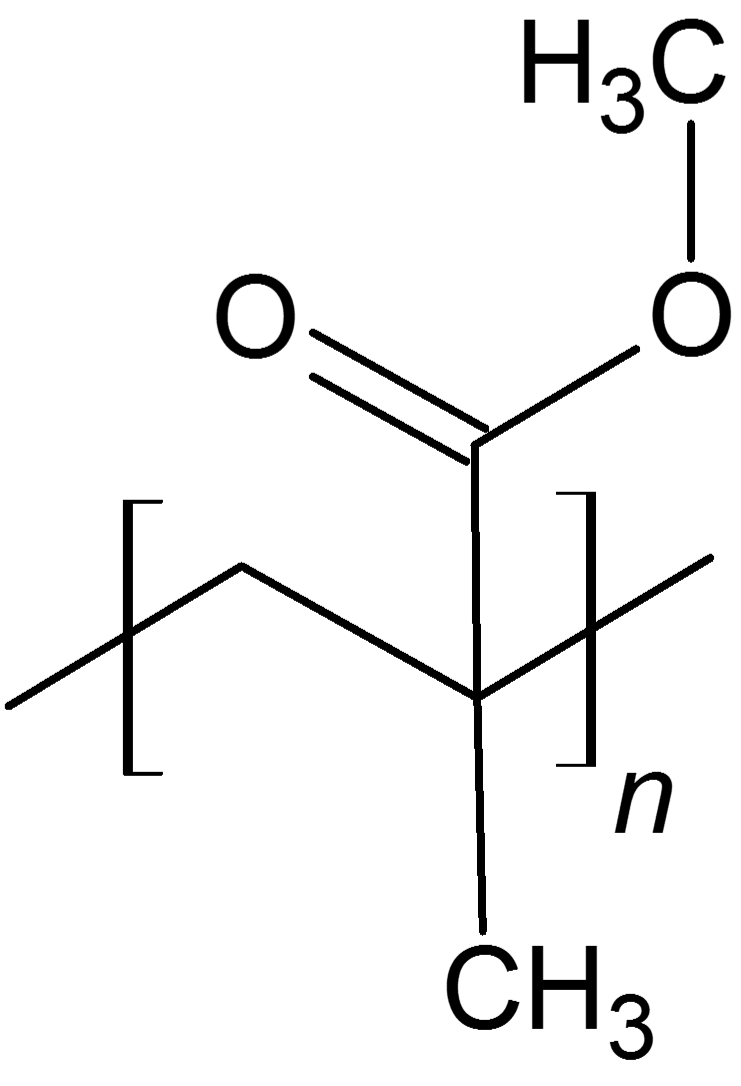}} & 100 & 50 & 47 & 2 & 598 & \cite{fetters2007chain} & -- &  & 13.6 & \cite{fetters2007chain} & 30 & \cite{fetters2007chain} & 387 & 8.2 (413) & \cite{fetters2007chain} & 1.9 & 41.2 \\
12 & {PS} & \parbox[c]{1.0truecm}{\includegraphics[width=1.0truecm]{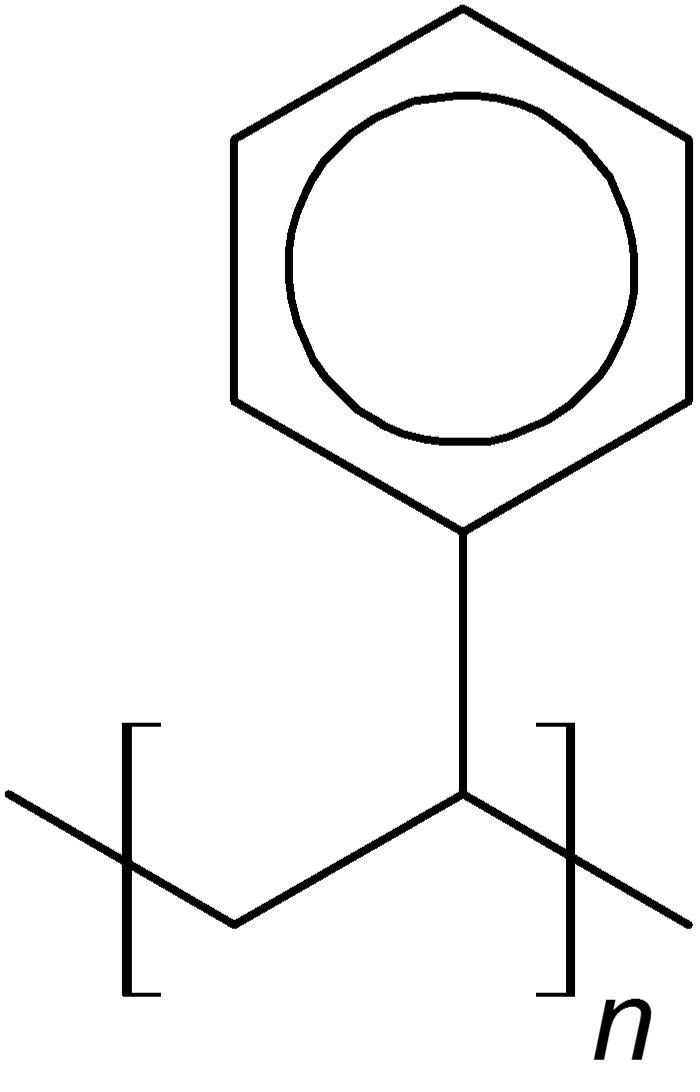}} & 104 & 52 & 54 & 58 & 720 & \cite{rubinstein2003polymer} & \parbox{0.8truecm}{850$^d$ \\ 5000} & \parbox{0.6truecm}{\cite{inoue1996role}\\\cite{Ding2004comment}} & 16.6 & \cite{fetters2007chain} & 35 & \cite{graessley1981entanglement} & 374 & 9.6 (413) & \cite{fetters1994connection} & 1.7 & 33.3 \\
14 & PAMS & \parbox[c]{1.0truecm}{\includegraphics[width=1.0truecm]{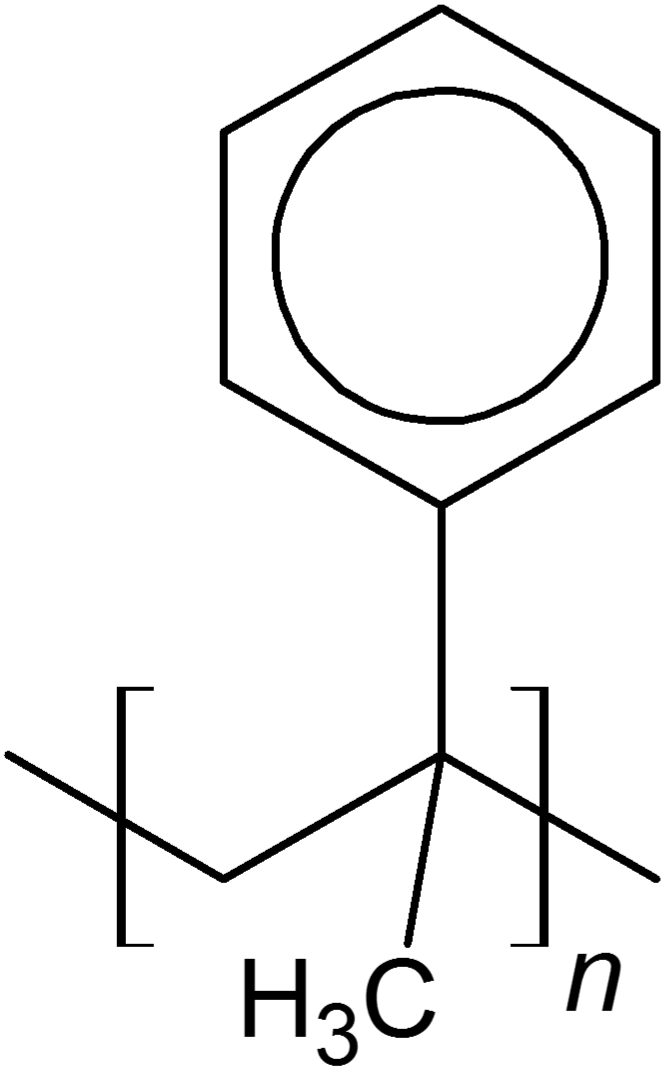}} & 118 & 59 & 62 & 2 & 960 & \cite{inoue1996role} & 730 & \cite{inoue1996role} & 13.3 & \cite{fetters2007chain} & 28 & \cite{fetters2007chain} & 438 & 10.1 (473) & \cite{graessley1981entanglement} & 1.9 & 17.9 \\
15 & {PC} &  & 254 & 85 & 60 & -- & 127 & \cite{fetters1994connection} & 490 & \cite{inoue1996role} & 1.3 & \cite{fetters1994connection} & -- &  & 426 & 2.4 (473) & \cite{wu1989chain} & 1.7 & 20.0 \\
 & \multicolumn{19}{l}{ \parbox[c]{3.0truecm}{\includegraphics[angle=0,width=3.0truecm]{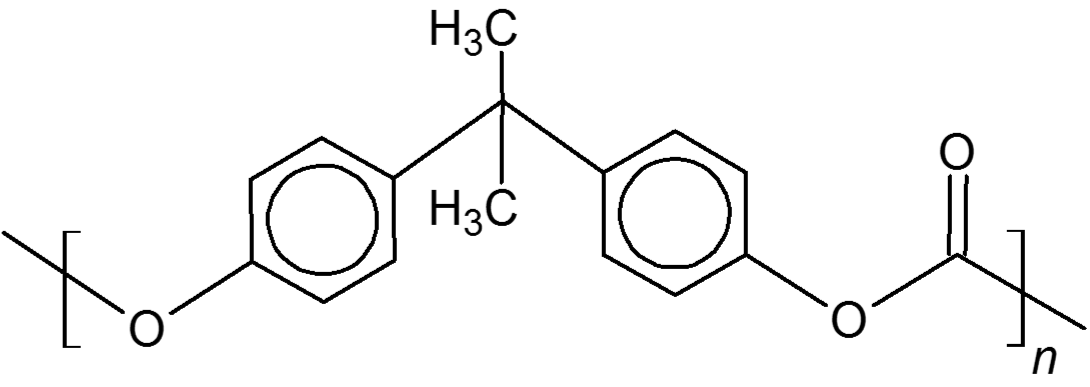}}} \\
17 & RADEL-R &  & 400 & 134 & 347 & -- & 113 & \cite{fetters2007chain} & -- &  & 1.6 & \cite{fetters2007chain} & -- &  & 502 & 2.0 (298) & \cite{roovers1990properties} & 2.8 & -- \\
 & \multicolumn{19}{l}{ \parbox[c]{3.0truecm}{\includegraphics[angle=0,width=3.0truecm]{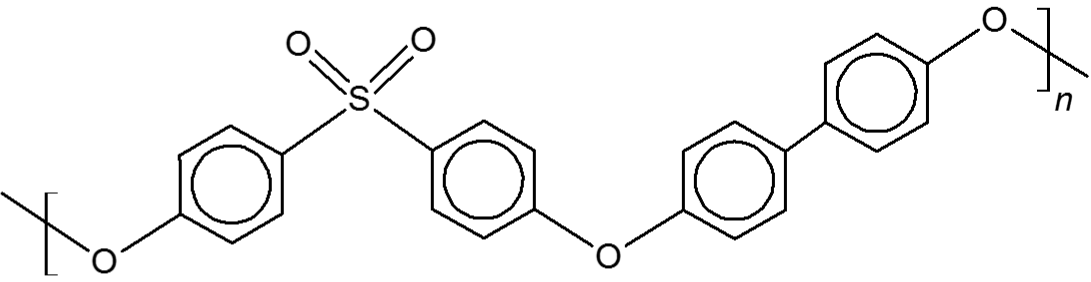}}} \\
26 & PDMS & \parbox[c]{1.6truecm}{\includegraphics[width=1.6truecm]{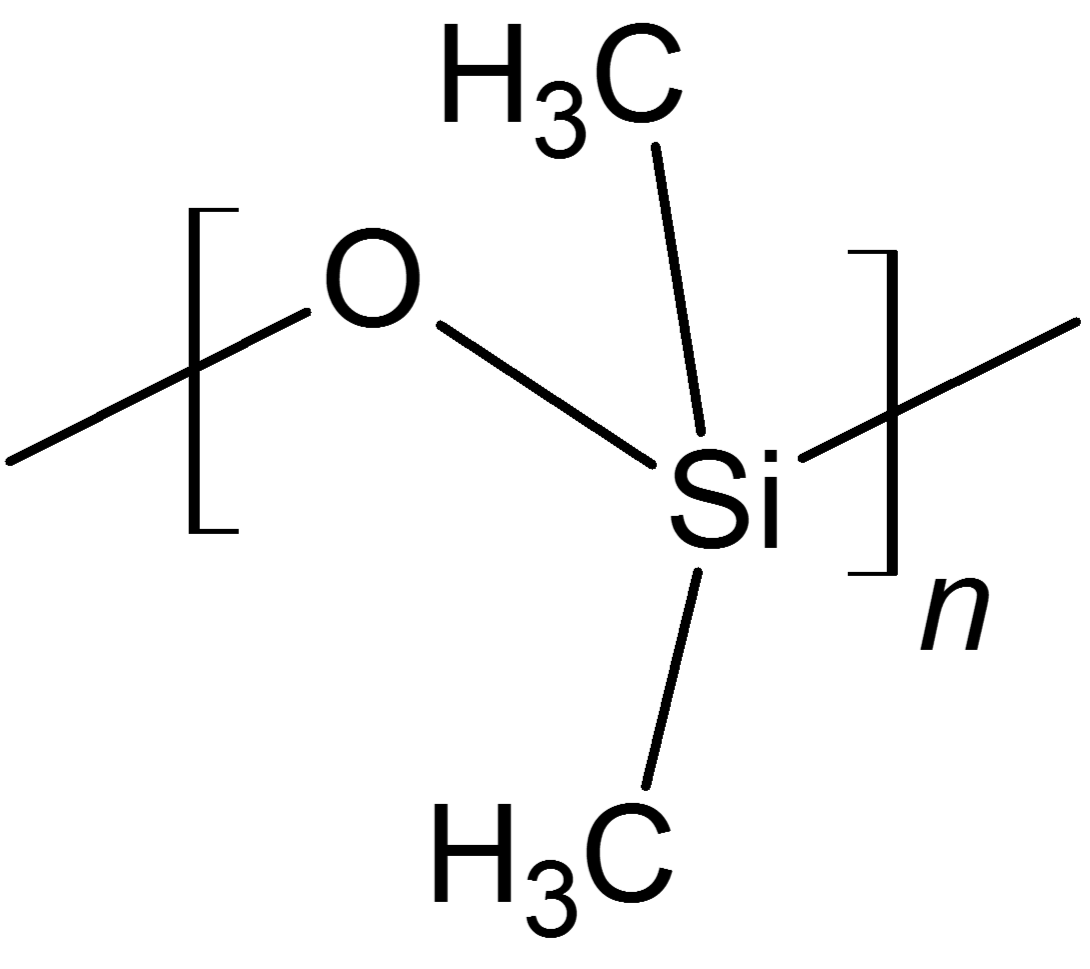}} & 74 & 37 & 38 & 162 & 381 & \cite{rubinstein2003polymer} & 600 & \cite{hofmann2012glassy} & 12 & \cite{fetters2007chain} & 25 & \cite{fetters2007chain} & 148 & \parbox{1.4truecm}{5.8~(298)  6.3~(413)}  & \cite{fetters1994connection} & 0.44 & 5.8 \\
\\\hline\hline
\label{tab:polymer_data}
\end{tabular}

\end{center}
\caption{Characteristic data for the polymers used in Figure 2. All parameters included in the table are described in the text. Masses are in g/mol  and are defined in Table~\ref{tab:MWs}. In each case the end group mass $M_{\textrm{end}}$ chosen is for a typical polymerization chemistry. $^a$Numbers in the first column correspond to the entries in Table~\ref{tab:fulllist}. $^{b}$References for $T_g^{\infty}$ are given in Table~\ref{tab:fulllist}. $^cC_{\infty}$ is given at the indicated temperatures. $^d$ Two values for the dynamic bead (Rouse) mass $M_R$ are given for PS, as reported in literature; 850 g/mol \cite{inoue1996role} or 5000 g/mol \cite{Ding2004comment}.} 
\label{tab:Ngai_params}
\end{table*}

The $T_g(M_n)$ literature data included in Fig. 2a-c are: PMMA  ($\color{orangeplot}\bullet$: \cite{Thompson1966dependence}, $\triangle$: \cite{o1991chain}, $\color{greenplot}\lozenge$: \cite{beevers1960physical}), PS ($\color{orangeplot}\bullet$: \cite{Hintermeyer2008molecular}, $\triangle$: \cite{claudy1983glass}, $\color{lightblueplot}\star$: \cite{Cowie1975some}, $\color{greenplot}{\blacklozenge}: $\cite{bartenev1988relaxational}, $\color{purpleplot}\blacktriangleleft$: \cite{santangelo1998molecular}) and PDMS ($\color{greenplot}\lozenge$: \cite{Hintermeyer2008molecular}, $\triangle$:  \cite{cowie1973molecular}, $\color{greyplot}\triangleright$: \cite{kirst1994molecular}). The $T_g(M)$ literature data included in Fig. 2d-e are: Radel-R \cite{roovers1990properties}, PC \cite{dobkowski1982influence,palczynski2017molecular}, PI \cite{abou2010rouse}, PIB \cite{greenberg1987evaluation}, PE \cite{miller1968kinetic}, PB \cite{colby1987melt,Hintermeyer2008molecular} and PPG-DME
\cite{mattsson2005influence} (for PPG-DME, the $T_g$-value for PPG of $M_n$=4000 g/mol ($n=69$) is included as a good approximation of the long-chain behavior for PPG-DME, since the influence of the hydroxyl end-groups is marginal for this high molecular weight \cite{mattsson2005influence,Bergman1998JNCS,Alper1976Polymer}). All data are number-averaged molecular weight $M_n$ except for Radel-R and one data set for PS \cite{bartenev1988relaxational} for which only $M_w$ is available.

Table~\ref{tab:Ngai_params} includes data for the eleven polymers included in the mastercurve in Fig.~2. The table includes the monomer repeat molecular weight $M_o$, the conformer molecular weight $M_{\phi}$ and volume $V_{\phi}$ (both defined below), the molecular weight of the chain-ends $M_{\textrm{end}}$, the Kuhn molecular weight $M_K$, the Rouse (or dynamic bead) molecular weight $M_R$, the entanglement molecular weight $M_e$, the critical molecular weight $M_c$ (at which entanglements are effective), the long-chain limit of the glass transition temperature, $T_g^{\infty}$, the long-chain limit of the Flory characteristic ratio, $C_{\infty}$, the molecular weight $M^{\star}$ that separates regimes I and II in the $T_g(M)$ behaviour, and the molecular weight $M^{\star\star}$ that separates regimes II and III. For the majority of polymers, where significant data were available in all three regimes, $M^{\star}$ was determined by fitting data within regimes I and II to the form $T_g=A_{I,II} + B_{I,II} \log_{10} M$, while $M^{\star\star}$ was determined as the molecular weight above which $T_g\simeq T_g^{\infty}$. For polymers where data were mainly, or only, available within two of the regimes (PI, PIB, PPG-DME and Radel-R), $M^{\star}$ was instead determined by optimisation to the mastercurve formed by the other polymers. For PE, data are only available within regime I and $M^{\star}$ was thus determined by optimisation to regime I data of the other polymers. The temperature at which $C_{\infty}$ was determined is noted in the table. For PAMS, PIB, PS and PC, the literature values for $M_R$ were determined from mechanical spectroscopy \cite{inoue1996role}, while for  PB, PDMS, PI and PPG-DME, the $M_R$ values were determined from Fast Field-Cycling Nuclear Magnatic Resonance (FFCNMR) \cite{hofmann2012glassy}. In both cases, the data were modeled as a superposition of $\alpha$-relaxation and Rouse relaxation spectrum contributions, where a linear superposition of either moduli or compliances (susceptibilities) were performed.

To calculate $M_{\phi}$ and $V_{\phi}$, we count the relevant number of conformational degrees of freedom (DOF), or conformers, per monomer, where we include the number of conformers $n_{\phi}$ that sweep out significant volume during a rearrangement. A dihedral rotation is counted as a conformer whether it is situated in the backbone or in a side-chain, and we also count an aromatic ring rotation, a cyclohexane group rotation, or a chair/boat conformational change as a conformer. However, we ignore groups whose motions displace small volumes, such as methyl groups, aromatic ring rotations within the backbone (such as in PET), and dihedrals  involving small groups such as CH=CH$_2$ in 1,2 PB, or O-CH$_3$ in PMMA. The mass per conformer $M_{\phi}$ is subsequently defined as the mass per monomer (or polymerization unit) $M_o$ divided by the total number of conformers per monomer $n_{\phi}$, as $M_{\phi}=M_o/n_{\phi}$. $M_{\phi}$ thus averages the conformational DOF within the monomer, representing a particular polymer chemistry. The average volume per conformer $V_\phi=V_{\textrm{mon}}/n_{\phi}$ is calculated from the sum $V_{\textrm{mon}}$ of the van der Waals volumes of all groups in the monomer, tabulated in Ref.~\cite{vankrevelenbook}.

Table~\ref{tab:fulllist} provides data for $M_{\phi}$, the number of conformers per monomer $n_{\phi}$, $M_o$, $V_\phi$, and $T_{g}^{\infty}$ for a wider range of polymers with C-, C-C-O-, Si- or Si-O-based backbones, as shown in the $T_g(M_{\phi})$ plot in Fig.~\ref{fig:Tg_values_PMMAPS}f. 


\begin{table*}[htbp]
\begin{tabular}{lllcccccccccc}
\hline\hline
\# & Polymer & Acronym & $M_{\phi}$ & $n_{\phi}$ & $M_0$ & $V_{\phi}$ & $T_g^{\infty}$ & Ref. & $p $ & $\ell_K $ & $T_{\textrm{char}}^a$ & Backbone \\
 &  &  &  &  &  & $\textrm{\AA}^3$ & K &  & \textrm{\AA} & \textrm{\AA} & K \\\hline
1 & poly(ethylene) & PE  & 14 & 2 & 28 & 17 & 200 & \cite{miller1968kinetic} & 1.39 & 1.54 & 298 & C \\
2 & 1,4-poly(butadiene) & 1,4-PB  & 18 & 3 & 54 & 63 & 175 & [$\ast$] & 2.44 & 8.28 & 298 & C \\
3 & poly(propylene)$^{b}$ & PP  & 21 & 2 & 42 & 51 & 266 & \cite{cowie1973glass} & 1.12 & 2.88 & 298 & C \\
4 & poly(vinylethylene)$^{b}$ & PVE  & 27 & 2 & 54 & 63 & 273 & \cite{shenogin2007dynamics} &  & 14 &  & C \\
5 & poly(isobutylene) & PIB & 28 & 2 & 56 & 68 & 210 & [$\ast$] & 3.18 & 12.50 & 298 & C \\
6 & poly(vinyl chloride)$^{b}$ & PVC  & 31 & 2 & 63 & 49 & 354 & \cite{ribelles1987glass} &  &  &  & C \\
7 & poly(ethylene terephthalate) & PET  & 32 & 6 & 192 & 161 & 346 & \cite{zhang2004glass} & 1.99 & 14.91 & 548 & C \\
8 & poly(vinylidene fluoride) & PVDF & 32 & 2 & 64 & 43 & 238 & \cite{mccrum1967anelastic} &  &  &  & C \\
9 & 1,4 poly(isoprene) & 1,4-PI & 34 & 2 & 68 & 79 & 213 & [$\ast$] & 2.69 & 9.34 & 298 & C \\
10 & poly(vinylidene chloride) & PVDC & 49 & 2 & 97 & 63 & 255 & \cite{mccrum1967anelastic} &  &  &  & C \\
11 & poly(methyl methacrylate)$^{b}$ & PMMA  & 50 & 2 & 100 & 95 & 387 & [$\ast$] & 3.77 & 15.30 & 413 & C \\
12 & poly(styrene)$^{b}$ & PS & 52 & 2 & 104 & 107 & 374 & [$\ast$] & 3.92 & 17.80 & 413 & C \\
13 & poly(phenylene sulfide) & PPS  & 54 & 2 & 108 & 94 & 348 & \cite{d1997poly} &  &  &  & C \\
14 & poly($\alpha$-methyl styrene)$^{b}$ & PAMS  & 59 & 2 & 118 & 124 & 438 & [$\ast$] & 3.61 & 20.43 & 473 & C \\
15 & poly(carbonate) of bisphenol A & PC  & 85 & 3 & 254 & 239 & 426 & [$\ast$] & 1.69 & 18.43 & 473 & C \\
16 & poly(ether ether ketone) & PEEK  & 92 & 3 & 276 & 257 & 437 & \cite{sandler2002carbon} &  &  &  & C \\
17 & poly(4,4$^{\prime}$-biphenol-\emph{alt}-dichlorodiphenyl sulfone) & Radel-R & 111 & 4 & 444 & 347 & 502 & [$\ast$] & 1.66 &  &  & C \\
18 & poly(phenyl ether) & PPE  & 120 & 1 & 120 & 81 & 484 & \cite{matsuoka1997entropy} &  &  &  & C \\\hline
19 & poly (ethylene glycol) & PEG  & 15 & 3 & 44 & 42 & 213 & \cite{tormala1974determination} & 1.95 & 9.71 & 353 & C-C-O \\
20 & poly (propylene glycol) dimethyl ether$^{b}$ & PPG-DME & 19 & 3 & 58 & 59 & 197 & [$\ast$] & 2.77 &  &  & C-C-O \\\hline
21 & poly(di-n-hexylsilane) & PDHS  & 18 & 11 & 198 & 243 & 221 & \cite{varma1991thermal} &  &  &  & Si \\
22 & poly(propylmethylsilane)$^{b}$ & PPrMS  & 29 & 3 & 86 & 107 & 245 & \cite{fujino1992preparation} &  &  &  & Si \\
23 & poly(trifluoropropylmethylsilane)$^{b}$ & PTFPrMS  & 47 & 3 & 140 & 120 & 270 & \cite{fujino1992preparation} &  &  &  & Si \\
24 & poly(cyclohexylmethylsilane)$^{b}$ & PCHMS  & 63 & 2 & 125 & 188 & 366 & \cite{fonseca1995plasma} &  &  &  & Si \\
25 & poly(phenylmethylsilane)$^{b}$ & PPMS  & 120 & 1 & 120 & 126 & 390 & \cite{fonseca1995plasma} &  &  &  & Si \\\hline
26 & poly(dimethylsiloxane) & PDMS  & 37 & 2 & 74 & 76 & 148 & [$\ast$] & 4.06 & 11.40 & 298 & Si-O \\
27 & poly(methylphenylsiloxane)$^{b}$ & PMPS  & 68 & 2 & 136 & 135 & 228 & \cite{sengupta1994ultrafast} &  &  &  & Si-O \\
\hline\hline
\end{tabular}
\caption{Table of molecular weights (in g/mol) and volumes per conformer, ordered in increasing mass per total conformer (\textit{i.e.} including the side groups but excluding methyl groups), and separated according to backbone chemistry.  $^a$Polymer packing length $p$ and Kuhn step $\ell_K$ have been characterized at temperature $T_{\textrm{char}}$ . $^b$Stereoisomeric  polymers are quoted for atactic materials. In some cases the tacticity is known and published, while in other cases it is not known. $T_g$ for PE was determined by extrapolation from Ref.~\cite{miller1968kinetic}; while for polymers with references noted as [$\ast$], $T_g^{\infty}$ was determined as the high-$M$ limit of data referenced and shown in Figure 2.}
\label{tab:fulllist}
\end{table*}

\begin{figure*}[htbp]
\includegraphics[width=1\textwidth]{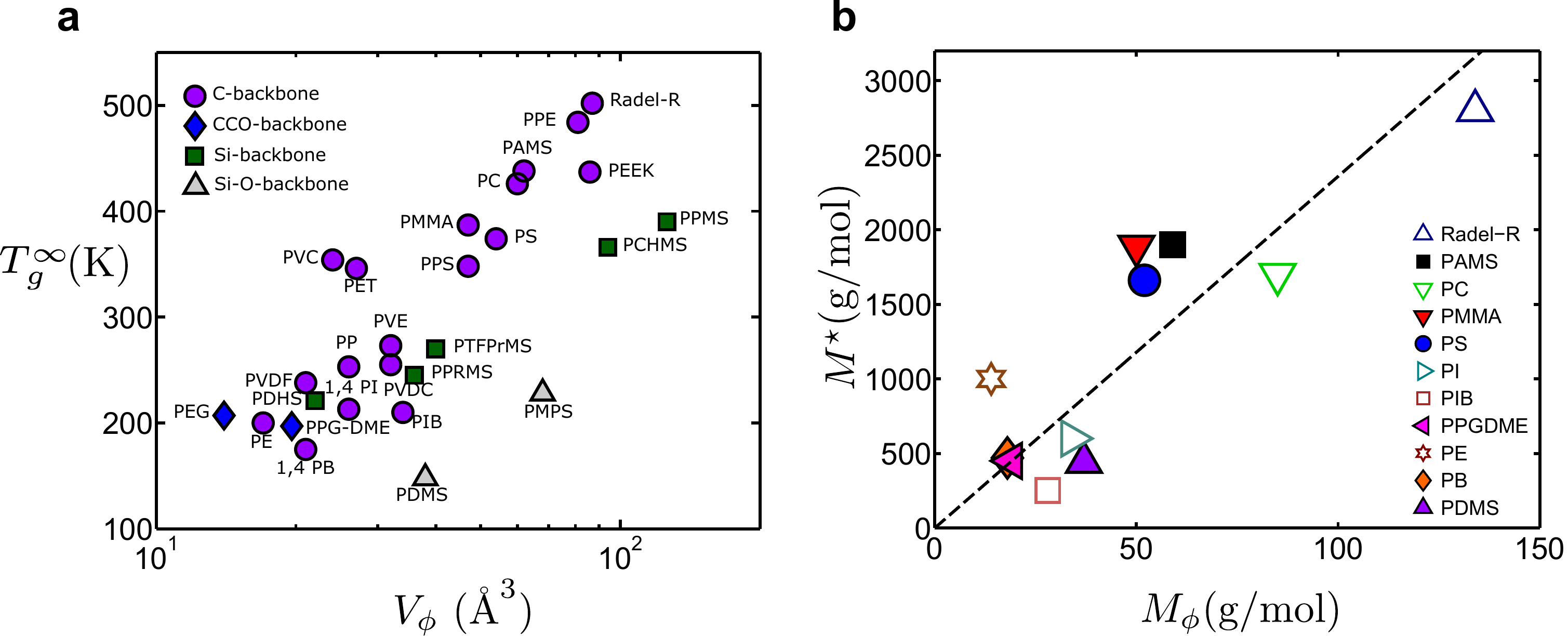}
\caption{(a) $T_{g}^{\infty}$ vs $V_{\phi}$ for the polymer systems in Table~\ref{tab:fulllist}; the different symbols/colors refer to different backbone chemistries. 
(b) $M^{\star}$ vs $M_{\phi}$. The dashed line is a linear fit yielding, $M^{\star} \simeq 24 M_{\phi}$. Open symbols in (b) denote polymers with less certainty in $M^{\ast}$ due to data that do not  cover all three regimes in $T_g(M)$. 
}
\label{fig:Tg_Vconf}
\end{figure*}

As a complement to Figure 2e, Figure \ref{fig:Tg_Vconf}a shows the dependence of $T_{g}^{\infty}$ on $V_{\phi}$ (in $\textrm{\AA}^3$) for the polymers in Table~\ref{tab:fulllist}, demonstrating a rough correlation  $T_{g}^{\infty} = T_V +  B_V\log_{10}(V_{\phi})$, where $T_V$ depends on the sequence and species of atoms in the polymer backbone, and $B_V\sim 300\,\textrm{K}$ for carbon-based backbones, but at least for the Si-based backbones appears to be somewhat smaller. $M^{\star}$ is plotted versus $M_{\phi}$ in Fig. \ref{fig:Tg_Vconf}b to investiate the inter-relationship between the two characteristic molecular weights. We find  $M^{\star}\approx 24 M_{\phi}$, consistent with the crossover between regime I and II occuring when the chain has reached a length corresponding to $\sim$24 conformers. 

\begin{figure*}[htbp]
\begin{center}
\includegraphics[angle=0,width=1\textwidth]{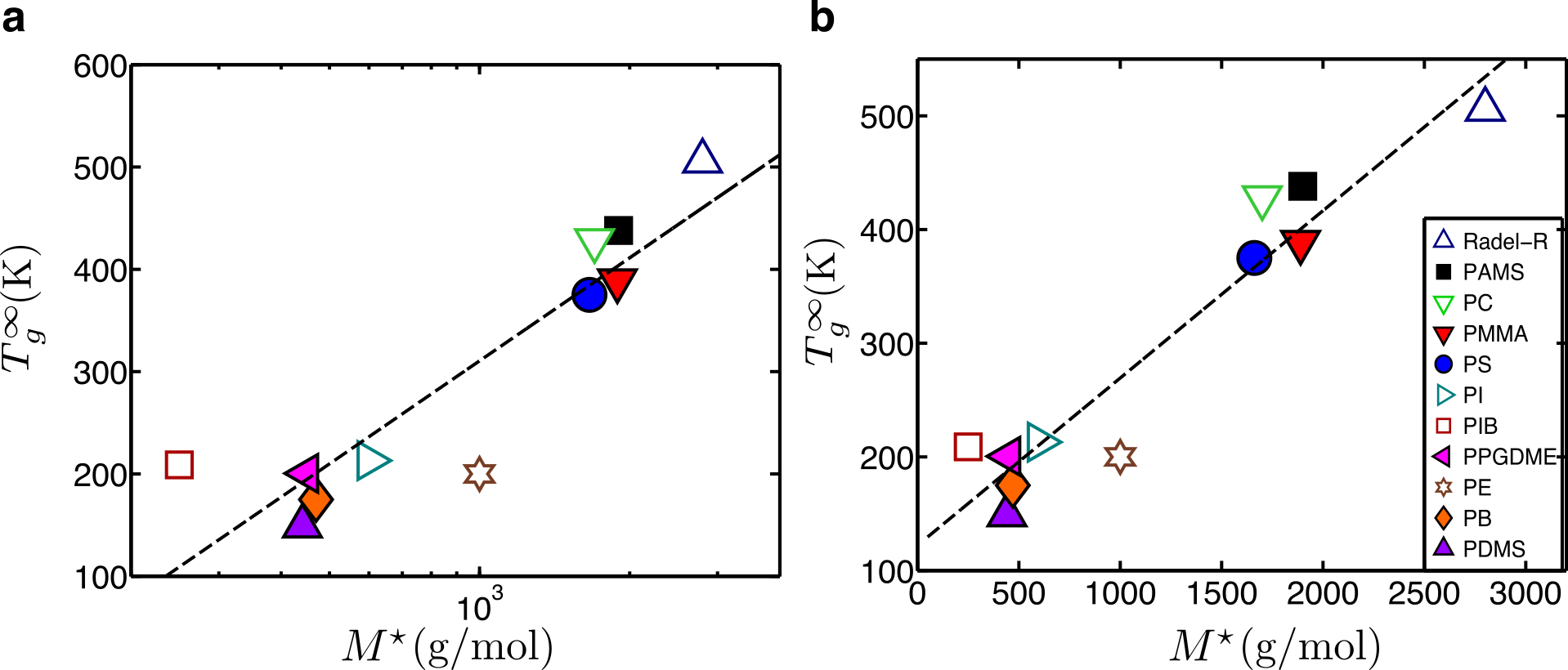}
\end{center}
\caption{$T_{g}^{\infty}$ vs $M^{\ast}$ in a (a) semi-logarithmic and (b) linear plot. Open symbols denote polymers with less certainty in $M^{\ast}$ due to data that do not  cover all three regimes in $T_g(M)$.}
\label{fig:spectra_fitlines}
\end{figure*}

Figure \ref{fig:spectra_fitlines}a shows $T_{g}^{\infty}(M)$ for the 11 polymers of Table~\ref{tab:Ngai_params}, either in a (a) semi-logarithmic, or (b) linear plot. The comparison between the two panels demonstrates that the relationship between $T_{g}^{\infty}$ and $M$ can, to a good approximation, be described using either a semi-logarithmic or linear form (see inset in Fig. 2e). {We have less confidence in $M^{\ast}$ for those polymers for which data covering all three regimes are not available (PE, PI, PIB, PC, and Radel-R); these data are shown as open symbols in Fig.~\ref{fig:Tg_Vconf}b and Fig.~\ref{fig:spectra_fitlines}.}

\begin{figure*}[htbp]
\begin{center}
\includegraphics[angle=0,width=1\textwidth]{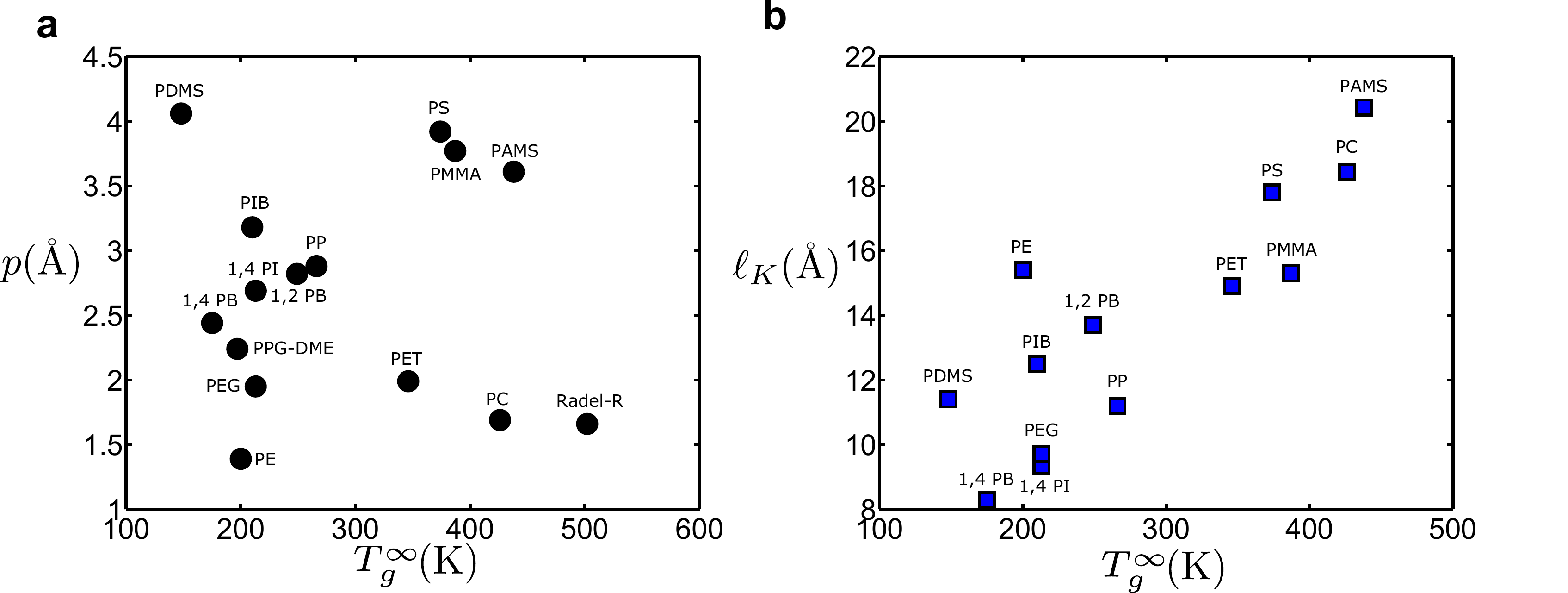}
\end{center}
\caption{Relation between $T_g^{\infty}$ and the packing length $p$ and the Kuhn step $\ell_K$ for: PE 1,4-PB, PP, 1,2-PB, PIB, PET, 1,4-PI,PMMA, PS, PAMS, PC, Radel-R (only $p$), PEG, PPG-DME (only $p$), and PDMS. }
\label{fig:packKuhn}
\end{figure*}

Fig.~\ref{fig:packKuhn}a shows the relation of the packing length $p$ or  Kuhn length $\ell_K$ vs $T_g^{\infty}$. The packing length is defined as the high molecular weight limit of $p=V/R_g^2$, where $V$ is the polymer volume. If we (naively) approximate a Kuhn volume as a cylinder of length $\ell_K$ and diameter $d$, we find $p\sim d^2/\ell_K$. The packing length quantifies the balance between intra- and inter-chain interactions, and has a strong correlation with  metrics such as the entanglement or critical molecular weights $M_e$ and $M_c$ \cite{fetters1994connection}. Fig.~\ref{fig:packKuhn}a shows no obvious correlation between $p$ and $T_g^{\infty}$. However, stiffer chains characterised by larger Kuhn lengths typically have higher $T_g^{\infty}$, as shown in Fig.~\ref{fig:packKuhn}b. 

\section{Data for non-polymeric `rigid' glass-formers}\label{app:rigid}

\begin{figure*}[htbp]
\begin{center}
{\includegraphics[width=1\textwidth]{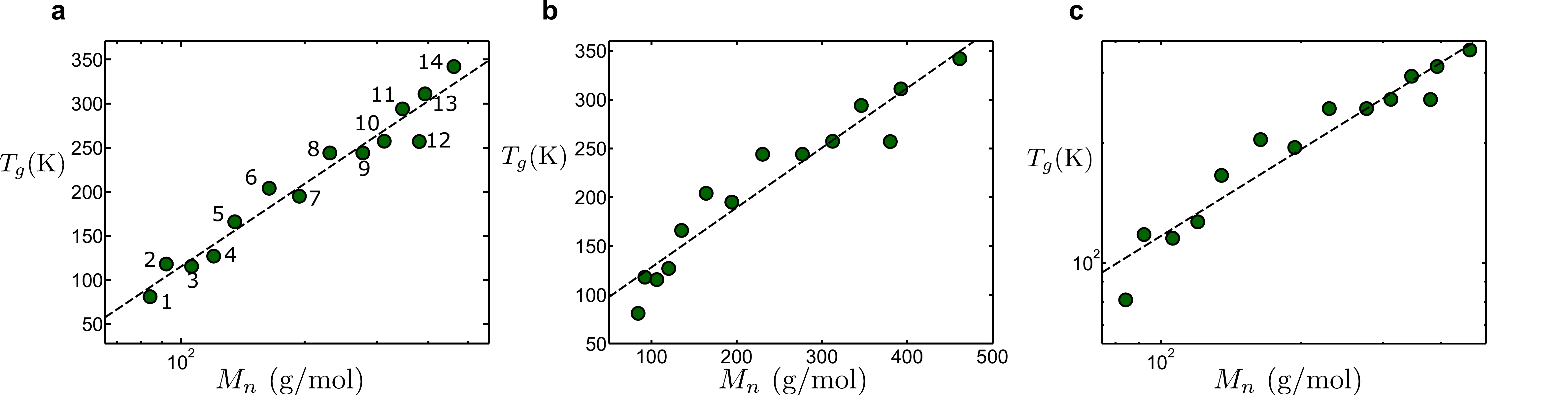}}
\end{center}
\caption{$T_g$ as a function of molecular weight $M$ for rigid molecules from Table~\ref{tab:larsen}, as shown in a semi-logarithmic (a), linear (b), and double logarithmic (c) representation. The dashed lines are linear fits to the data in each representation and provide a guide-to-the-eye for the evaluation of the degree of linearization provided in each case.}
\label{fig:rigiddiffscaling}
\end{figure*}

To investigate the molecular weight dependent $T_g$-behaviour for non-polymeric `rigid' glass-formers with as few conformational degrees of freedom as possible, we follow Ref.~\cite{larsen_effect_2011} and choose a series of mainly aromatic, carbon-based molecules, which do not contain alkane chains of more than three carbons. We expect all the chosen systems to interact in a similar manner, which allows for direct comparisons. The $T_g$-values were taken from Ref.~\cite{larsen_effect_2011} (with the addition of bisphenol A diacetate, number 10 in Table~\ref{tab:larsen}). Table \ref{tab:larsen} contains the molecular structure, chemical name, molecular weight $M$, and $T_{g}$. Fig.~\ref{fig:rigiddiffscaling} shows $T_{g}(M)$ for the `rigid' molecules in a semi-logarithmic (a), linear (b), or double logarithmic (c) representation; the  semi-logarithmic plot (Fig.~\ref{fig:rigiddiffscaling}a) provides the best fit. We stress that even though a semi-logarithmic fit describes our chosen data best, this is not necessarily the case for other series of `rigid' molecules. 

Novikov and R\"ossler studied $T_g(M)$ \cite{novikov2013correlation} for a wide range of non-polymeric (and some oligomeric) glass-formers of different chemistries and interactions. Their entire data set could be fit to a power law $T_g(M)\sim M^{\alpha}$, with  $\alpha\approx 0.5$. However, a subset of aromatic molecules was best fit by $\alpha\approx 0.7$. For comparison, our data (Fig.~\ref{fig:rigiddiffscaling}c) yield $\alpha\approx 0.7$.

\begin{table*}[htbp]
\begin{center}
\begin{tabular}{cclcl}
 \hline\hline\\[-8truept]
 & Structure & Name & $M$ & $T_g$\\
 &&&g/mol&K\\\hline
1 & \parbox[c]{0.7truecm}{\includegraphics[width=0.7truecm]{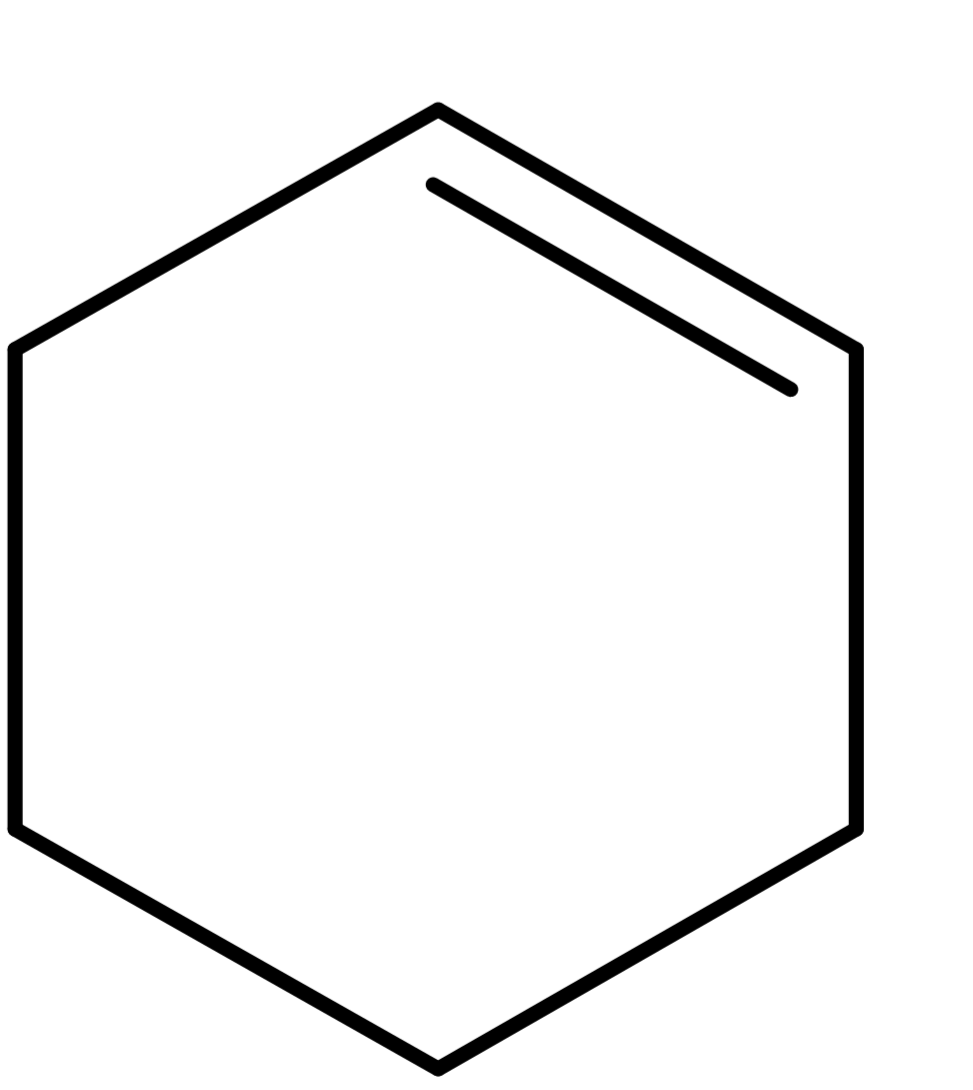}} & cyclohexene  & 84.2 & 81 \\
2 & \parbox[c]{1.5truecm}{\includegraphics[angle=-10,width=1.5truecm]{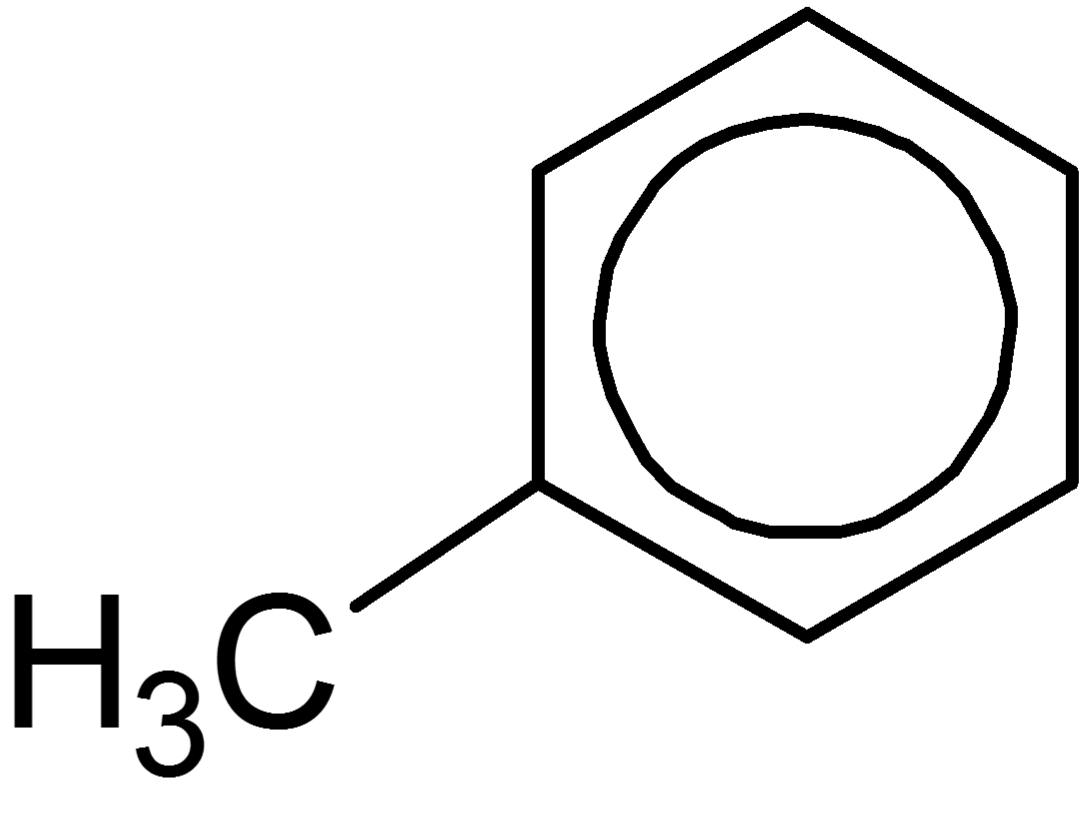}} & toluene  & 92.1 & 113 \\
3 & \parbox[c]{1.5truecm}{\includegraphics[width=1.5truecm]{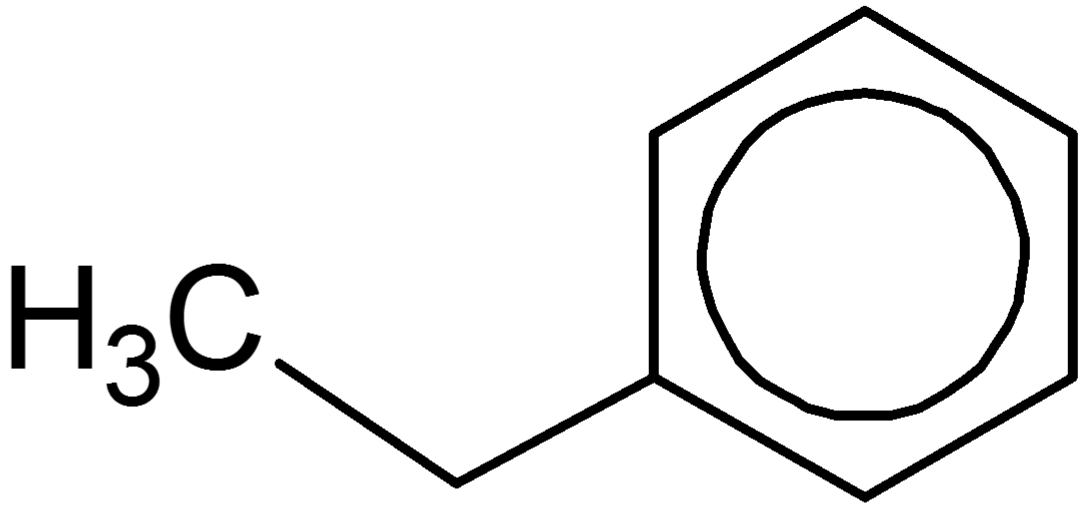}} & ethylbenzene  & 106 & 111 \\
4 & \parbox[c]{1.5truecm}{\includegraphics[width=1.5truecm]{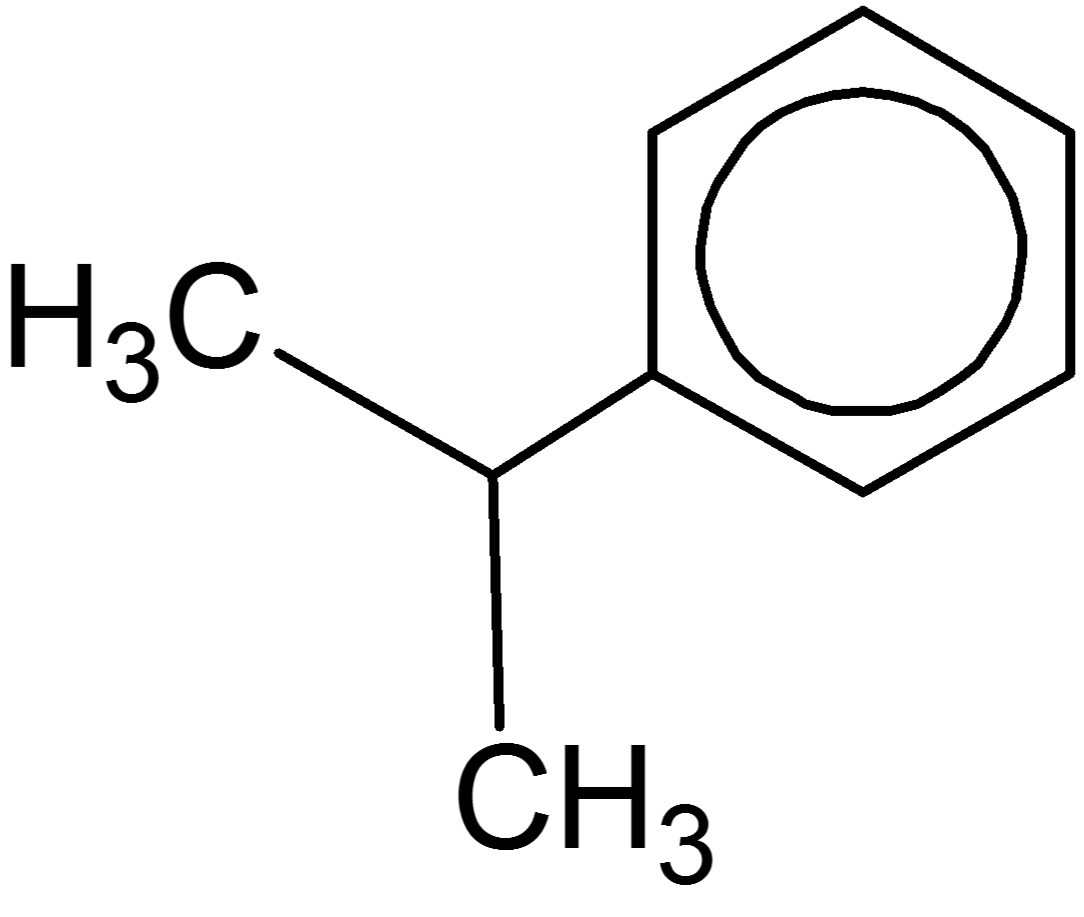}} & iso-propylbenzene  & 120 & 127 \\
5 & \parbox[c]{1.5truecm}{\includegraphics[width=1.5truecm]{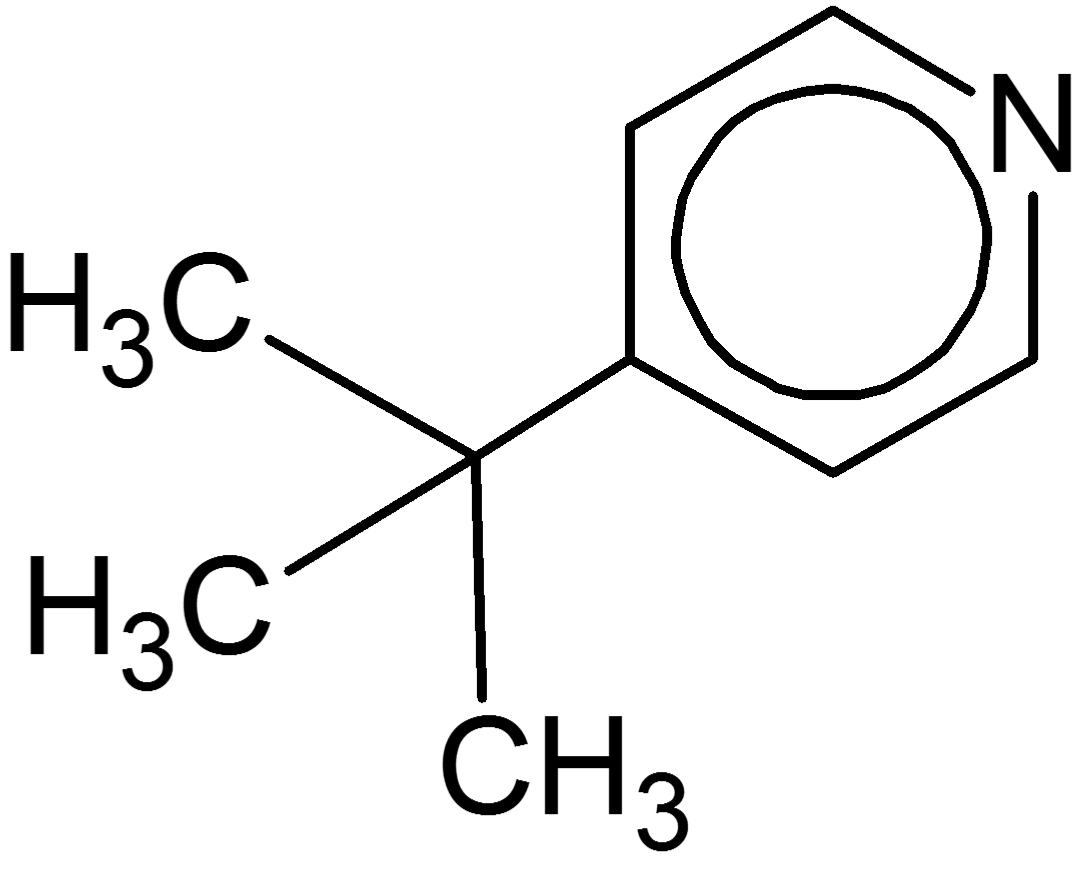}} & 4-tert-butyl-pyridine (4-TBP)  & 135 & 166 \\
6 & \parbox[c]{2.3truecm}{\includegraphics[width=2.3truecm]{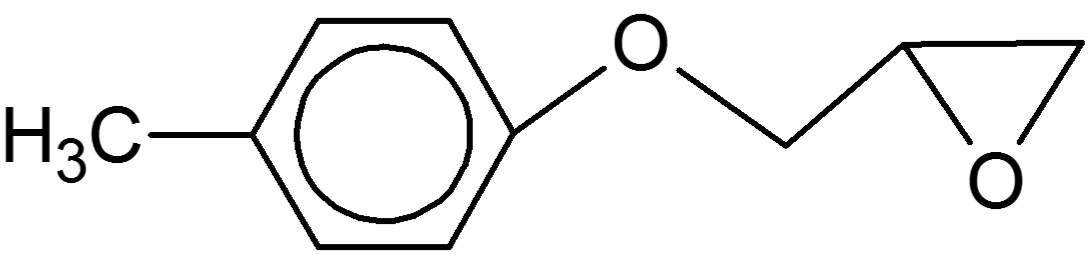}} & cresyl-glycidyl-ether (CGE)  & 164 & 204 \\
7 & \parbox[c]{1.8truecm}{\includegraphics[width=1.8truecm]{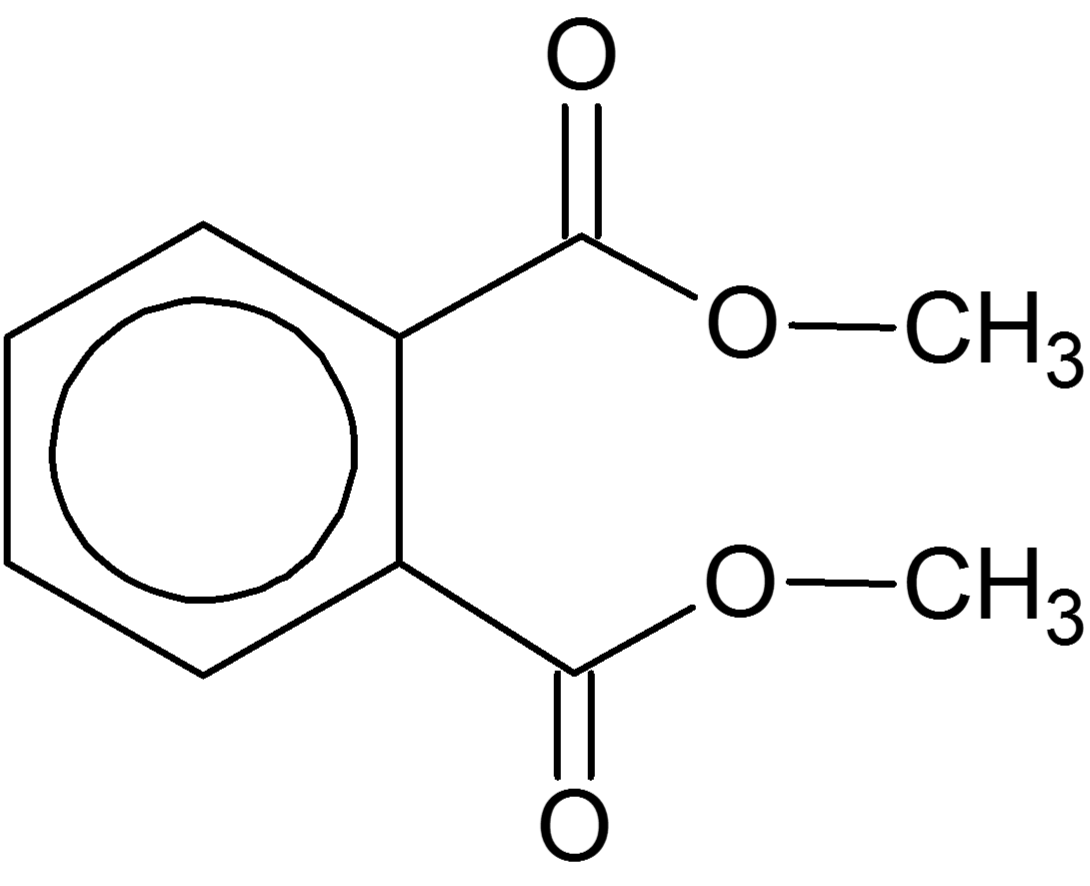}} & dimethylphthalate (DMP)  & 194 & 195 \\
8 & \parbox[c]{1.5truecm}{\includegraphics[width=1.5truecm]{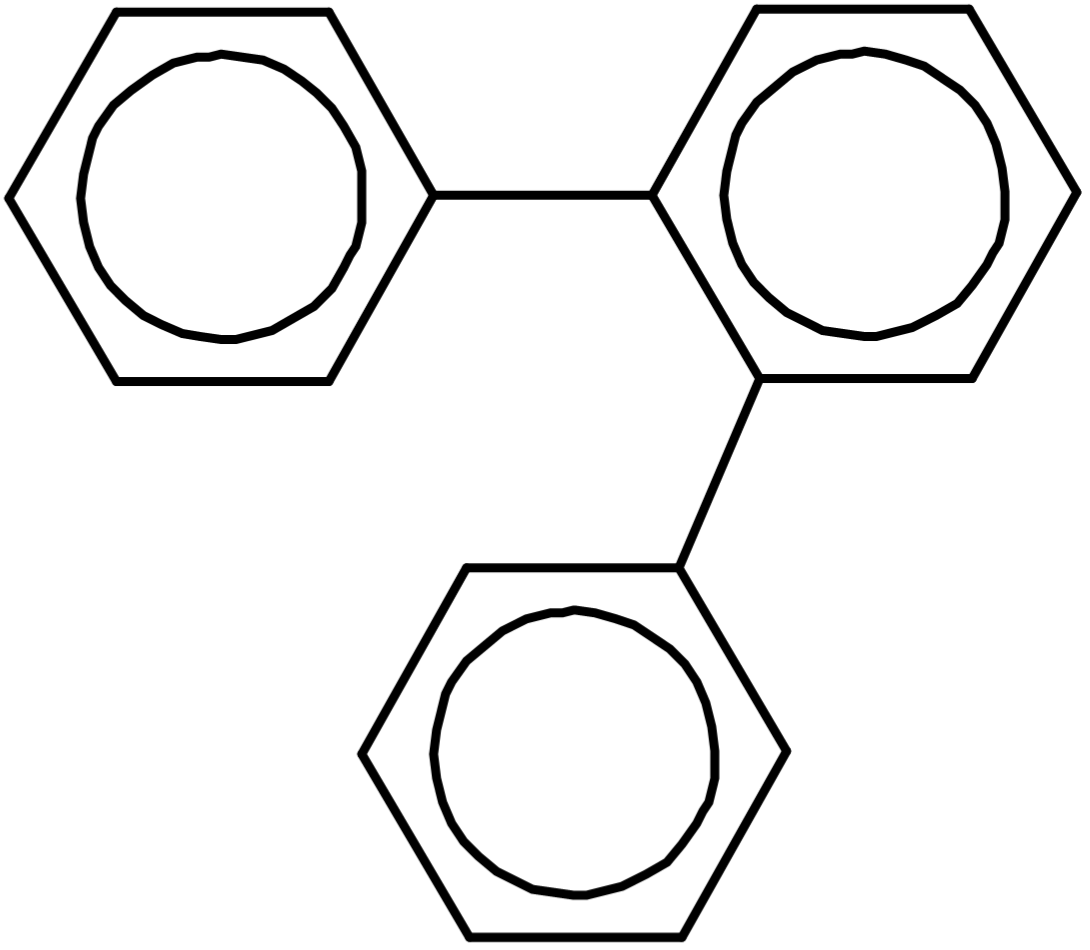}} & ortho-terphenyl  & 230 & 244 \\
9 & \parbox[c]{1.9truecm}{\includegraphics[width=1.9truecm]{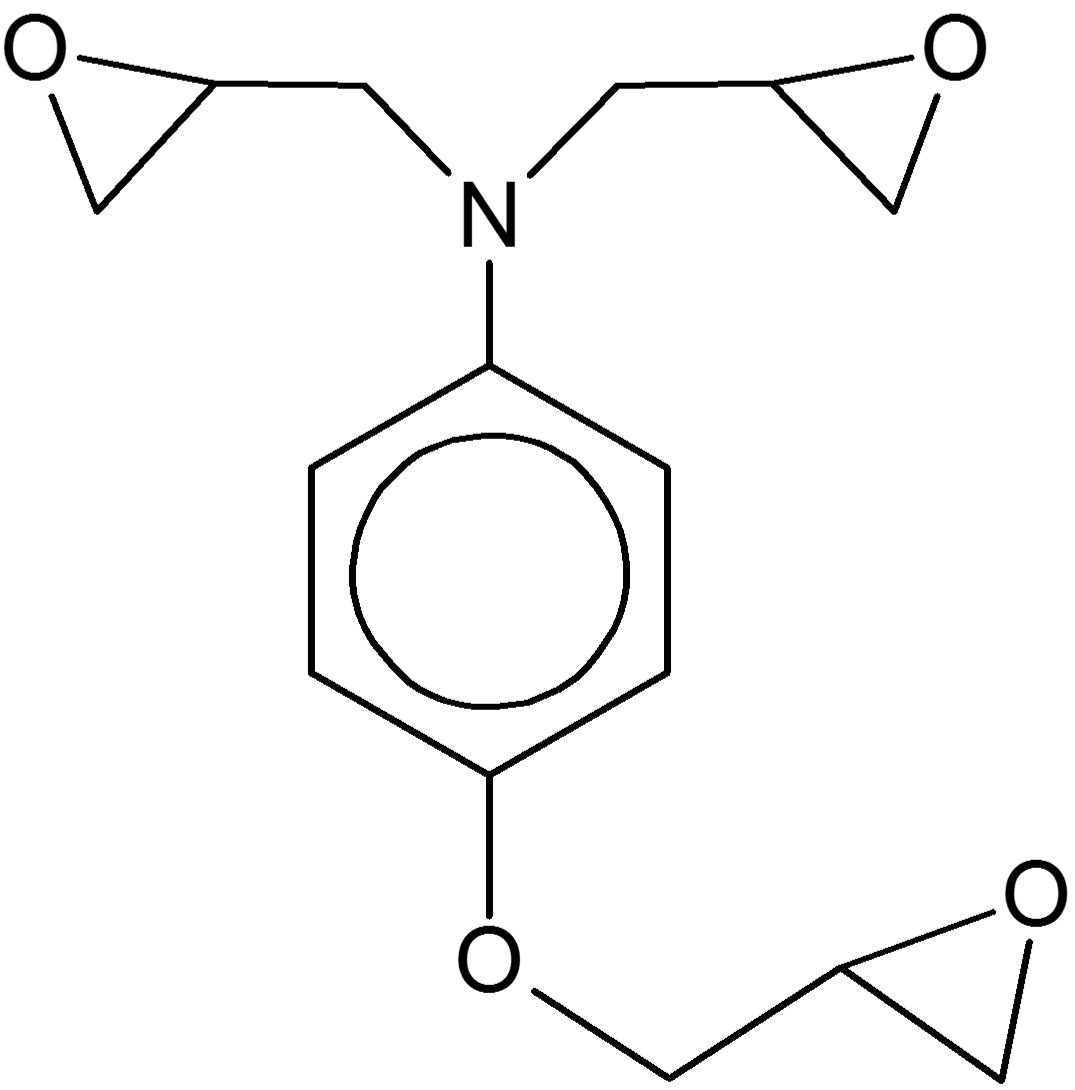}} & triepoxide N-N-diglycidyl-4-glycidyloxyaniline (DGGOA)  & 277 & 244 \\
10 & \parbox[c]{3.3truecm}{\includegraphics[width=3.3truecm]{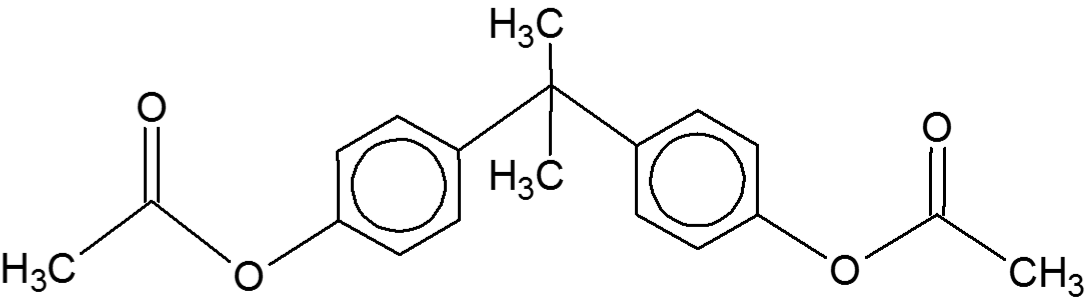}} & bisphenol A diacetate & 312 & 257 \\
11 & \parbox[c]{2.5truecm}{\includegraphics[width=2.5truecm]{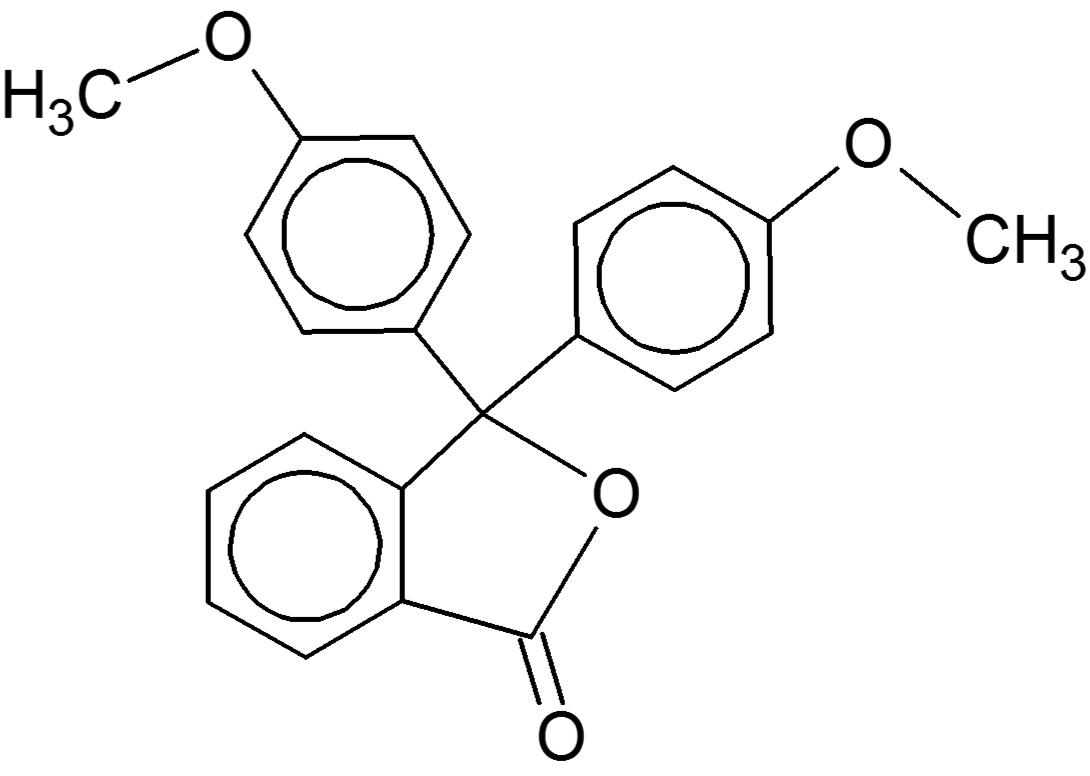}} &  phenolphthalein-dimethylether (PDE)  & 340 & 294 \\
12 & \parbox[c]{2.5truecm}{\includegraphics[width=2.5truecm]{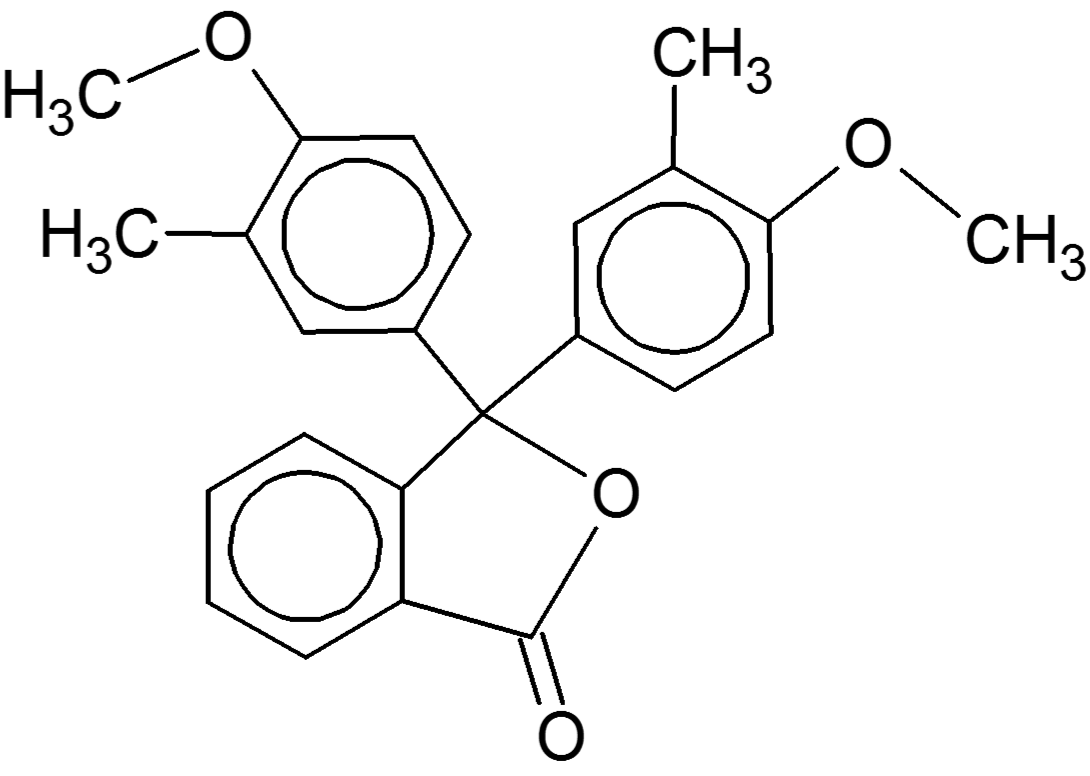}} & kresolphtalein-dimethylether (KDE)  & 376 & 311 \\
13 & \parbox[c]{2.5truecm}{\includegraphics[width=2.5truecm]{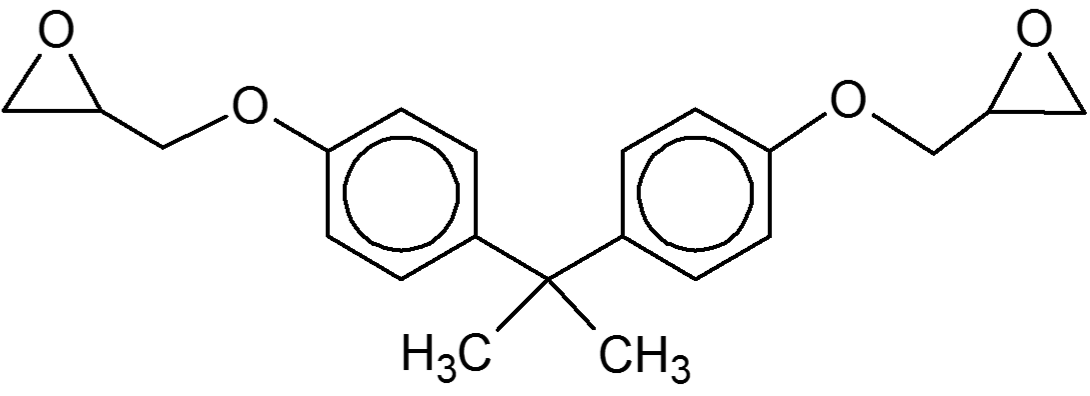}} & diglycyl ether of bisphenol A (DGEBA)  & 380 & 257 \\
14 & \parbox[c]{2.5truecm}{\includegraphics[width=2.5truecm]{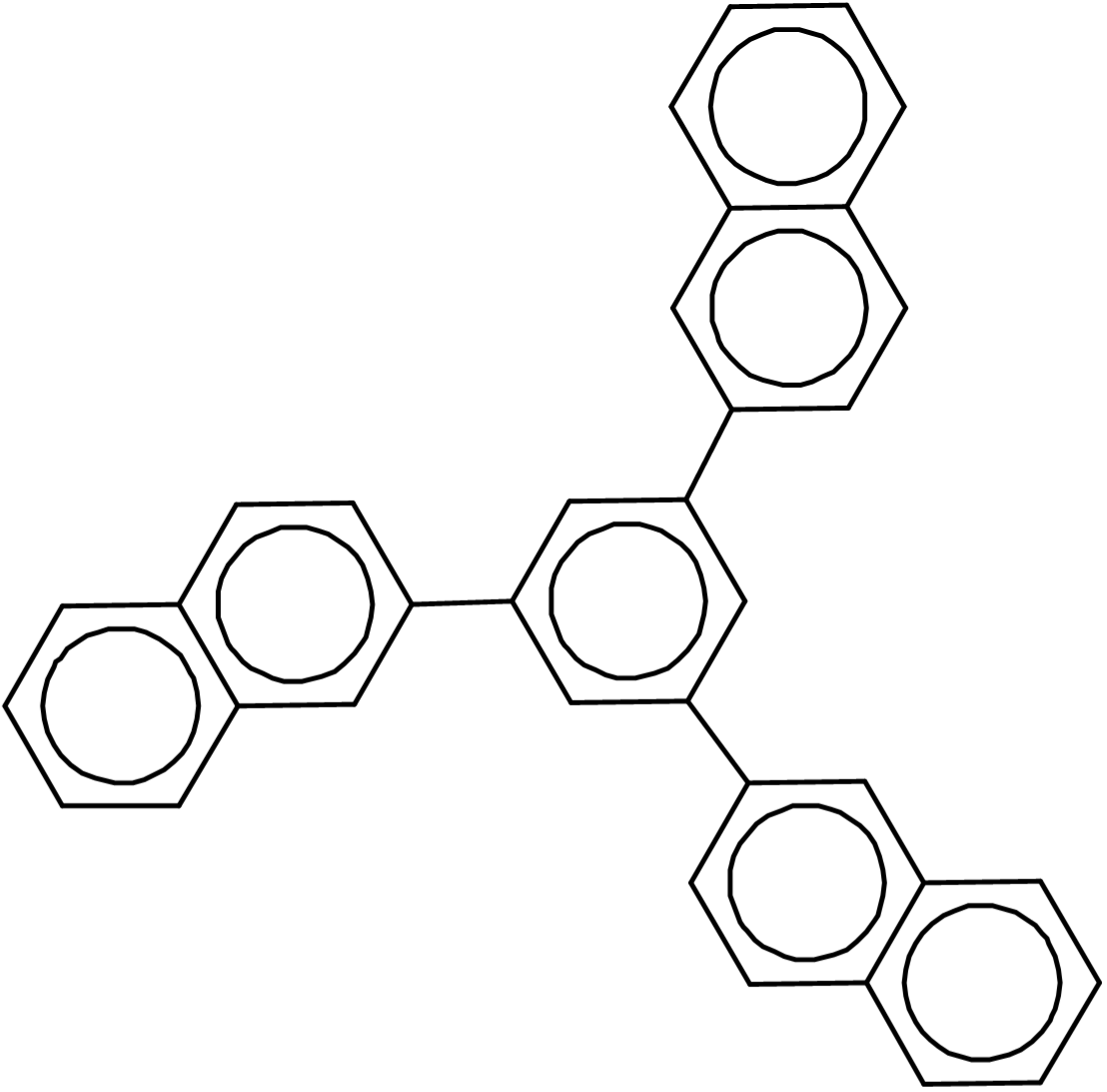}} & 1,3,5-tri-1-naphthyl benzene  & 456 & 342 \\
 \\[8truept]
\hline\hline
\end{tabular}
\end{center}
\caption{Table of data compiled by Larsen and Zukowski \cite{larsen_effect_2011}, with the addition of molecule 10, for rigid, mainly aromatic, small molecular glass formers. These are ordered according to  increasing molecular weight.}
\label{tab:larsen}
\end{table*}

\section{Secondary $\beta$ and $\gamma$ relaxations}\label{app:secondary}
\subsection{General}
\begin{figure}
\includegraphics[width=0.5\textwidth]{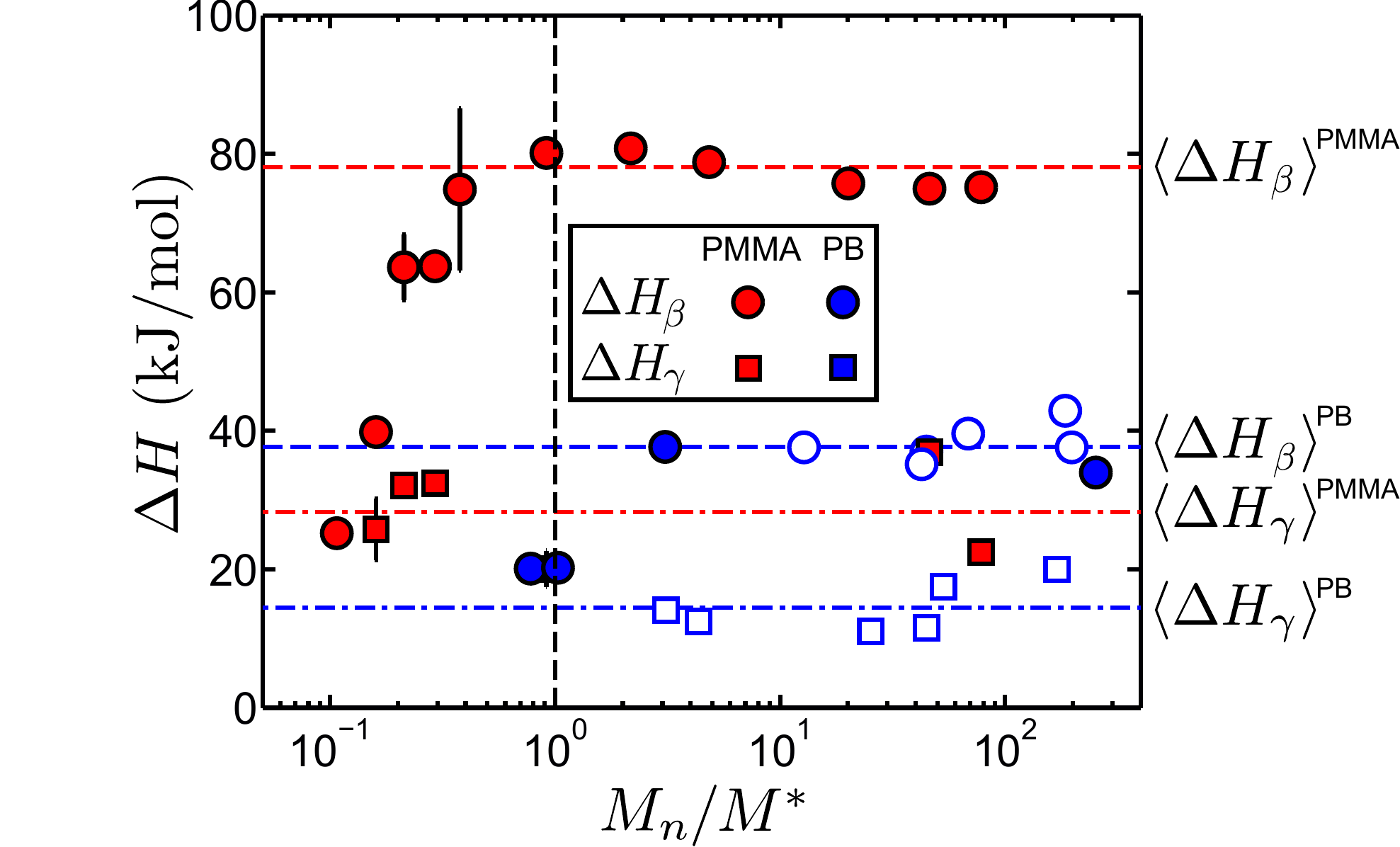}
\caption{Activation enthalpies $\Delta H_{\beta,\gamma}(M)$ for PMMA and PB.  Circles denote $\Delta H_{\beta}$ and squares denote $\Delta H_{\gamma}$. PB data from literature \cite{arbe1996study,hofmann1996secondary,deegan1995dielectric,richter1992decoupling,korber2017nature,lusceac2005secondary} are shown in open symbols.}
\label{fig:fig5new}
\end{figure}

Secondary relaxations are generally observed in polymer glasses. For the 11 polymer systems investigated in detail in this work, secondary relaxations have been experimentally reported in all except Radel-R for which only very few detailed spectroscopic investigations have been performed \cite{roovers1990properties,Roovers1990HighPerfPol_2}; other polysulfones show secondary relaxations \cite{Fried1990polymer} suggesting that they would be found also in Radel-R by detailed experimental investigation. The other 10 polymers all demonstrate secondary relaxations, see e.g. PAMS \cite{Crissman1990JPolPhysA}, PC \cite{alegria2007JNCS,Alegria2006macromol,Jho1991macromol}, PMMA \cite{kremer2012broadband,Schmidt-Rohr1994Macromol,Kuebler1997Macromol}, PS \cite{Arrese-Igor2011Macromol}, PI \cite{Kolodziej2018Macromol,Roland2004Macromol}, PIB \cite{Arbe1998Macromol}, PPG-DME \cite{Mattsson2009PRB}, PE \cite{Boyd1974Macromol,Atvars1993EurPolJ}, PB \cite{Casalini2000JPolSci,korber2017nature,lusceac2005secondary}, PDMS\cite{bershtein1994role,kirst1994molecular}. 

Depending on the particular polymer chemistry and the experimental technique used (e.g. Broadband Dielectric Spectroscopy (BDS), Nuclear Magnetic Resonance (NMR), Neutron Spin Echo (NSE) and Dynamic Light Scattering (DLS)), the experimental sensitivity to specific molecular motions can vary significantly, However, secondary relaxations in glassy polymers are typically assigned to molecular rearrangements that include both backbone and side-group rotations \cite{kremer2012broadband,Schmidt-Rohr1994Macromol} even though the exact rearrangements are often difficult to determine and literature assignments often vary depending on the experimental or computational technique used or the dynamic range investigated. A lot of work has focussed long-chain PMMA and PB. Detailed NMR and BDS studies on PMMA \cite{kremer2012broadband,Schmidt-Rohr1994Macromol,Kuebler1997Macromol} concluded that the molecular motions involved in the $\beta$ relaxation are complex and involve coupled small- and large-angle motions of both the side and main-chains. For PB, a combined NSE and BDS study found evidence for cooperative rotations involving several units along the chain, and concluded that the $\beta$ relaxation originated in intramolecular rotational motion of \textit{cis} and \textit{trans} chain units \cite{kremer2012broadband}.

The role of conformational dihedral transitions in the relaxation dynamics of polymers was studied by computer simulations \cite{Paul2004RepProgPhys,Smith2007JPolSci,Bedrov2005PRE}. Atomistic MD simulations of PB \cite{Smith2007JPolSci,Bedrov2005PRE} demonstrated the strong link between conformational dihedral rotation and the $\beta$ relaxation; and simulations of PPG-DME \cite{vogel2008macromol} demonstrated the importance of intramolecular dihedral reorientations in controlling relaxation dynamics. 

We present a detailed study of the molecular weight dependent secondary relaxation behaviour of PMMA and PB, since for these two polymer systems we can access the secondary $\beta$ and $\gamma$ relaxations across the full molecular weight range.

\subsection{Activation enthalpies for $\beta$ and $\gamma$ relaxations}

The $M$-dependent activation enthalpies for $\beta$ (circles) and $\gamma$ (squares) relaxations within the glassy state are shown in Fig.~\ref{fig:fig5new} for PMMA and PB. As described in Section \ref{sec:activation},  the $\beta$ and $\gamma$ relaxation enthalpy data for PB are obtained from BDS measurements, complemented with literature data \cite{arbe1996study,hofmann1996secondary,deegan1995dielectric,richter1992decoupling,korber2017nature,lusceac2005secondary}, shown in Fig.~5 as $\Delta H/\left<\Delta H_{\gamma}\right>$. In this representation PMMA and PB behave similarly, despite the  different chain flexibilities of PMMA and PB. The more flexible nature of PB is reflected in a smaller activation enthalpy (Fig.~\ref{fig:fig5new}). Note that in Ref.~\cite{lusceac2005secondary}, the observed secondary
\mathchardef\mhyphen="2D 
$\gamma$ relaxations are termed $\gamma_A$ (between $T=80\textrm{--}100\,\textrm{K}$) and $\gamma_B$ (between $T=50\textrm{--}65\,\textrm{K}$).


Table~\ref{tab:VFT_arr_params} summarizes the molecular weights and polydispersities for the PMMA samples studied, as well as the the fitting parameters for the VFT and Arrhenius fits of the $\alpha$, $\beta$ and $\gamma$ relaxations shown in Fig.~1a.
\begin{table*}[htbp]
\centering
\begin{tabular}{rcccccccc}
\hline\hline
$M_{w}$ & PDI & $\mathrm{log}_{10}(\tau_0^{\alpha}/\textrm{s})$ & $D$ & $T_0$  & $\mathrm{log}_{10}(\tau_0^{\beta}/\textrm{s})$ & $\Delta H_{\beta}$& $\mathrm{log}_{10}(\tau_0^{\gamma}/\textrm{s})$ & $\Delta H_{\gamma}$ \\ 
g/mol & & & & K && kJ/mol &&kJ/mol \\ 
\hline
 202 &  1 &  -16.1  & 13.5  & 126.0  & -14.2 & 25.2  & - &  - \\ 
 302 &  1 &  -14.3  & 10.0 & 165.7  & -13.1 & 39.8  & -12.0 & 26.0  \\ 
 402 &  1 &  -13.5 & 8.7  & 192.0  & -16.0 & 64.0 & -13.4  & 32.1  \\ 
 660 &  1.21 &  -12.5 & 8.0  & 205.0  & -14.5  & 63.8  & -12.9 & 32.4 \\ 
 840 &  1.44 &  -12.7 & 9.3  & 217.0 & -16.0  & 75.0 & - &  - \\ 
 1900 &  1.10 &  -12.9  & 6.9  & 278.0  & -16.0  & 80.1  & -10.6  & 20.0  \\ 
 4300 &  1.05 &  -14.0  & 6.6  & 310.0  & -15.9  & 80.8  & - &  - \\ 
 9590 &  1.05 &  -12.4  & 4.7  & 330.0  & -15.5  & 78.8 & - &  - \\ 
 39500 &  1.04 &  -10.9  & 2.1  & 366.0  & -15.0  & 75.7 & - &  - \\ 
 90600 &  1.04 &  -11.0  & 2.2  & 366.0  & -14.9 & 75.0 &  -12.6  & 36.9  \\ 
\hline\hline
\end{tabular}
\caption{Weight-averaged molecular weight $M_w$, polydispersity index (PDI=$M_w/M_n$, where $M_n$ is the number-averaged molecular weight),  fit parameters from Vogel-Fulcher-Tammann (VFT) fits of $\tau_\alpha$ data to $\tau_\alpha=\tau_0^{\alpha}\exp{DT_0/(T-T_0)}$, and fit parameters from Arrhenius fits of $\tau_\beta$ and $\tau_\gamma$ data  to $\tau_{\beta}=\tau_0^{\beta}\exp{\Delta H_{\beta}/RT}$; for the PMMA data in Figure 1.}
\label{tab:VFT_arr_params}
\end{table*}

\bibliographystyle{naturemagPDO}
\bibliography{all_references}

\end{document}